\documentclass[twocolumn]{emulateapj}
\usepackage{url}
\usepackage{rotating}

\begin{document}
\newcommand{\gsim}{\lower.7ex\hbox{$\;\stackrel{\textstyle>}{\sim}\;$}}
\newcommand{\lsim}{\lower.7ex\hbox{$\;\stackrel{\textstyle<}{\sim}\;$}}
\newcommand{\lcdm}{$\Lambda$CDM}
\newcommand{\etal}{et {al.}}
\newcommand{\om}{\Omega_{\rm M}}
\newcommand{\ola}{\Omega_{\Lambda}}
\newcommand{\BV}{$B-V {\rm|}_{t=B_{\rm max}}+0.057$}
\newcommand{\mbmax}{m_B^{\rm max}}
\newcommand{\mbmaxp}{m_B^{ \prime}}
\newcommand{\betared}{1.45\pm0.15}
\newcommand{\betablue}{3.55\pm0.46}
\newcommand{\Ntotal}{307}
\newcommand{\sint}{\sigma_{\rm intr}}
\newcommand{\snlist}{SN 1999aa, SN 1999ao, SN 1999ar, SN 1999aw, SN 1999bi, SN 1999bm, SN 1999bn and SN 1999bp}

\title{Improved Cosmological Constraints from New, Old and Combined Supernova Datasets}
\author{
M.~Kowalski\altaffilmark{1},
D.~Rubin\altaffilmark{2,3},
G.~Aldering\altaffilmark{2},
R.~J.~Agostinho\altaffilmark{4},
A.~Amadon\altaffilmark{5},
R.~Amanullah\altaffilmark{6},
C.~Balland\altaffilmark{7},
K. ~Barbary\altaffilmark{2,3},
G.~Blanc\altaffilmark{8},
P.~J.~Challis\altaffilmark{9},
A.~Conley\altaffilmark{10},
N.~V.~Connolly\altaffilmark{11},
R.~Covarrubias\altaffilmark{12},
K.~S.~Dawson\altaffilmark{2},
S.~E.~Deustua\altaffilmark{13},
R.~Ellis\altaffilmark{14},
S.~Fabbro\altaffilmark{15},
V.~Fadeyev\altaffilmark{16},
X.~Fan\altaffilmark{17},
B.~Farris\altaffilmark{18},
G.~Folatelli\altaffilmark{12},
B.~L.~Frye\altaffilmark{19},
G.~Garavini\altaffilmark{20},
E.~L.~Gates\altaffilmark{21},
L.~Germany\altaffilmark{22},
G.~Goldhaber\altaffilmark{2,3},
B.~Goldman\altaffilmark{23},
A.~Goobar\altaffilmark{20},
D.~E.~Groom\altaffilmark{2},
J.~Haissinski\altaffilmark{24},
D.~Hardin\altaffilmark{7},
I.~Hook\altaffilmark{25},
S.~Kent\altaffilmark{26},
A.~G.~Kim\altaffilmark{2},
R.~A.~Knop\altaffilmark{27},
C.~Lidman\altaffilmark{28},
E.~V.~Linder\altaffilmark{6},
J.~Mendez\altaffilmark{29,30},
J.~Meyers\altaffilmark{2,3},
G.~J.~Miller\altaffilmark{31},
M.~Moniez\altaffilmark{24},
A.~M.~Mour\~ao\altaffilmark{15},
H.~Newberg\altaffilmark{32},
S.~Nobili\altaffilmark{20},
P.~E.~Nugent\altaffilmark{2},
R.~Pain\altaffilmark{7},
O.~Perdereau\altaffilmark{24},
S.~Perlmutter\altaffilmark{2,3},
M.~M.~Phillips\altaffilmark{33},
V.~Prasad\altaffilmark{2},
R.~Quimby\altaffilmark{14},
N.~Regnault\altaffilmark{7},
J.~Rich\altaffilmark{5},
E.~P.~Rubenstein\altaffilmark{34},
P.~Ruiz-Lapuente\altaffilmark{30},
F.~D.~Santos\altaffilmark{35},
B.~E.~Schaefer\altaffilmark{36},
R.~A.~Schommer\altaffilmark{37},
R.~C.~Smith\altaffilmark{38},
A.~M.~Soderberg\altaffilmark{14},
A.~L.~Spadafora\altaffilmark{2},
L.-G.~Strolger\altaffilmark{39},
M. ~Strovink\altaffilmark{2,3},
N.~B.~Suntzeff\altaffilmark{40},
N.~Suzuki\altaffilmark{2},
R.~C.~Thomas\altaffilmark{2},
N.~A.~Walton\altaffilmark{41},
L.~Wang\altaffilmark{40},
W.~M.~Wood-Vasey\altaffilmark{9},
J.~L.~Yun\altaffilmark{4}
(Supernova Cosmology Project)
}
\altaffiltext{1}{Institut f\"ur Physik, Humboldt-Universit\"at zu Berlin, Newtonstrasse 15, Berlin 12489, Germany}
\altaffiltext{2}{E. O. Lawrence Berkeley National Laboratory, 1 Cyclotron Rd., Berkeley, CA 94720, USA }
\altaffiltext{3}{Department of Physics, University of California Berkeley, Berkeley, 94720-7300 CA, USA}
\altaffiltext{4}{Centro de Astronomia e Astrof\'isica da Universidade de Lisboa, Observat\'orio Astron\'omico de Lisboa, Tapada da Ajuda, 1349-018 Lisbon, Portugal}
\altaffiltext{5}{DSM/DAPNIA, CEA/Saclay, 91191 Gif-sur-Yvette Cedex, France}
\altaffiltext{6}{Space Sciences Laboratory, University of California Berkeley, Berkeley, CA 94720, USA}
\altaffiltext{7}{LPNHE, CNRS-IN2P3, University of Paris VI \& VII, Paris, France }
\altaffiltext{8}{APC, Universit\'e Paris 7, 10 rue Alice Domon et L\'eonie Duquet, 75205 Paris Cedex 13, France}
\altaffiltext{9}{Center for Astrophysics, Harvard University, 60 Garden Street, Cambridge, MA 02138, USA}
\altaffiltext{10}{Department of Astronomy and Astrophysics, University of Toronto, 60 St. George St., Toronto, Ontario M5S 3H8, Canada}
\altaffiltext{11}{Department of Physics, Hamilton College, Clinton, NY 13323, USA}
\altaffiltext{12}{Observatories of the Carnegie Institution of  Washington,  813 Santa Barbara St., Pasadena,  CA 9110, USA}
\altaffiltext{13}{American Astronomical Society,  2000 Florida Ave, NW, Suite 400, Washington, DC, 20009 USA}
\altaffiltext{14}{California Institute of Technology, E. California Blvd, Pasadena,  CA 91125, USA}
\altaffiltext{15}{CENTRA e Dep. de Fisica, IST, Avenida Rovisco Pais, 1049 Lisbon, Portugal}
\altaffiltext{16}{Department of Physics, University of California Santa Cruz, Santa Cruz, CA 95064, USA}
\altaffiltext{17}{Steward Observatory, the University of Arizona, Tucson , AZ 85721, USA}
\altaffiltext{18}{Department of Physics, University of Illinois at Urbana-Champaign, 1110 West Green, Urbana, IL 61801-3080, USA}
\altaffiltext{19}{Department of Physical Sciences, Dublin City University, Glasnevin, Dublin 9, Ireland}
\altaffiltext{20}{Department of Physics, Stockholm University,  Albanova University Center, S-106 91 Stockholm, Sweden}
\altaffiltext{21}{Lick Observatory, P.O. Box 85, Mount Hamilton, CA 95140, USA}
\altaffiltext{22}{Centre for Astrophysics and Supercomputing, Swinburne University of Technology, John St., Hawthorn, VIC, 3122, Australia}
\altaffiltext{23}{M.P.I.A., K\"onigstuhl~17, 69117 Heidelberg, Germany}
\altaffiltext{24}{Laboratoire de l'Acc\'elerateur Lin\'eaire, IN2P3-CNRS, Universit\'e Paris Sud, B.P. 34, 91898 Orsay Cedex, France}
\altaffiltext{25}{Sub-Department of Astrophysics, University of Oxford, Denys Wilkinson Building, Keble Road, Oxford OX1 3RH, UK}
\altaffiltext{26}{Fermi National Accelerator Laboratory, P.O. Box 500, Batavia, IL 60510, USA}
\altaffiltext{27}{Department of Physics and Astronomy, Vanderbilt University, Nashville, TN 37240, USA}
\altaffiltext{28}{European Southern Observatory, Alonso de Cordova 3107, Vitacura, Casilla 19001, Santiago 19, Chile }
\altaffiltext{29}{Isaac Newton Group, Apartado de Correos 321, 38780 Santa Cruz de La Palma, Islas Canarias, Spain}
\altaffiltext{30}{Department of Astronomy, University of Barcelona, Barcelona, Spain }
\altaffiltext{31}{Southwestern College,  Department of Astronomy, 900 Otay Lakes Road, Chula Vista, CA 91910}
\altaffiltext{32}{Rensselaer Polytechnic Institute, Physics Dept., SC1C25, Troy NY 12180, USA}
\altaffiltext{33}{Las Campanas Observatory, Carnegie Observatories, Casilla 601, La Serena, Chile}
\altaffiltext{34}{Advanced Fuel Research, Inc., 87 Church Street, East Hartford, CT 06108}
\altaffiltext{35}{Department of Physics, Faculty of Sciences, University of Lisbon, Ed. C8, Campo Grande, 1749-016 Lisbon, Portugal}
\altaffiltext{36}{Louisiana State University, Department of Physics and Astronomy,
Baton Rouge, LA, 70803, USA}
\altaffiltext{37}{Deceased}
\altaffiltext{38}{Cerro Tololo Inter-American Observatory, Casilla 603, La Serena, Chile}
\altaffiltext{39}{Department of Physics and Astronomy, Western Kentucky University, Bowling Green, KY, USA}
\altaffiltext{40}{Department of Physics, Texas A\&M University, College Station, TX 77843, USA}
\altaffiltext{41}{Institute of Astronomy, Madingley Road, Cambridge CB3 0HA, UK }

%

\begin{abstract}

We present a new compilation of Type Ia supernovae 
(SNe Ia),  a new dataset of low-redshift nearby-Hubble-flow SNe 
and new analysis
procedures to work with these heterogeneous compilations.
This ``Union'' compilation of 414 SN Ia, which reduces to \Ntotal{} 
SNe after selection cuts, includes the recent large samples of SNe Ia 
from the Supernova Legacy Survey and ESSENCE Survey, the older datasets, as 
well as the recently extended dataset of distant 
supernovae observed with HST.   A single, consistent and blind analysis 
procedure is used for all the various SN Ia subsamples, and a new procedure is 
implemented that consistently weights the heterogeneous data sets and 
rejects outliers.   We present the 
latest results from this Union compilation and discuss the cosmological constraints from this new compilation and its combination with other cosmological measurements (CMB and BAO). The constraint we obtain  from supernovae 
on the dark energy density is $\ola= 0.713^{+0.027}_{-0.029} {\rm (stat)}^{+0.036}_{-0.039} {\rm (sys)} $, for a flat, \lcdm{} Universe. Assuming a constant equation of state 
parameter, $w$, the combined constraints from SNe, BAO and CMB give $w=-0.969^{+0.059}_{-0.063}{\rm (stat)}^{+0.063}_{-0.066} {\rm (sys)}$.
While our results are consistent with a cosmological constant, we obtain only 
relatively weak constraints on a $w$ that varies with redshift. In particular, 
the current SN data do not yet significantly constrain $w$ 
at $z>1$.  
With the addition of our new nearby Hubble-flow SNe Ia, these resulting 
cosmological constraints  are 
currently the tightest available.
\end{abstract}
\section{Introduction}

The evidence for dark energy has evolved from the first hints, for the case of a flat Universe \citep{perl98,garnavich98,schmidt98}, through the 
more definite evidence for the general case of unconstrained curvature \citep{riess98,perl99}, to  the current work which aims to explore the properties of dark energy (for a  review see \citeauthor{schmidt_perl}~\citeyear{schmidt_perl}).
Several new cosmological measurement techniques and several new 
Type Ia 
supernova (SN Ia) datasets have helped begin the laborious 
process of narrowing in on the 
parameters that describe the cosmological model.   
The SN Ia measurements remain a 
key ingredient in all current determinations of cosmological parameters 
(see, e.g., the recent CMB results \citep{dunkley08}). It is therefore 
necessary to understand 
how the current world dataset of SN Ia measurements is constructed, and how it
 can be used coherently, particularly since no one SN Ia sample by itself
 provides an accurate cosmological measurement.

Until recently, the SN Ia compilations (e.g., \citeauthor{riess98}~\citeyear{riess98}, \citeauthor{perl99}~\citeyear{perl99}, \citeauthor{tonry03}~\citeyear{tonry03}, \citeauthor{knop03}~\citeyear{knop03}, \citeauthor{astier05}~\citeyear{astier05} and \citeauthor{essence_wv_07}~\citeyear{essence_wv_07}) primarily consisted of a relatively 
uniform high-redshift ($z \sim 0.5$) dataset from a single study put together 
with 
a low-redshift $(z \sim 0.05)$ sample collected in a different study or studies.  
However, once there
 were several independent datasets at high redshift it became more important
 and interesting to see the cosmological constraints obtainable by combining 
several groups' work.   \citet{riess04,riess06} provided a first compilation 
analysis of this kind, drawing on data chosen from \citet{perl99,riess98,schmidt98,knop03,tonry03} and \citet{barris04}.
  Many of the subsequent cosmology studies have used this compilation 
 as the representation of the SN Ia sample, in particular the 
selection of supernovae that \citet{riess04,riess06} nicknamed the ``Gold'' 
sample. Other recent compilations that have been used are that of
\citet{essence_wv_07} and \citet{davis07}.

At present a number of updates should be made to the SN Ia datasets, and 
a number of analysis issues  should be addressed, including several that 
will recur with every future generation of SN compilations.  These include the
 following major goals:  
\begin{itemize}
\item[(1)]
 It is important to add a new low-redshift SN Ia sample to 
complement the large and rapidly growing number of distant SNe.  
Especially valuable are the SNe in the smooth, nearby Hubble-flow 
($z$ above $\sim 0.02$). Since 
this part of the Hubble diagram is currently not well constrained, 
new nearby SNe lead to a relatively large incremental improvement 
\citep{2006PhRvD..74j3518L}. It is interesting to note, that the largest contribution in this redshift range still comes from 
the landmark Calan/Tololo survey (Hamuy et al, 1993).
 
\item[(2)] The analysis should reflect the heterogeneous nature of the data 
set. In particular, it is important that a sample 
of poorer quality will not 
degrade the impact of the higher quality data, such as the 
 Supernova Legacy Survey (SNLS) and ESSENCE high-redshift 
datasets which have recently been published. 

\item[(3)]  The different supernova datasets should be analyzed with the same 
analysis procedure.  The previous compilations combined measurements 
and peak-magnitude fits that were obtained with 
disparate lightcurve fitting functions and analysis procedures, particularly 
for handling the color correction for both extinction and any intrinsic color 
luminosity relation.

\item[(4)]  A reproducible, well-characterized approach to selecting the good 
SNe Ia, and rejecting the questionable and outlier SNe, should be used.
Previous compilations  relied to a large extent on the heterogeneous
classification information provided by the original authors. 
The selection process was somewhat
subjective: The Gold compilation of \citet{riess04,riess06} excluded SNe
that \citet{knop03} considered comparably well-confirmed SNe Ia. 

\item[(5)]  To the extent possible, 
the analysis should not introduce biases into the fit, 
including some that have only recently been recognized as being present 
in methods of determining extinction properties of SNe Ia.

\end{itemize}

To reach the goal of carrying out these improvements, we present in this paper
	 a new SN compilation, a new nearby-Hubble-flow SN Ia
dataset, and new analysis procedures.   Several 
	 additional smaller enhancements are also presented.

With respect to Goal (1), it is important to note that both nearby and distant 
supernovae are needed to  measure cosmological parameters.
The brightness of nearby supernovae in the Hubble flow  
is compared to that of
high redshift supernovae, which --- following the dynamics of the 
Universe --- might appear dimmer or brighter than expected for a 
reference cosmology.
Nearby SNe 
lightcurves typically have  better observational coverage and 
{ signal-to-noise~ratio} (SNR) than their high-redshift counterparts. 
However, they are significantly 
more difficult to discover since
vast amounts of sky have to be searched 
to obtain a sizable number of supernovae, due to the small volume of the low redshift Universe. 
We present lightcurves  from the Supernova Cosmology Project (SCP) 
Spring 1999 Nearby Supernova 
 Campaign \citep{aldering2000}, 
which consisted primarily of wide-field magnitude limited searches and extensive photometric and 
spectroscopic 
follow-up observations using a large number of ground-based telescopes.
We provide BVRI lightcurves for 8 nearby supernovae in the Hubble flow.

We then address Goal (2) by combining the new data sample with published data 
of nearby and distant supernovae to construct the largest Hubble diagram to 
date (but presumably not for long). In this combination we adjust the weight of SNe belonging to a 
sample to reflect the 
dispersion we determine for the sample. With our prescription, SN samples 
with significant  unaccounted-for statistical or systematic uncertainties are 
effectively deweighted.

All SN lightcurves are fitted consistently in the observer frame 
system using the spectral-template-based fit method 
of \citet{salt} (also known as SALT). Where possible, the original band pass 
functions are used (Goal (3)).

To address Goal (4) we adopt a robust analysis technique based on outlier 
rejection which we show is resilient against contamination. 
The analysis strategy was developed to limit the influence of human 
subjectivity. Spectroscopic classification is arguably the most 
subjective component of SN cosmology  (primarily because of the 
observational 
challenges associated with high redshift supernova spectroscopy) and we avoid 
decisions whether to include a specific SN that are 
based on  spectroscopic 
features that go beyond that of the authors' classification.

Following 
\citet{conley06}, the full analysis chain was
developed in a blind fashion, that is, hiding the best fitting cosmological 
parameters until the analysis was finalized. This helps resist the impulse 
to stop searching for systematic effects once the ``right'' answer is 
obtained.  
We derive constraints  on the cosmological parameters,  taking care 
to test and remove  possible sources of bias introduced in the fitting 
procedure (Goal (5)).

The paper is organized as follows.
In Section \ref{sec:nb99}  we methodically present the data reduction 
and photometric calibration 
of the lightcurves from the SCP Nearby 1999 Supernova Campaign: the reader
more interested in Goals (2-5) and the subsequent 
cosmological analysis might want to only skim this section.
In Section \ref{sec:ltcvs} we combine the new supernovae  with a large set 
of nearby and high redshift supernovae from the literature and fit the 
full set of lightcurves in a consistent manner. 
We then proceed to determine stringent constraints on the dynamics of the 
Universe. Section \ref{sec:par_estimate} explains the methods employed for 
cosmological parameter estimation, which includes blinding the analysis and 
using robust statistics.  We evaluate the systematic errors of the measurements
 in Section \ref{sec:systematics} and summarize the  
resulting constraints on $\om$, $\ola$, $w$, and other parameters in Section 
\ref{sec:cosmo}.

\section{A new sample of nearby supernovae}
\label{sec:nb99}
The SN lightcurve data presented in this paper were obtained as part of
the SCP Nearby 1999 Supernova Campaign \citep{aldering2000}.  The search
portion of this campaign was designed to discover Type~Ia supernovae in
the smooth nearby Hubble flow, and was performed in collaboration with
a number of wide-field CCD imaging teams: EROS-II \citep{blanc04}, NGSS
\citep{ngss}, QUEST-I \citep{questI}, NEAT \citep{neat} and Spacewatch
\citep{spacewatch}.  In some cases the wide-field searches were focused
entirely on supernova discovery (EROS-II and NGSS), while in other cases
the primary data had different scientific goals, such as discovery
of near-earth objects (NEAT, Spacewatch), quasars or micro-lenses
(QUEST-I). The wide-field cameras operated in either point and track
(NGSS, NEAT, EROS-II) or driftscan (QUEST-I, Spacewatch) modes, and
in total covered hundreds of square degrees per night.  Over a two
month period beginning in February 1999, a total of more than 1300
square degrees was monitored for SNe. Since the search was magnitude
limited---no specific galaxies where targeted---it resembles typical
searches for high redshift supernovae. This is important because common
systematics effects, such as Malmquist bias, are then expected to more
nearly cancel when comparing low redshift with high redshift supernovae.

A total of 32 spectroscopically-confirmed SNe were discovered by
the search component of this campaign.  Of these, 22 were of Type~Ia and 14
(of these) were discovered near maximum light, making them useful
for cosmological studies.  In addition, early alerts of potential
SNe by LOTOSS \citep{filippenko01} and similar galaxy-targeted
searches, and the WOOTS-I \citep{galyam} and MSACS \citep{germany04}
cluster-targeted searches, provided a supplement to the primary
sample as the wide-area searches ramped up.  Extensive spectroscopic
screening and follow-up was obtained using guest observer time on
the CTIO~4-m, KPNO~4-m, APO~3.5-m, Lick~3-m, NOT, INT, MDM~2.4-m, ESO~3.6m,
and WHT~4.2-m telescopes. The results of these observations
have been reported elsewhere: \cite{iau_7117,iau_7122,iau_99ar,
iau_99aw,iau_7131,iau_7133,iau_7134,iau_99bi,blanc04,strolger_aw,
spec99aa,spec99ac,folatelli04,garavini07}.  Photometric follow-up
observations were obtained with the LICK~1-m, YALO~1-m, CTIO~0.9-m,
CTIO~1.5-m, MARLY, Danish~1.5-m, ESO~3.6m, KPNO~2.1m, JKT~1-m, CFHT~3.6-m,
KECK-I~10-m, WIYN~3.5-m and MLO~1-m telescopes.  These consist of $UBVRI$
photometry with a nominal cadence of 3-7 days.  The follow-up observations
were performed between February and June 1999 and additional reference
images to determine the contribution of host galaxy light contamination
were obtained in spring 2000.

From this campaign we present $BVRI$ lightcurves for the eight Type~Ia SNe, that fall into 
 the redshift range $0.015$\lsim z \lsim 0.15 and for which we where able to 
obtain enough photometric follow-up data:
SN~1999aa \citep{iau_99aa_a,iau_99aa_b}, SN~1999ao \citep{iau_99ao},
SN~1999ar \citep{iau_99ar}, SN~1999aw \citep{iau_99aw} and  SN~1999bi,
SN~1999bm, SN~1999bn, SN~1999bp \citet{iau_99bi}.  Further
information on these SNe is summarized in Table \ref{tb:sn_coord}.
Photometric data on SN~1999aw have already been published by
\citet{strolger_aw}; here we present a self-consistent reanalysis of
that photometry.


\begin{deluxetable}{c c c c c }
\tablecaption{Summary of SNe coordinates and redshifts. The heliocentric redshift was determined using narrow host-galaxy features for all but one SN. In case of 
SN 1999aw: due to the faintness of its host, the redshift was determined 
from the SN spectra.\label{tb:sn_coord}}
\tablehead{
\colhead{Name}& \colhead{R.A. (J2000)} & \colhead{Decl. (J2000)} &\colhead{Redshift} & \colhead{IAUC}}

\startdata
SN1999aa & 08:27:42.03 & +21$^{\circ}$29'14''8  & 0.0142 & 7180,7109\\
SN1999ao   &06:27:26.37&  $-$35$^\circ$50'24"2 &  0.0539    & 7124 \\
SN1999aw   &11:01:36.37 &$-$06$^{\circ}$06'31''6  & 0.038 & 7130 \\
SN1999ar   & 09:20:16.00&+00$^{\circ}$33'39''6  & 0.1548 & 7125\\
SN1999bi   &11:01:15.76 &$-$11$^{\circ}$45'15''2  & 0.1227 & 7136 \\
SN1999bm   & 12:45:00.84 &$-$06$^{\circ}$27'30''2  & 0.1428 & 7136 \\
SN1999bn   & 11:57:00.40 & $-$11$^{\circ}$26'38''4  &0.1285  & 7136  \\
SN1999bp   & 11:39:46.42& $-$08$^{\circ}$51'34''8 & 0.0770 & 7136\\
\enddata

\end{deluxetable}


\subsection{Data reduction \& photometric calibration} 

The data were preprocessed using standard algorithms for bias and flat
field correction. Additionally, images that showed significant fringing
were corrected by subtracting the structured sky residuals obtained from
the median of fringing-affected images. Reflecting an original goal
of this program --- to obtain data for nearby SNe~Ia matching that of
the high redshift data of \citet{perl99} --- we have employed aperture
photometry to measure the SN lightcurves.  For measurement of moderately
bright point sources projected onto complex host galaxy backgrounds in
fields sparsely covered by foreground stars, aperture photometry has
higher systematic accuracy, but slightly lower statistical 
precision, than PSF fitting.  We used
an aperture radius equal to the FWHM of a point source, as determined
from the field stars in the image. The aperture correction, which is
defined as the fraction of total light which is outside the FWHM radius,
is determined by approximating an infinite aperture by a 4$\times$FWHM
radius aperture.  The aperture correction for a given image is then
obtained by a weighted average for all the stars in the field.

In all, photometric observations employed a total of 12 different
telescopes and 14 different detector/filter systems. This presented
the opportunity to obtain a more accurate estimate of systematic
errors induced by different instrumental setups --- which might
otherwise be masked by apparent internal consistency --- and thereby
come closer to achieving calibration on a system consistent with that of
high-redshift SNe as required for accurate measurement of the cosmological
parameters. Of course the need to account for the specific characteristics
of these many different instruments, and their cross-calibration, made
the calibration a particularly challenging component of this analysis,
which we have addressed in a unique fashion.

Our photometric calibration procedure is subdivided into three parts:
\begin{itemize}
\item[1)]{Determination of zero points, color terms and atmospheric
extinction for photometric nights on telescopes at high-quality sites,
simultaneously employing observations of both \citet{landolt} standard
stars and SN field tertiary standard stars.}
\item[2)]{Use of the tertiary standard stars to simultaneously
determine color terms for all other instruments, and zero points for
all other images.} 
\item[3)]{Determination of SN magnitudes. This includes the SN host
subtraction and photometric correction necessary for non-standard band
passes.}
\end{itemize}
In steps 1 and 2 the robustness of the fits was ensured by heavily
de-weighting significant outliers, using an automated iterative
prescription.

Elaborating further on step~1, the instrumental magnitudes were converted
to magnitudes on the standard $BV(RI)_{KC}$ system using the relation:
\begin{equation} 
m_x = \tilde{m}_x + m_{zp} + k_x \chi + c_x (m_x-m_y),
\label{eq:inst}
 \end{equation}
where $ \tilde{m}_x$ is the instrumental magnitude measured in band
$x$, $m_x$ and $m_y$ are the apparent magnitudes in bands $x$ and $y$,
$\chi$ is the airmass, $m_{zp}$ is the zero point and  $k_x$ and $c_x$
are the atmospheric extinction and filter correction terms for band
$x$.  A simultaneous fit in two bands of standard stars cataloged by
\cite{landolt} and our SN field stars allowed determination of $m_{zp}$,
$k_x$, $\chi$ and $c_x$, as well as $m_x$ and $m_y$ for our tertiary standard
stars.  In total 125 Landolt standards, spread across 16 photometric 
nights, were used for calibration in $B$, $V$, $R$ and $I$, respectively. 
Accordingly, the uncertainties on the night and telescope dependent terms 
$k_x$ and $c_x$ are typically very small. Their covariance with the other 
parameters is properly accounted for.    The catalog of tertiary standard stars
generated as a by-product of this procedure are reported in
Appendix~\ref{sec:field_stars}.

Then, in step~2, the apparent magnitudes from the tertiary standard stars
were used to determine color terms for all remaining instruments and
zero points for all images.  Since $BVRI$ do not require airmass-color
cross terms over the range of airmasses covered by our observations,
and since absorption by any clouds present would be grey, it was
possible to absorb the atmospheric extinction into the zero point of
each image. The catalog of  tertiary standard
stars includes both rather blue and red stars, therefore allowing reliable 
determination of the color terms.
 The color terms obtained for all instruments are summarized
in Table~\ref{tb:beta_term} of Appendix \ref{sec:instruments}.

In order to determine the counts from the SN in a given aperture, the
counts expected from the underlying host galaxy must be subtracted.
In our approach, the image with the SN and the reference
images without SN light are first convolved to have matching point-spread
functions. Stars in the images are used to approximate the PSF as a
Gaussian, which for the purposes of determining the convolution kernel
needed to match one PSF to another is usually adequate.  The instrumental
magnitudes of objects (including galaxies) in the field are then used
to determine the ratio of counts between the images. For a given image,
the counts due to the SN are obtained by subtracting the counts from
the reference image scaled by the ratio of counts averaged over all
objects. Note that with this approach the images are never spatially
translated, thereby minimizing pixel-pixel correlations due to resampling.

\begin{figure*}[t]
\begin{center}
\includegraphics[width=0.6\textwidth]{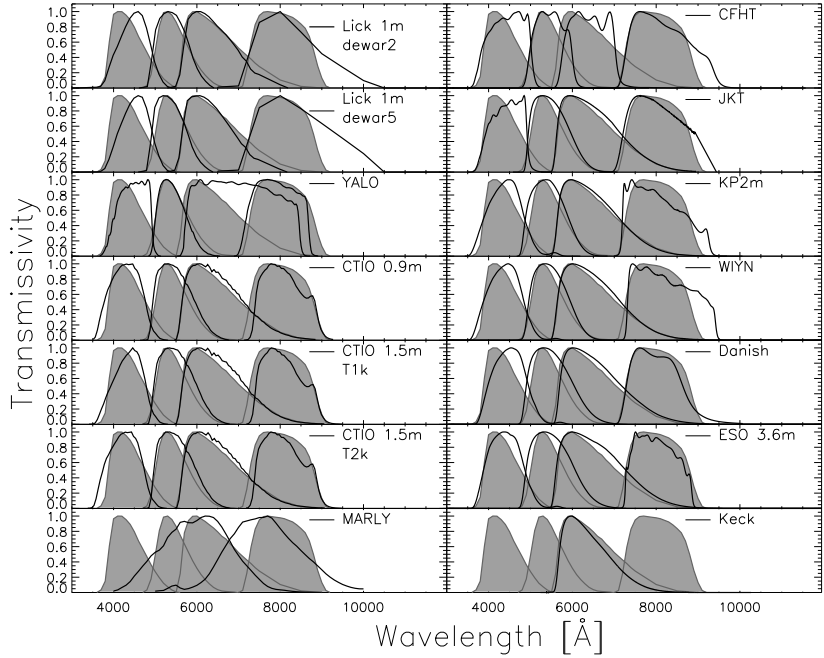}

\caption{Band passes for the various instruments used in the Spring 1999 Nearby Supernova Campaign. 
For comparison, the filled regions represent the pass band 
transmission functions of the \citet{bessel} system.}
\end{center}
\label{bandpass}
\end{figure*}

Several contributions to the uncertainty were evaluated, and added in
quadrature: photon statistics, uncertainties in the image zero points and
uncertainties in the scaling between reference and SN image. Additionally,
possible systematic errors introduced during sky subtraction and
flat-fielding are evaluated using field stars. The variance of field star
residuals is used to rescale all uncertainties. Additionally, an error
floor is determined for all instruments by investigation of the variance
of the residuals as a function of the calculated uncertainty. Such an
error might occur due to large scale variation in the flat-field. An
appropriate error floor was found to be typically 1-2~\% of the signal
counts.

\subsection{Band pass determination}
The band passes for all telescopes have to be established in 
order to correct for potential mismatches to the Landolt/Bessell system \citep{landolt,bessel}.

The band pass is the product of the quantum efficiency of the CCD, 
the filter transmission curve, the atmospheric transmission and the 
reflectivity of the telescope mirrors. 
Figure \ref{bandpass} shows the band pass curves for the various instruments 
used in this work. The relevant data were obtained either from the instrument 
documentation or through private communication.

We test for consistency of the band passes using synthetic photometry (for a related study see \citet{stritzinger}).
For this, stellar spectra which best match the published $UBVRI$ colors of
our standard stars are selected from the catalog of \citet{spec_catalog}.
The spectra that best match the published colors of the standard stars
are further adjusted using cubic splines to exactly match the published
colors.  For instruments without standard star observations, a second
catalog is generated using our determination of $BVRI$ magnitudes of field
stars. With the spectra of standard and field stars at hand we perform
synthetic photometry for the various band pass functions. The band pass
functions are then shifted in central wavelength by $\Delta\lambda$
until they optimally reproduce the observed instrumental magnitudes.
The change in color-term, $c_x$, when shifting the pass band is ${\rm d}c_x /
{\rm d} \lambda \approx 0.001,0.0008,0.0005,0.0003 $ [1/\AA] for $B$,
$V$, $R$ and $I$, respectively.  An alternative procedure is to evaluate
the color terms for a given band pass in an analogous way as for the
observed magnitudes (see  equation \ref{eq:inst}). We  then determine
the wavelength shift to apply to the band passes in order to reproduce
the instrumental color terms. The two approaches agree on average to
within 1~\AA\ with an RMS of about 20~\AA. The results are summarized
in  the Table \ref{tb:shift} of Appendix \ref{sec:instruments}. The associated 
systematic uncertainty on the photometric zero point due to this shift 
depends on the color of the object and for $B-V \le 1$ will remain 
below 0.02 mag.

\newcommand{\pfracd}{0.32}
\begin{figure*}[htb]
\begin{center}

\includegraphics[width=\pfracd\textwidth]{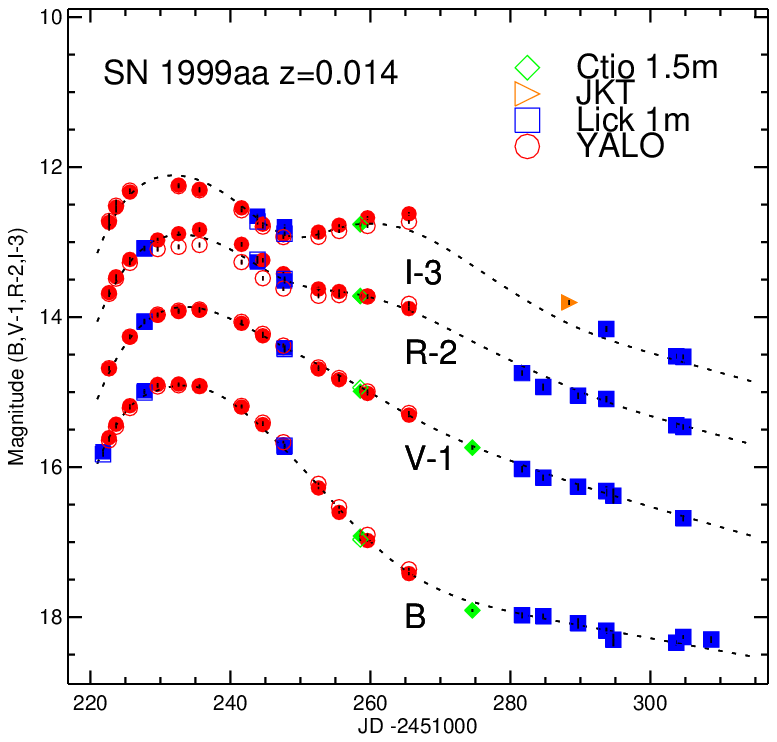}
\includegraphics[width=\pfracd\textwidth]{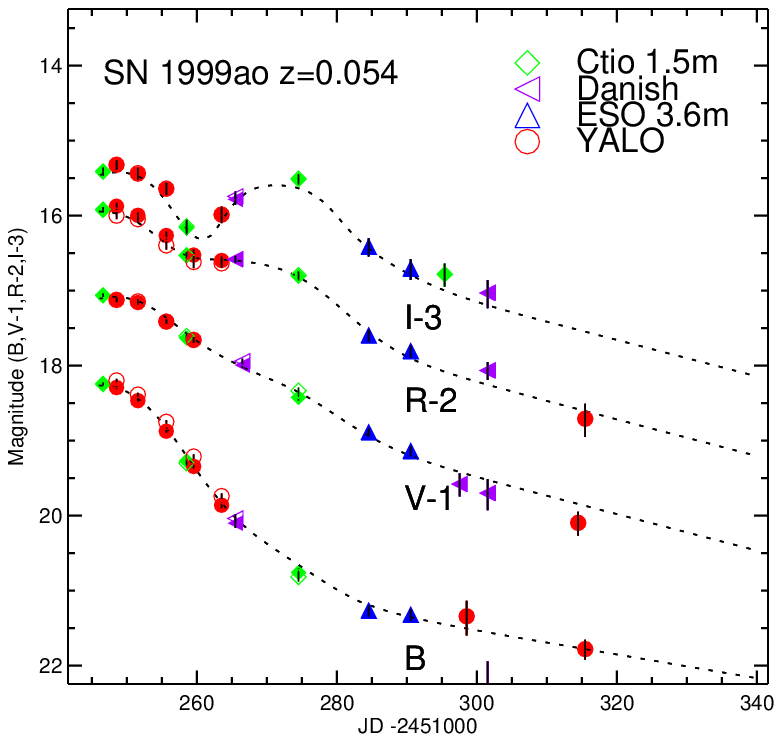}
\includegraphics[width=\pfracd\textwidth]{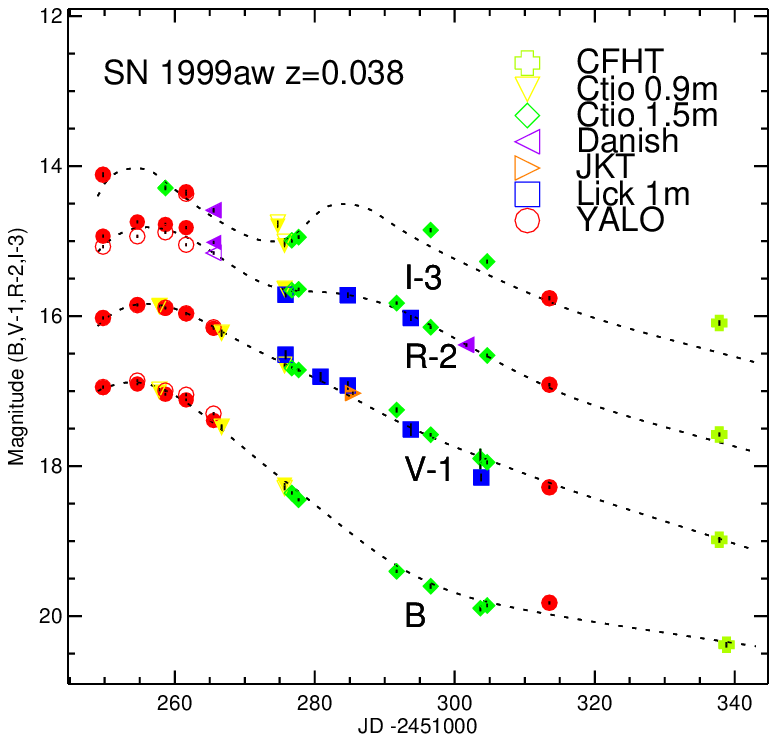}

\vspace{0.2cm}

\includegraphics[width=\pfracd\textwidth]{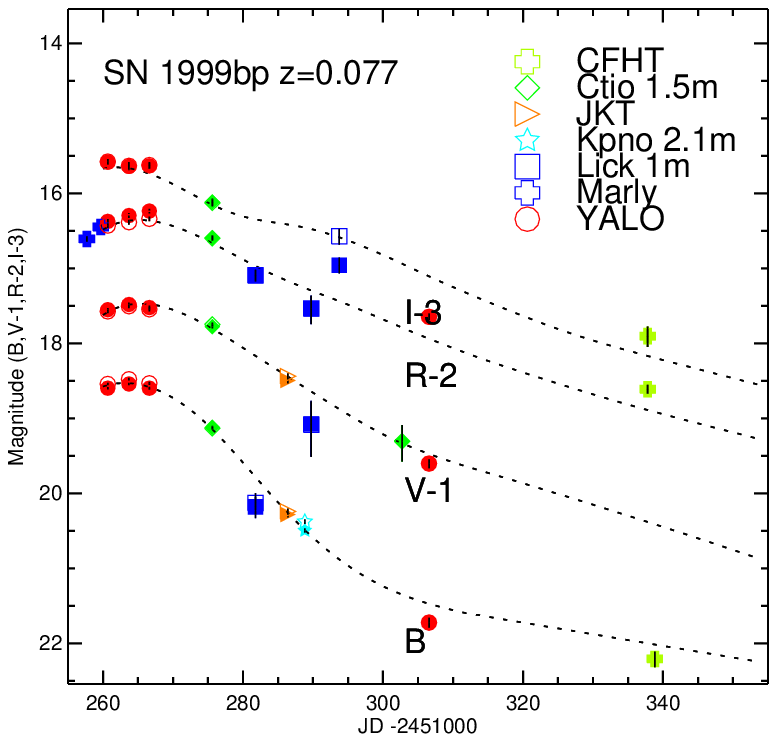}
\includegraphics[width=\pfracd\textwidth]{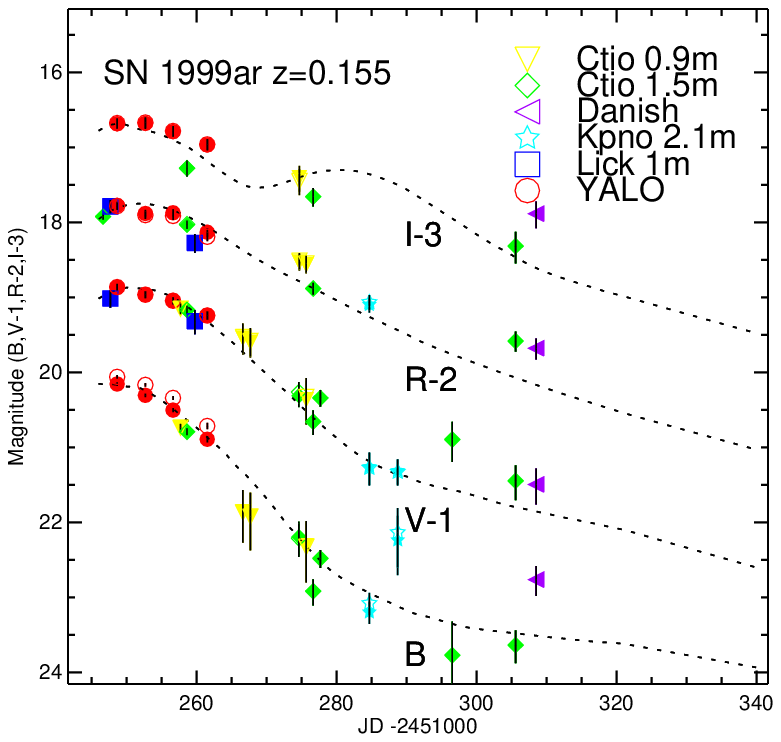}
\includegraphics[width=\pfracd\textwidth]{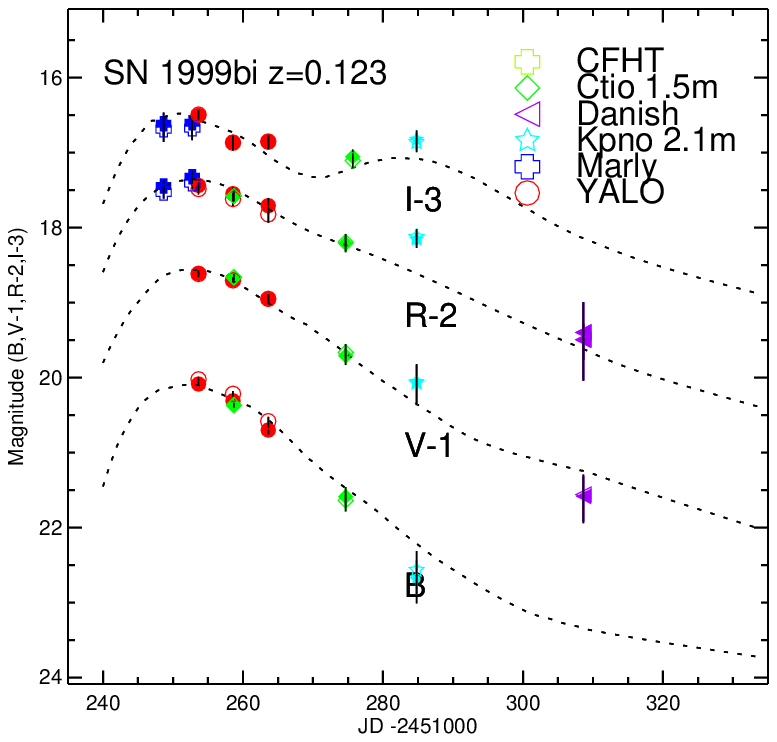}

\vspace{0.2cm}
\hspace{-6cm}
\includegraphics[width=\pfracd\textwidth]{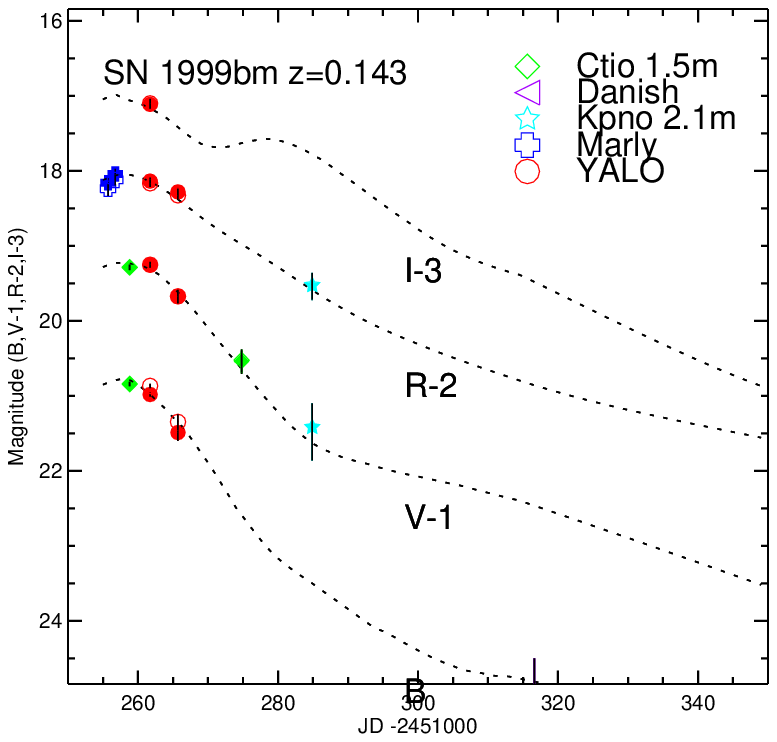}
\includegraphics[width=\pfracd\textwidth]{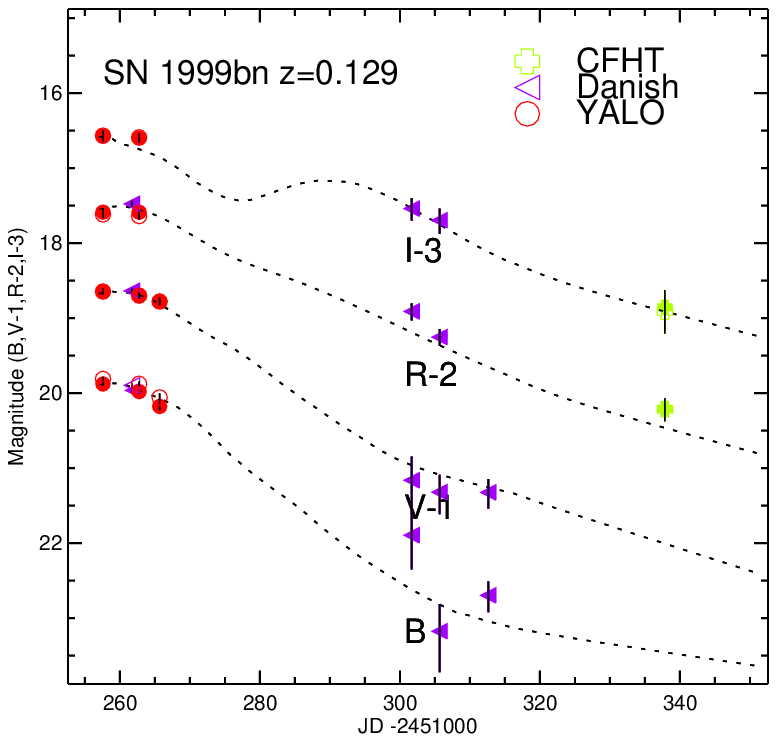}

\end{center}
\caption{SNe lightcurves of the SCP Nearby 1999 campaign. The filled symbols represent the S-corrected data, the empty symbols the raw photometric data. Both the 
S-corrected data as well as the model parameterization (dashed line) are 
shown to guide the eye only and are not used any further in the 
remaining paper.}
\label{fig:ltcv}
\end{figure*}

\newcommand{\xdiv}{0.5}
\newcommand{\twidth}{0.78}
\begin{figure*}
\begin{minipage}[b]{0.5\linewidth} 
\begin{center}
\includegraphics[width=\twidth\textwidth]{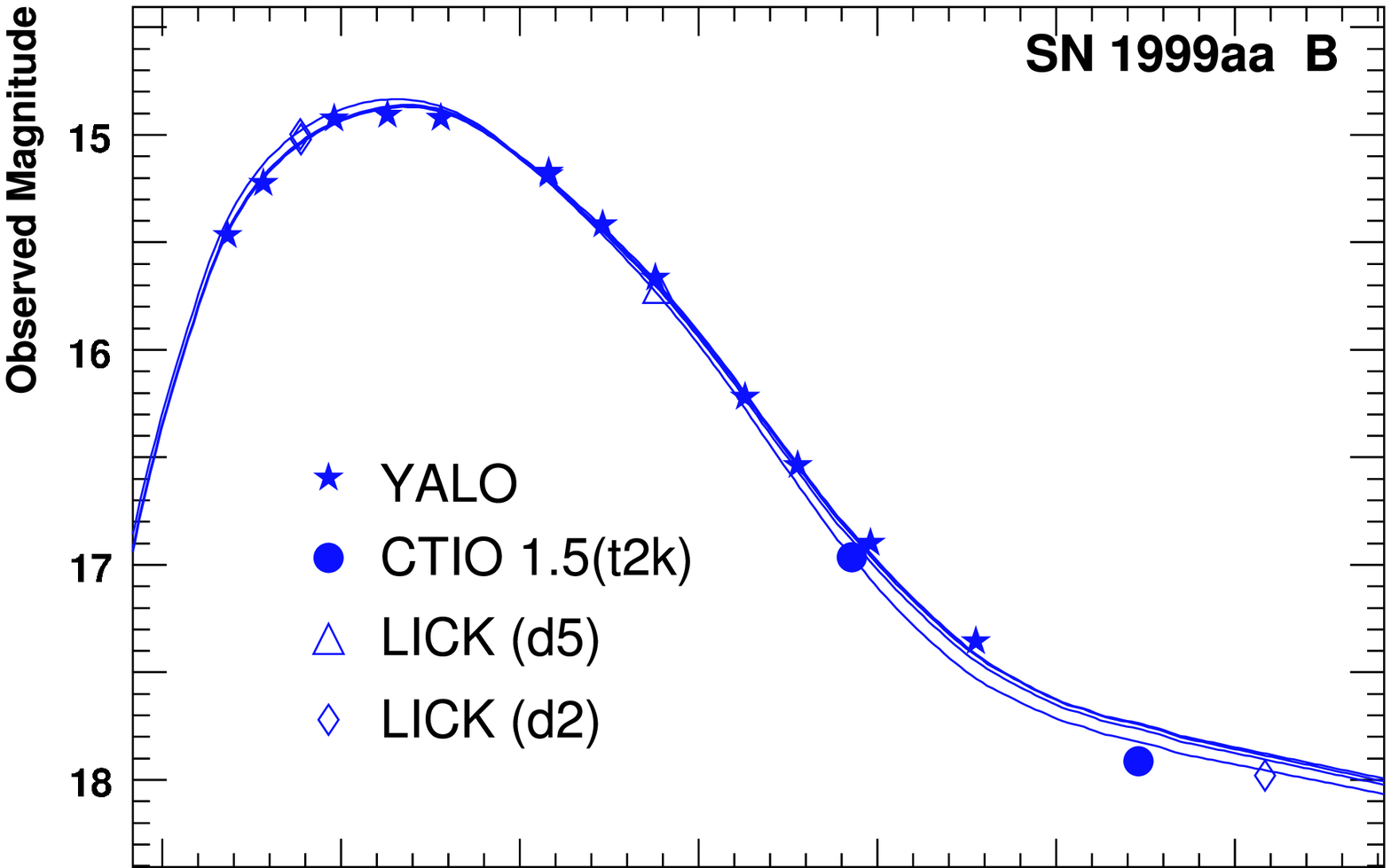}
\end{center}
\end{minipage}
\begin{minipage}[b]{0.5\linewidth} 
\begin{center}
\includegraphics[width=\twidth\textwidth]{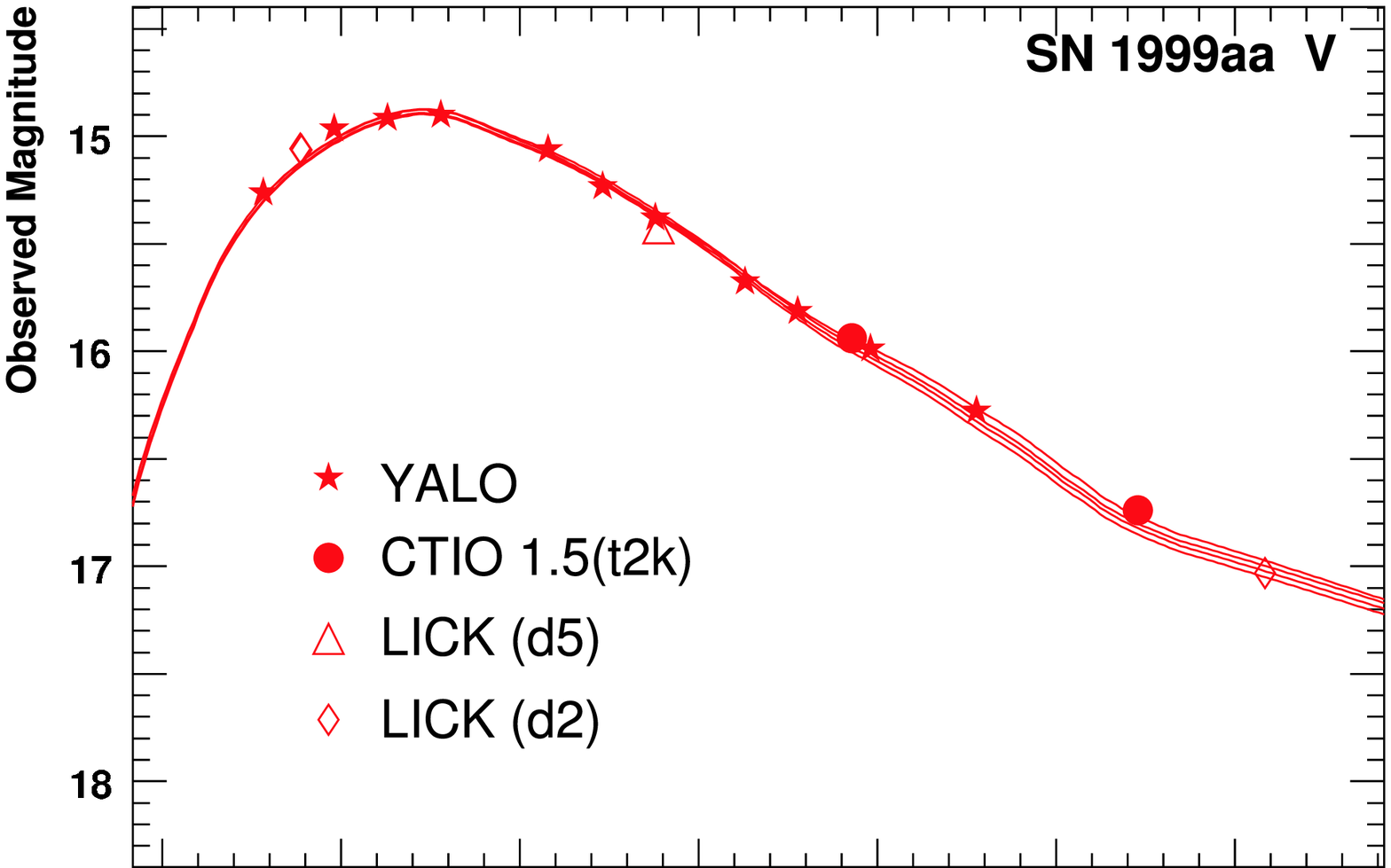}
\end{center}
\end{minipage}
\begin{minipage}[b]{0.5\linewidth} 
\begin{center}
\includegraphics[width=\twidth\textwidth]{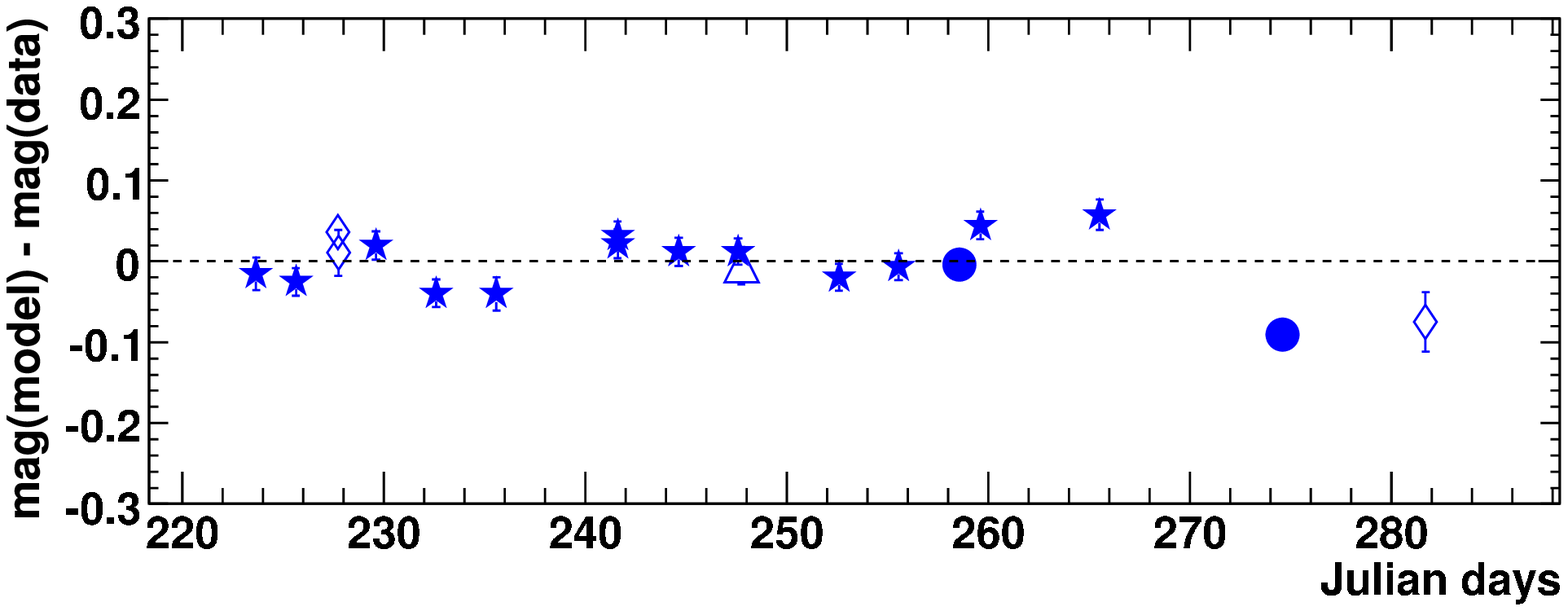}
\end{center}
\end{minipage}
\begin{minipage}[b]{0.5\linewidth} 
\begin{center}
\includegraphics[width=\twidth\textwidth]{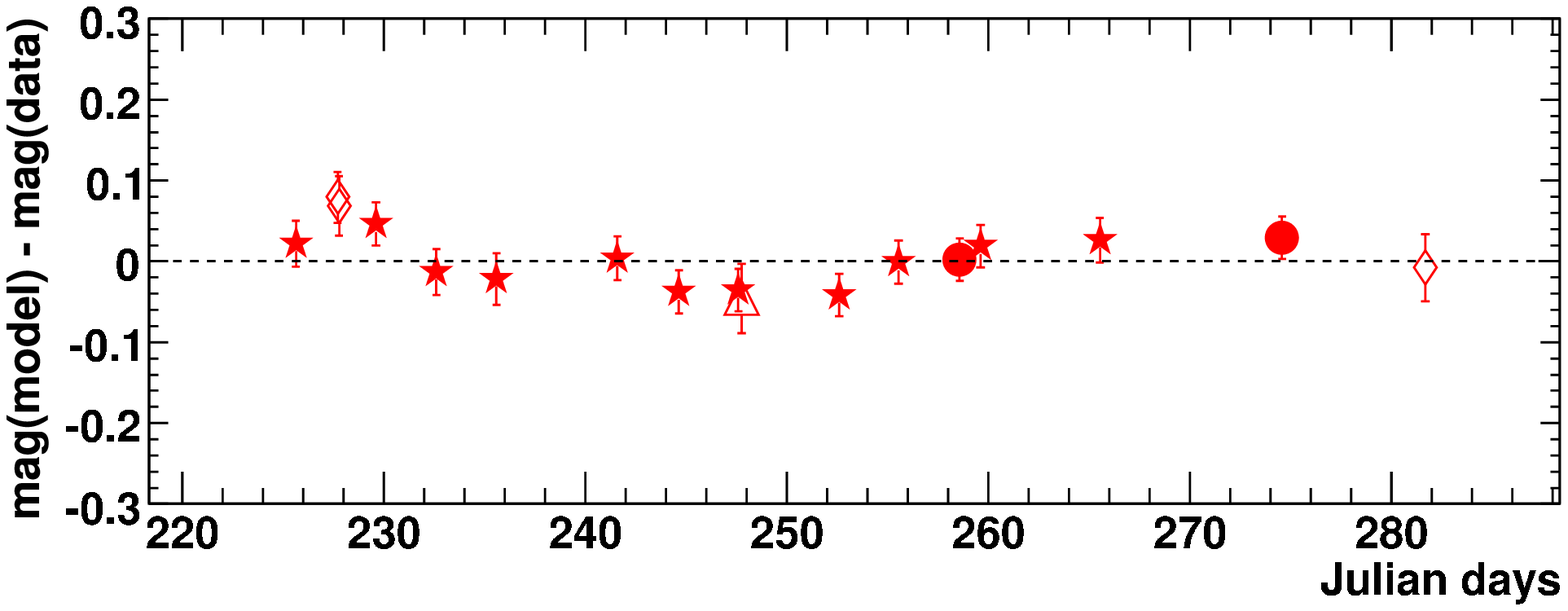}
\end{center}
\end{minipage}
\end{figure*}

\begin{figure*}
\begin{minipage}{\xdiv\linewidth} 
\begin{center}
\includegraphics[width=\twidth\textwidth]{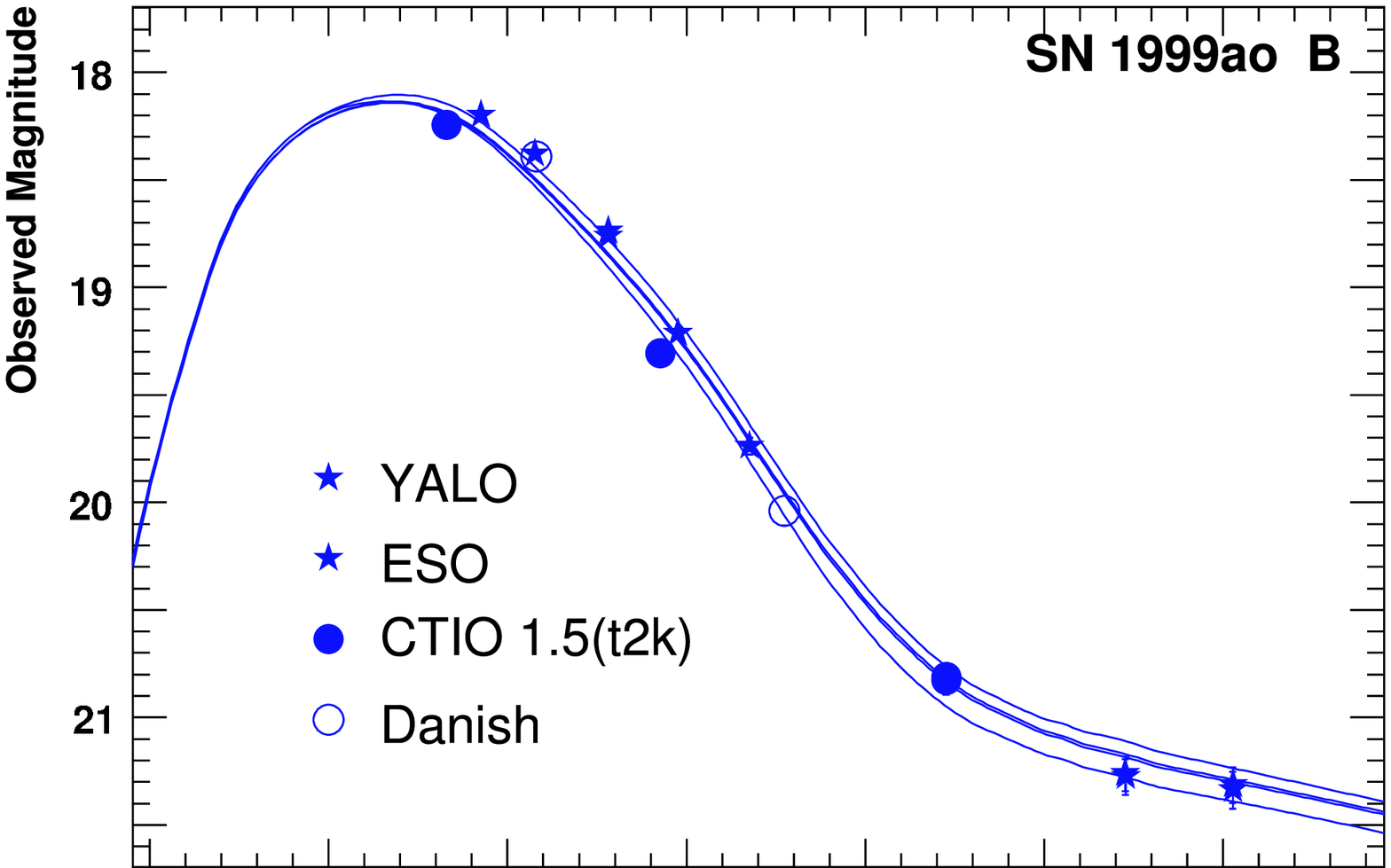}
\end{center}
\end{minipage}
\begin{minipage}{\xdiv\linewidth} 
\begin{center}
\includegraphics[width=\twidth\textwidth]{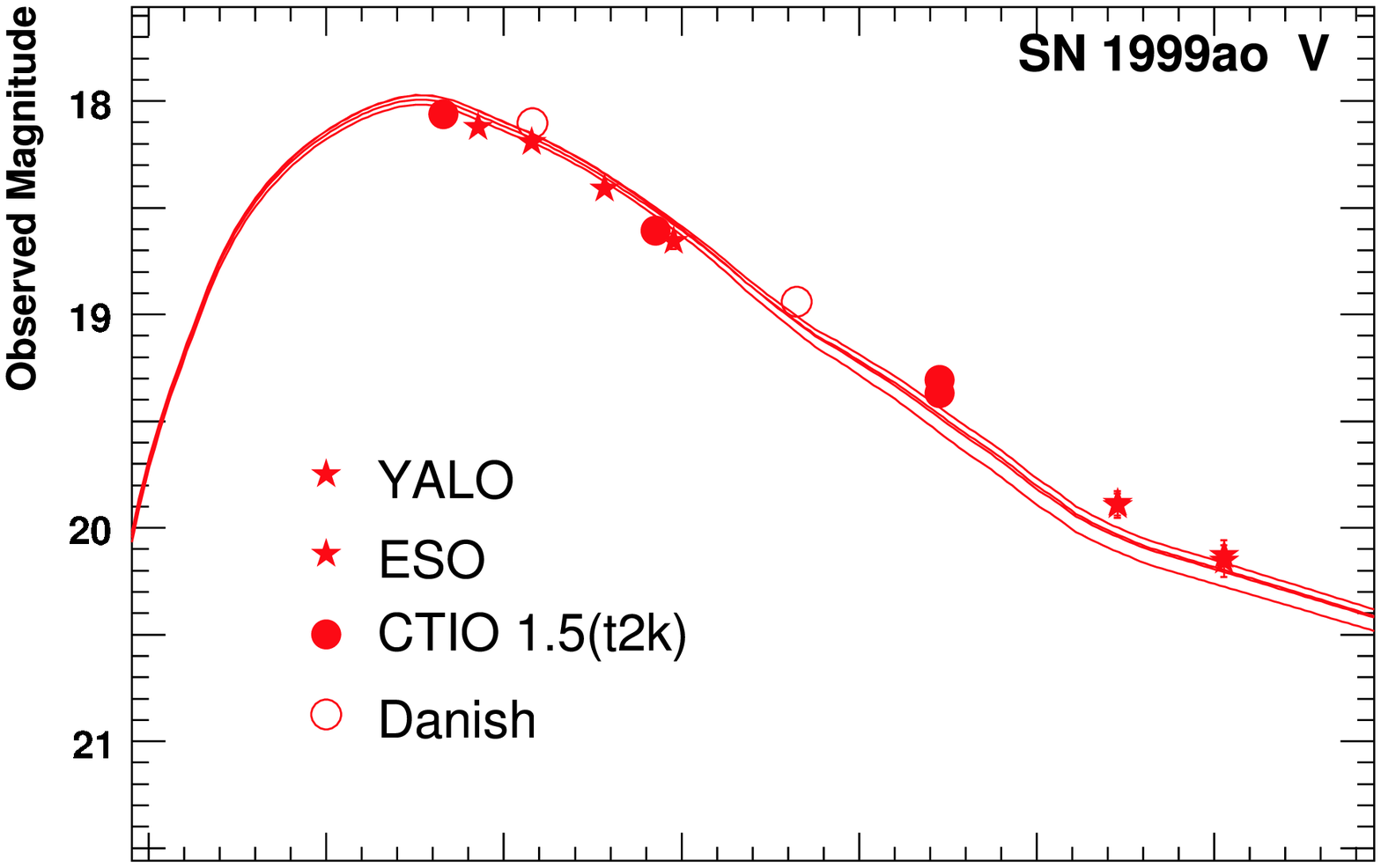}
\end{center}
\end{minipage}
\begin{minipage}{0.5\linewidth} 
\begin{center}
\includegraphics[width=\twidth\textwidth]{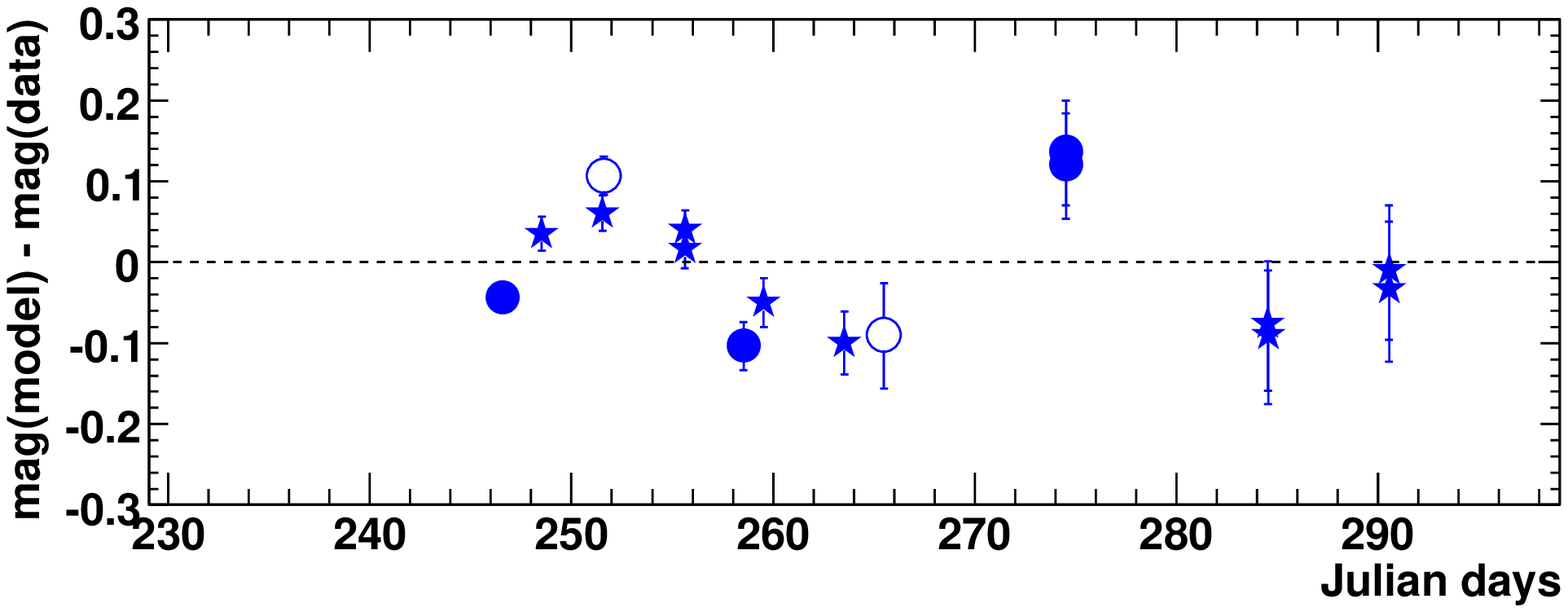}
\end{center}
\end{minipage}
\begin{minipage}{0.5\linewidth} 
\begin{center}
\includegraphics[width=\twidth\textwidth]{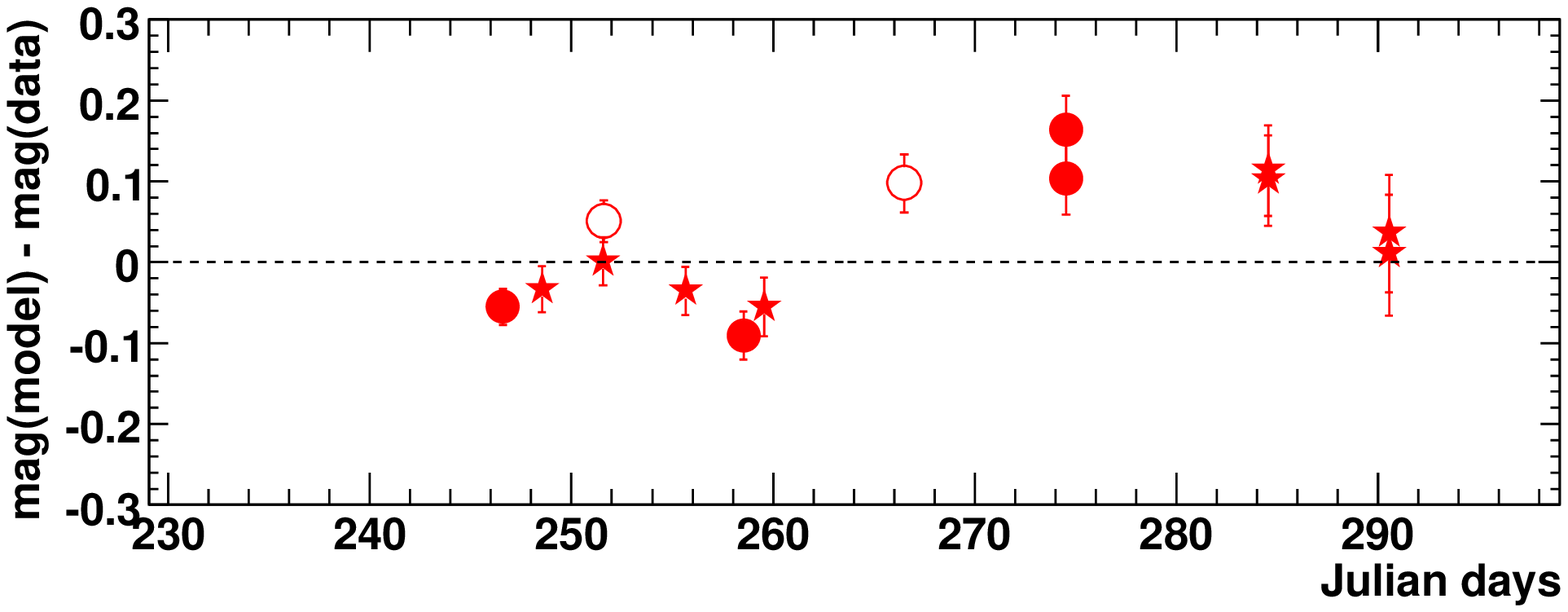}
\end{center}
\end{minipage}
\end{figure*}

\begin{figure*}
\begin{minipage}{0.5\linewidth} 
\begin{center}
\includegraphics[width=\twidth\textwidth]{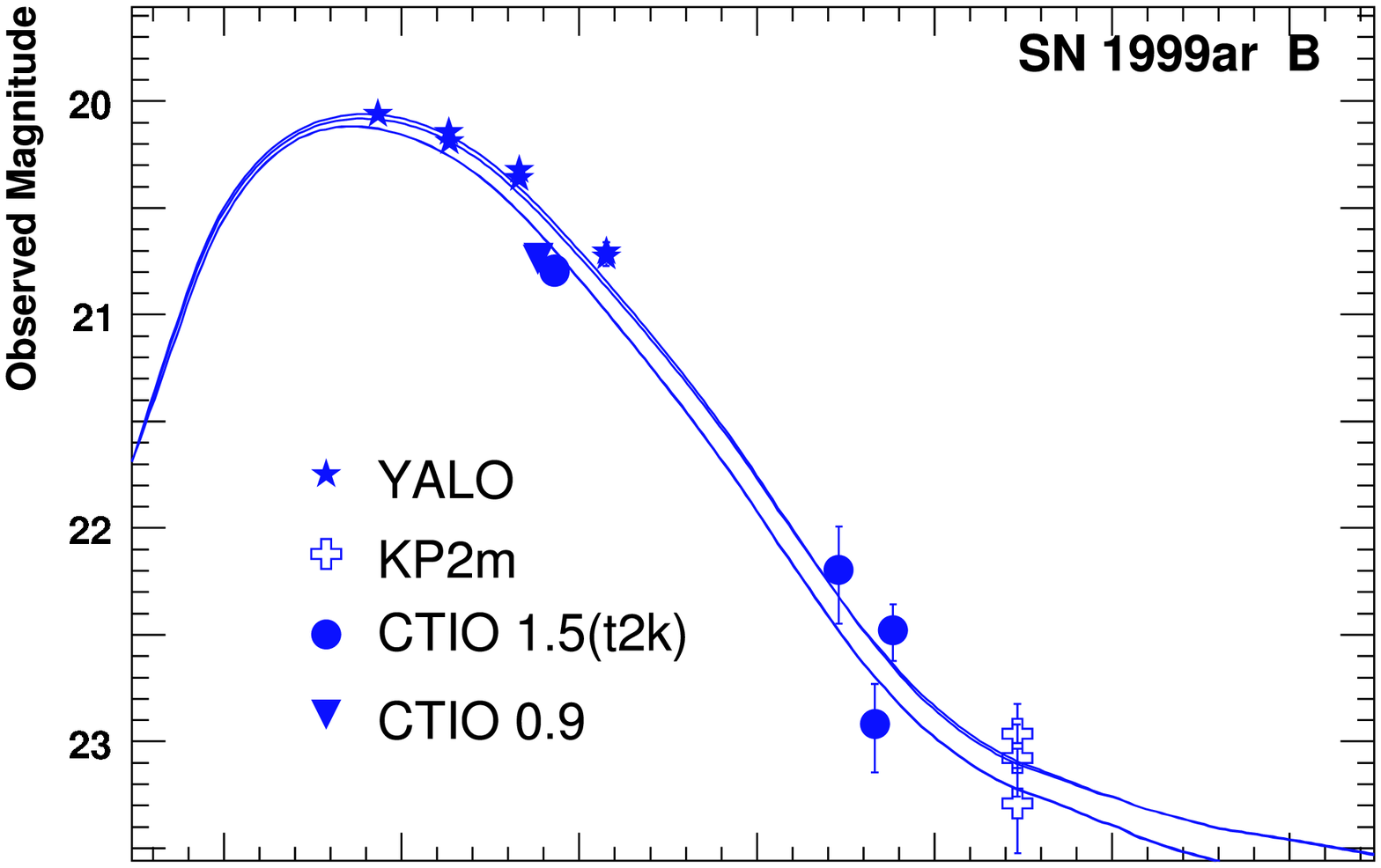}
\end{center}
\end{minipage}
\begin{minipage}{0.5\linewidth} 
\begin{center}
\includegraphics[width=\twidth\textwidth]{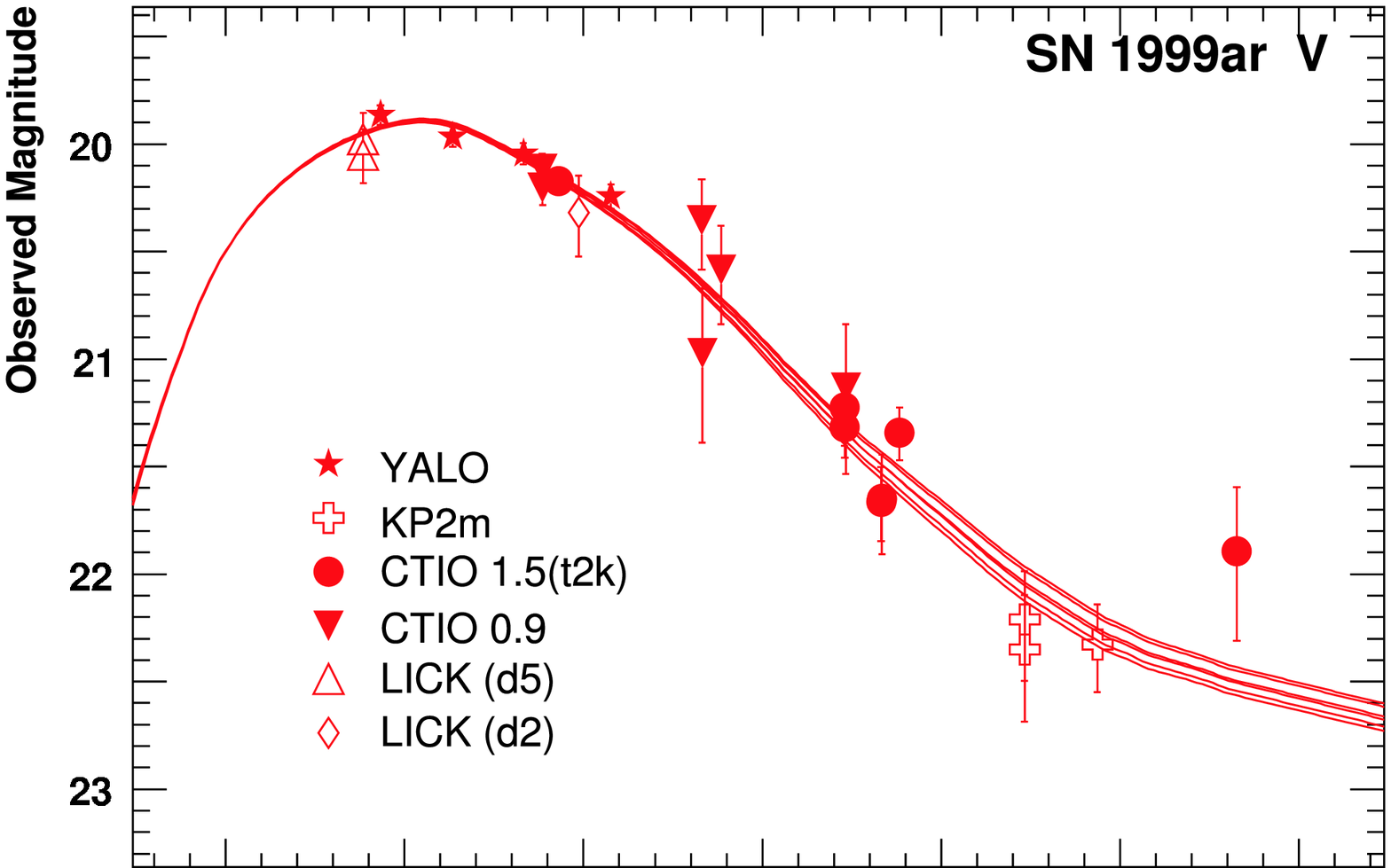}
\end{center}
\end{minipage}

\begin{minipage}{0.5\linewidth} 
\begin{center}
\includegraphics[width=\twidth\textwidth]{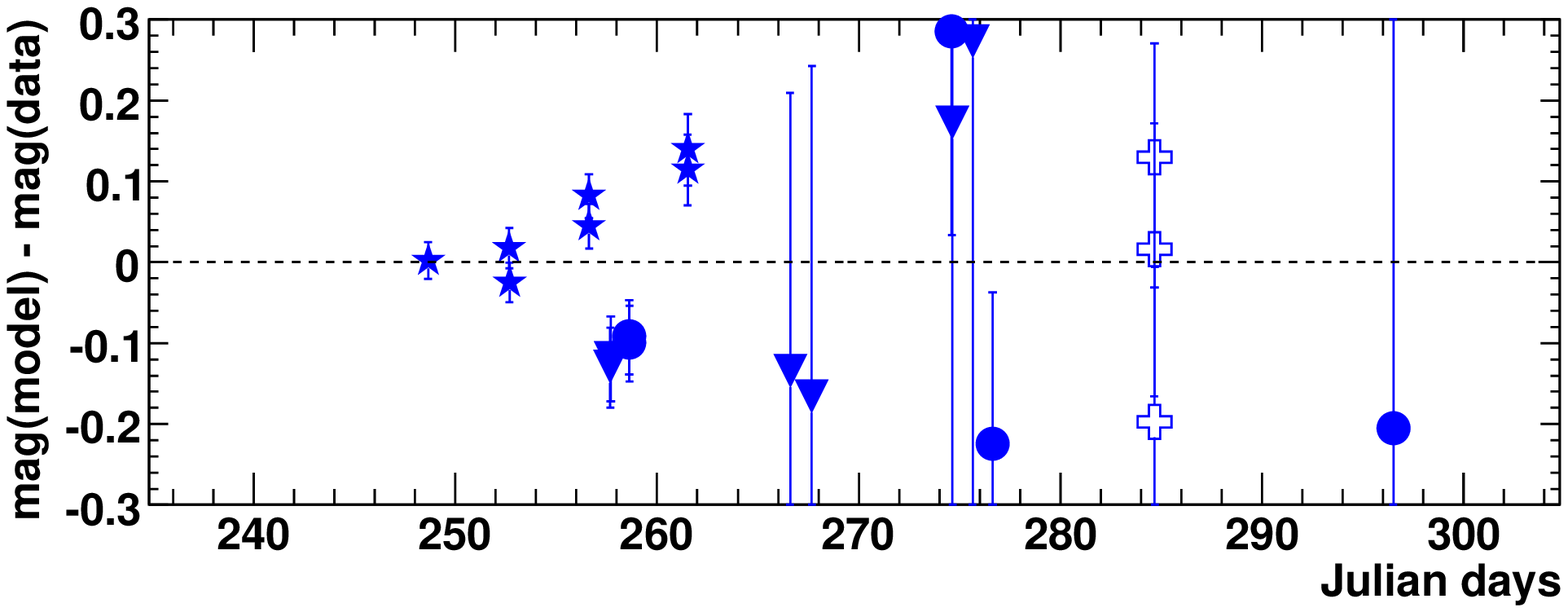}
\end{center}
\end{minipage}
\begin{minipage}{0.5\linewidth} 
\begin{center}
\includegraphics[width=\twidth\textwidth]{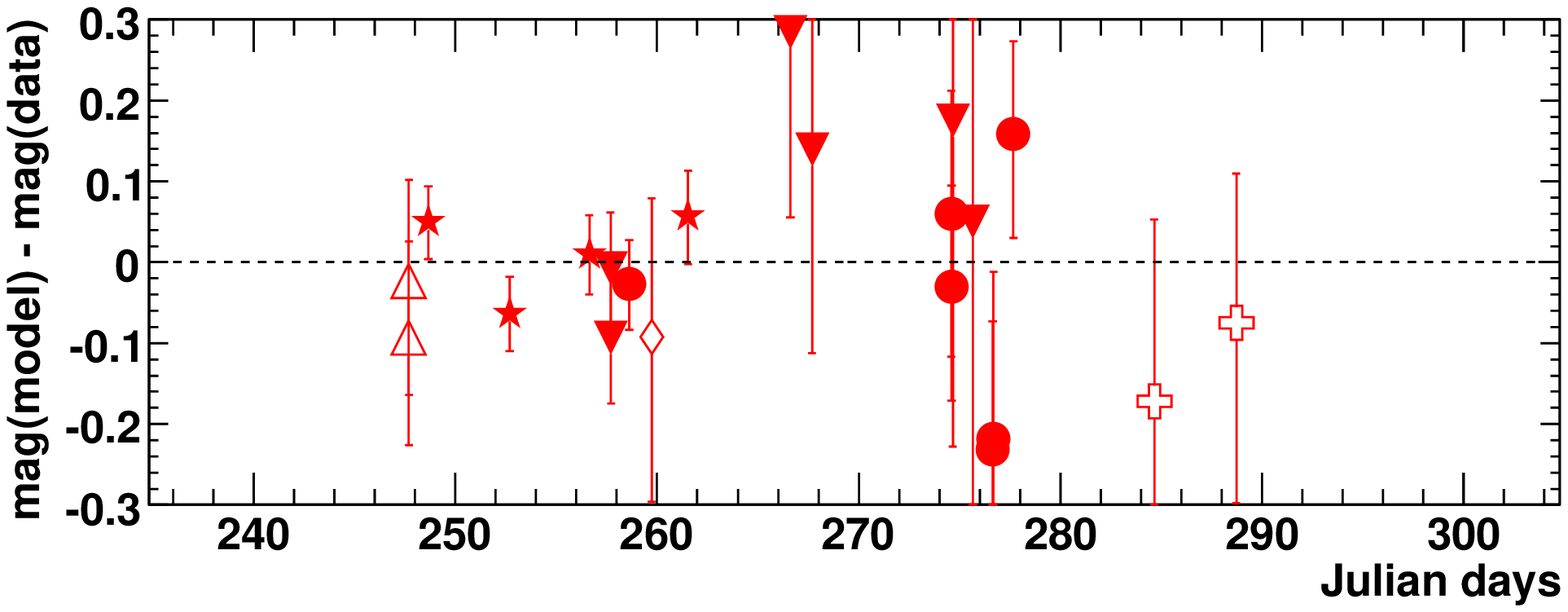}
\end{center}
\end{minipage}
\caption{B and V lightcurves and residuals. The multiple curves  
represent the model predictions for the different band passes, 
 and are obtained by integrating the product of passband and the 
redshifted spectral-template. \label{fig:saltlc}}
\end{figure*}

\begin{figure*}
\begin{minipage}{0.5\linewidth} 
\begin{center}
\includegraphics[width=\twidth\textwidth]{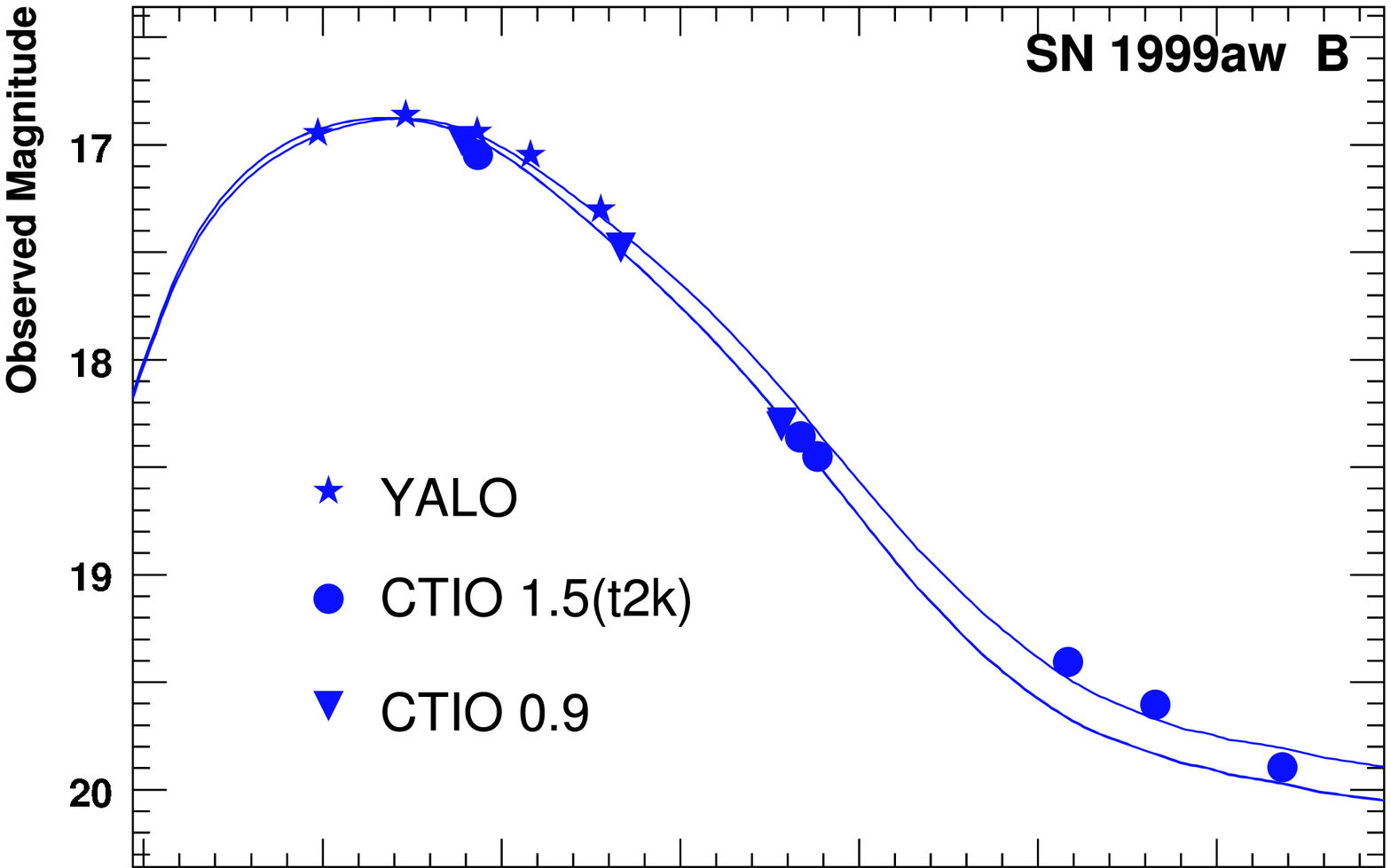}
\end{center}
\end{minipage}
\begin{minipage}{0.5\linewidth} 
\begin{center}
\includegraphics[width=\twidth\textwidth]{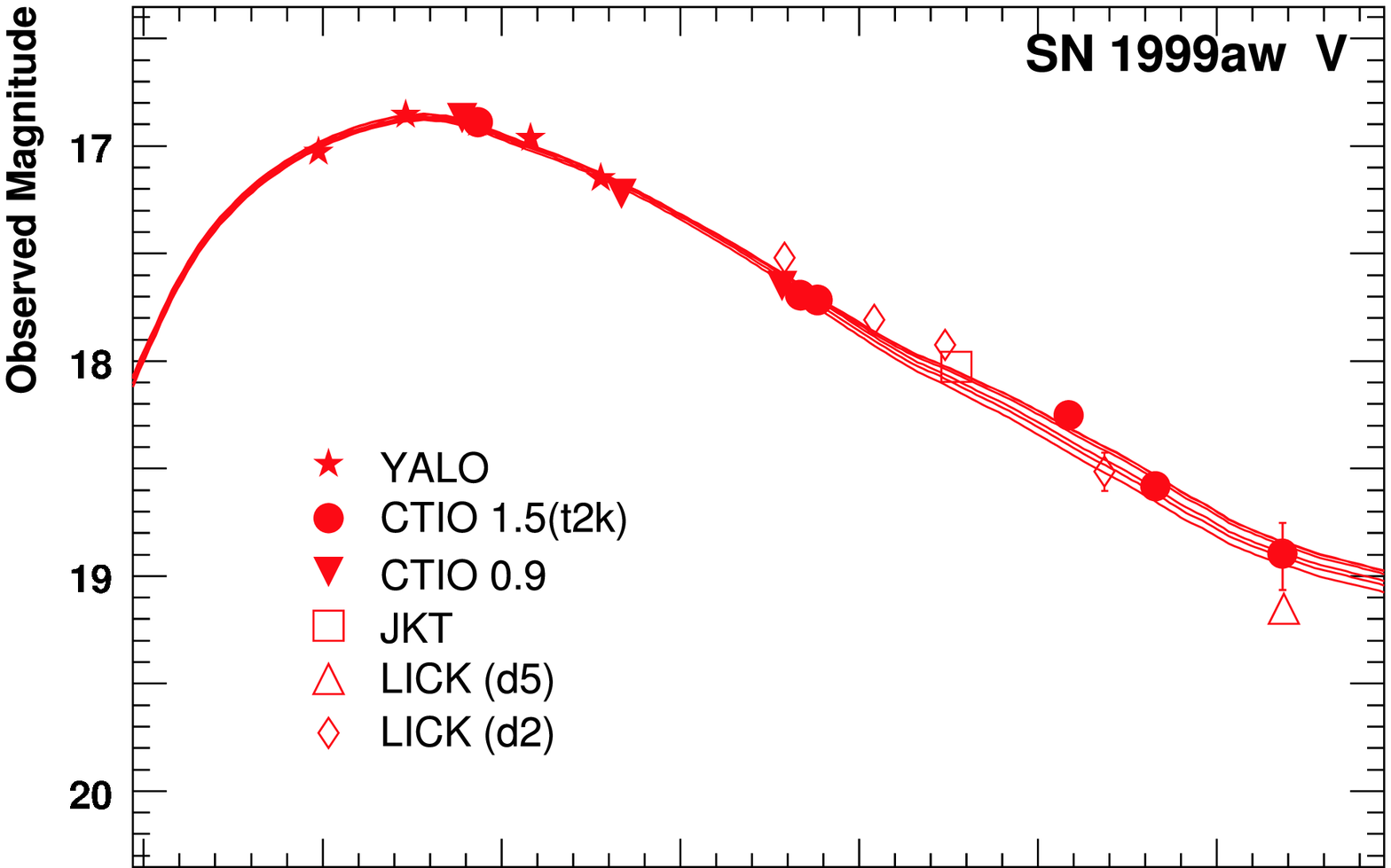}
\end{center}
\end{minipage}

\begin{minipage}{0.5\linewidth} 
\begin{center}
\includegraphics[width=\twidth\textwidth]{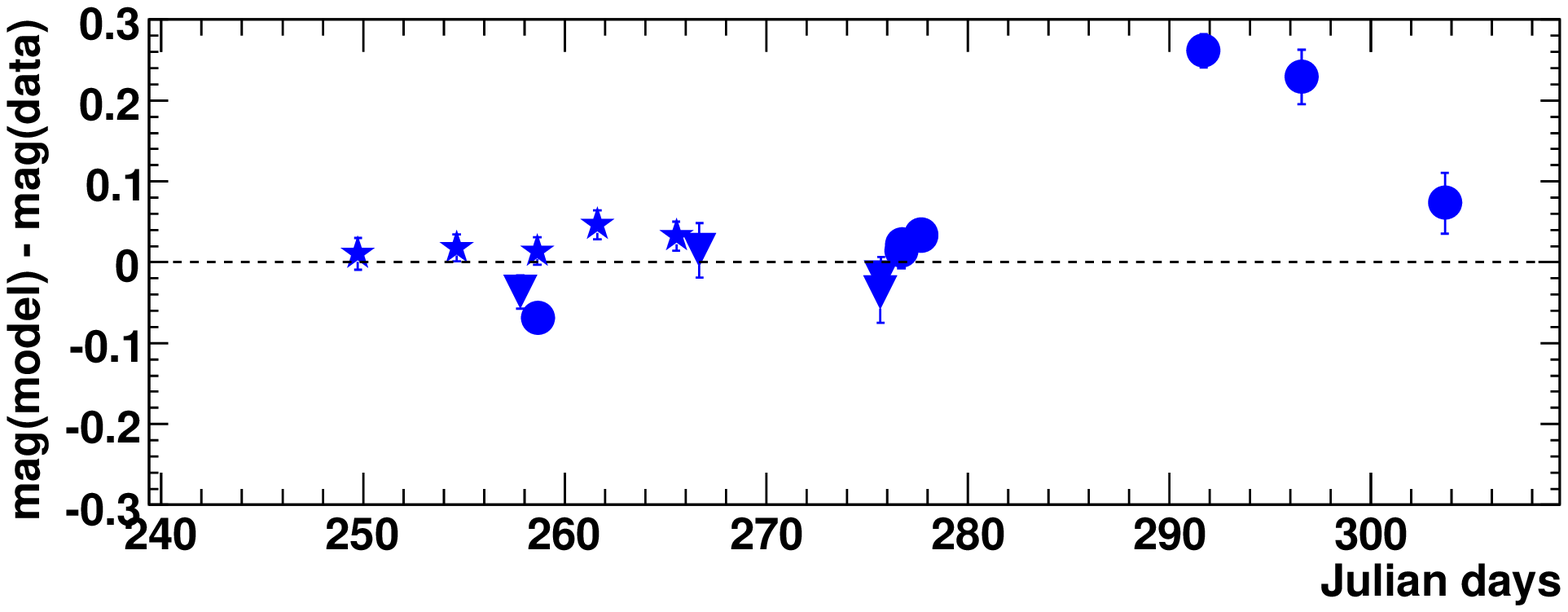}
\end{center}
\end{minipage}
\begin{minipage}{0.5\linewidth} 
\begin{center}
\includegraphics[width=\twidth\textwidth]{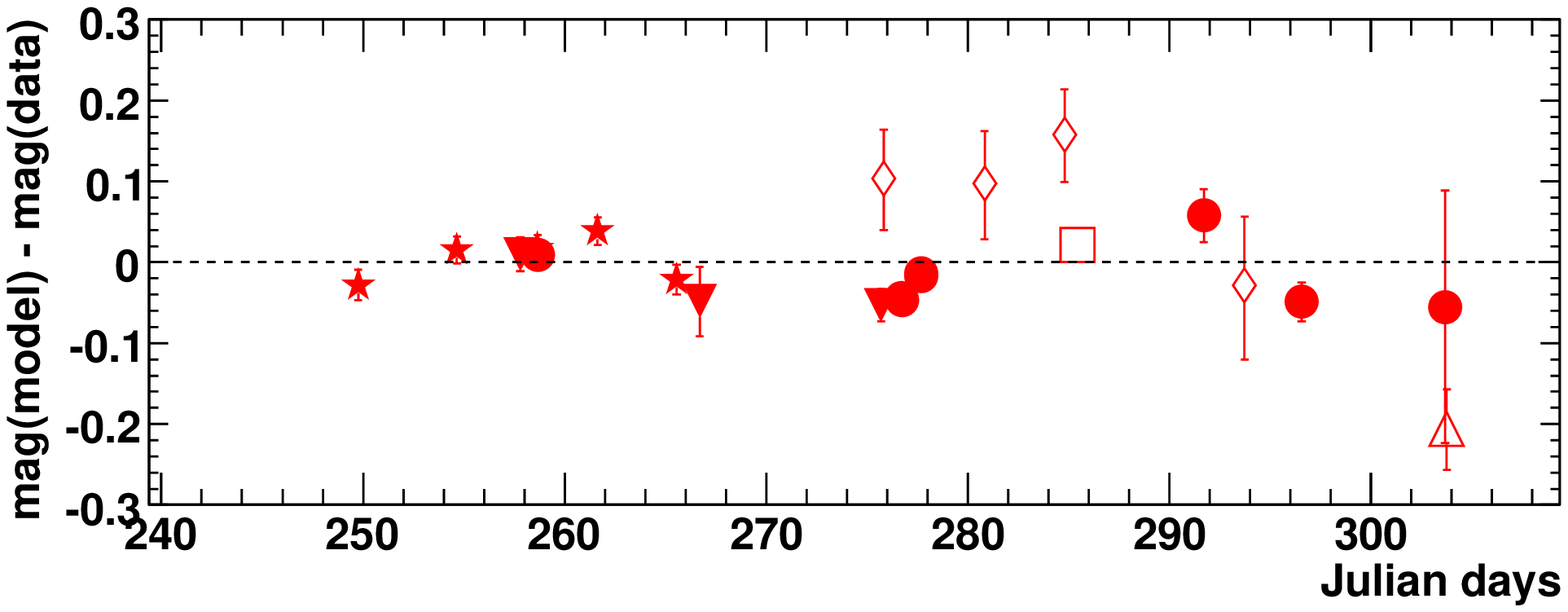}
\end{center}
\end{minipage}
\end{figure*}

\begin{figure*}
\begin{minipage}{0.5\linewidth} 
\begin{center}
\includegraphics[width=\twidth\textwidth]{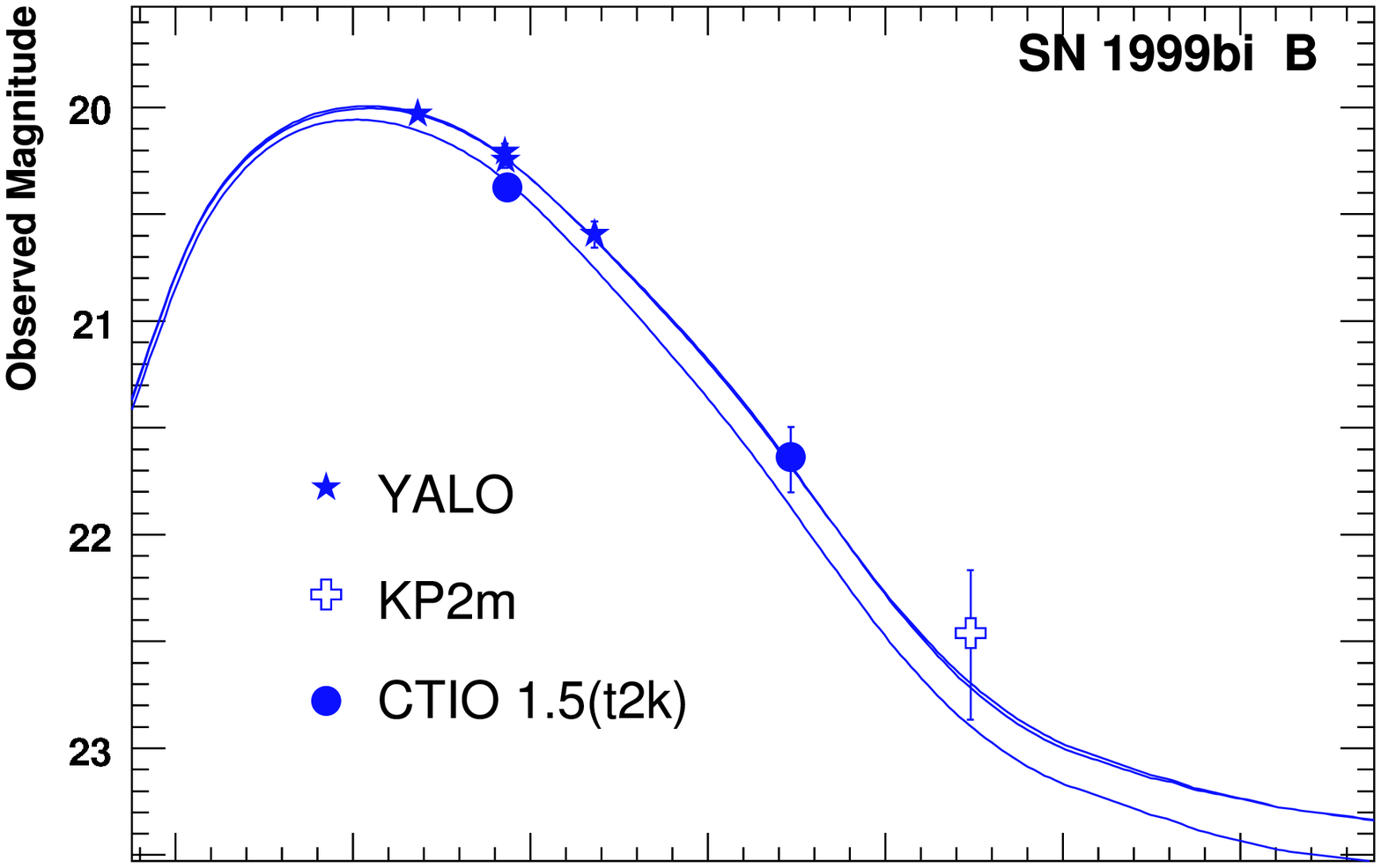}
\end{center}
\end{minipage}
\begin{minipage}{0.5\linewidth} 
\begin{center}
\includegraphics[width=\twidth\textwidth]{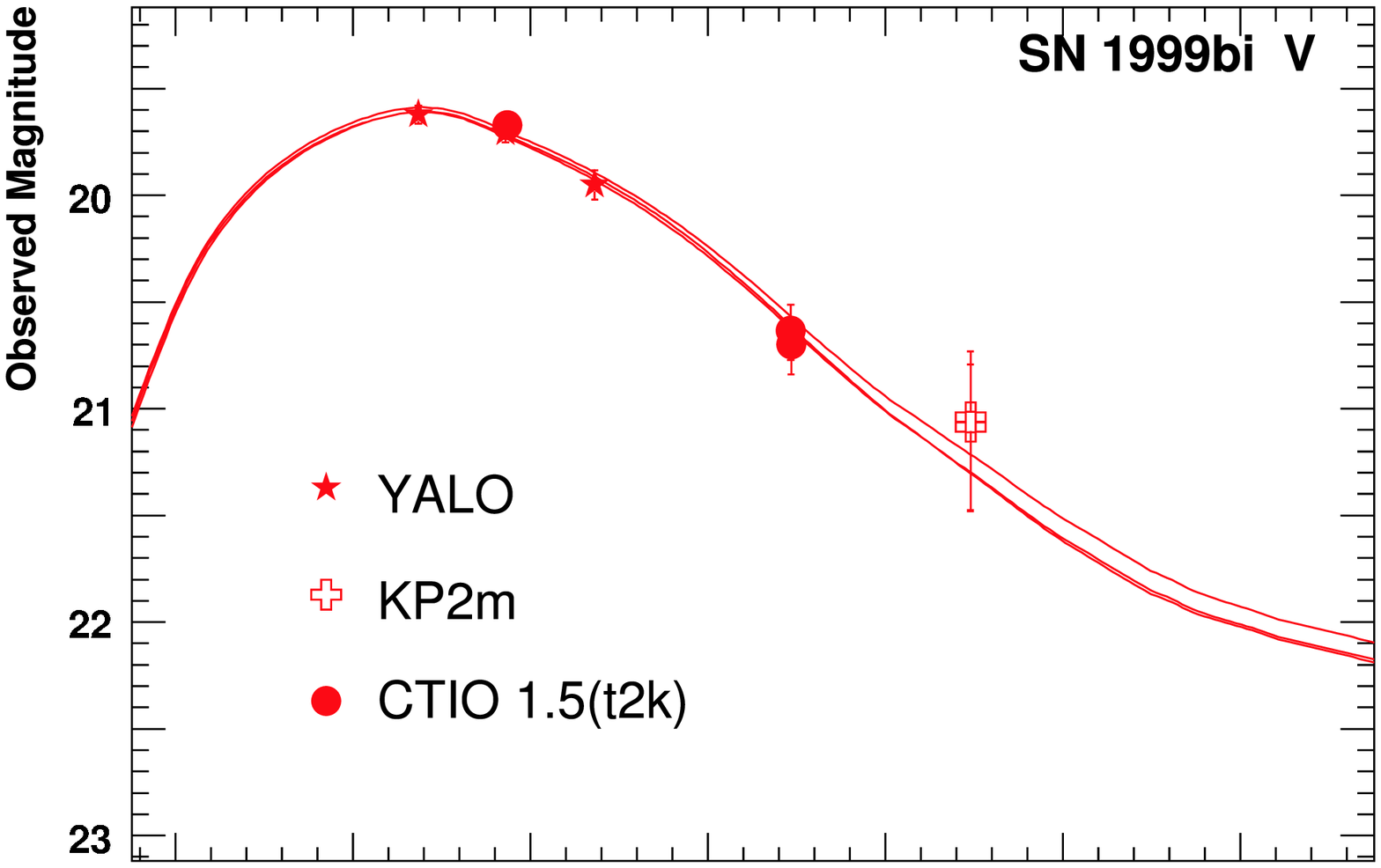}
\end{center}
\end{minipage}

\begin{minipage}{0.5\linewidth} 
\begin{center}
\includegraphics[width=\twidth\textwidth]{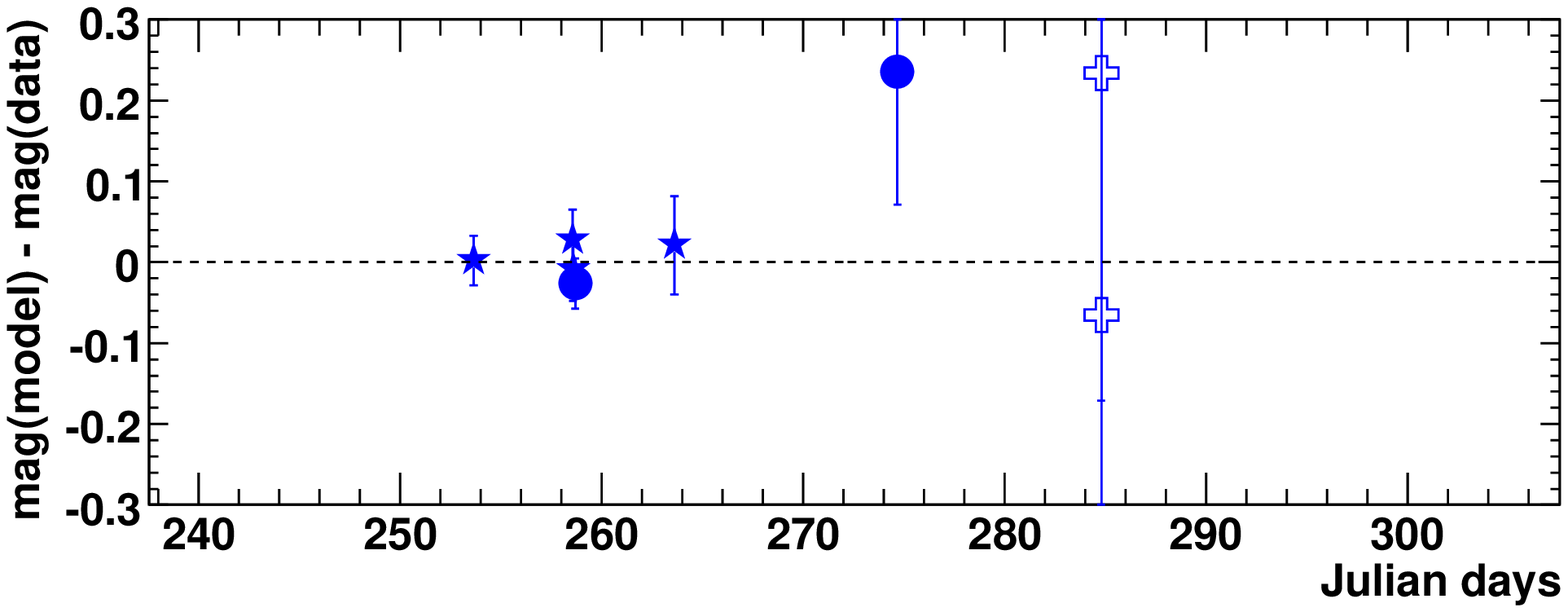}
\end{center}
\end{minipage}
\begin{minipage}{0.5\linewidth} 
\begin{center}
\includegraphics[width=\twidth\textwidth]{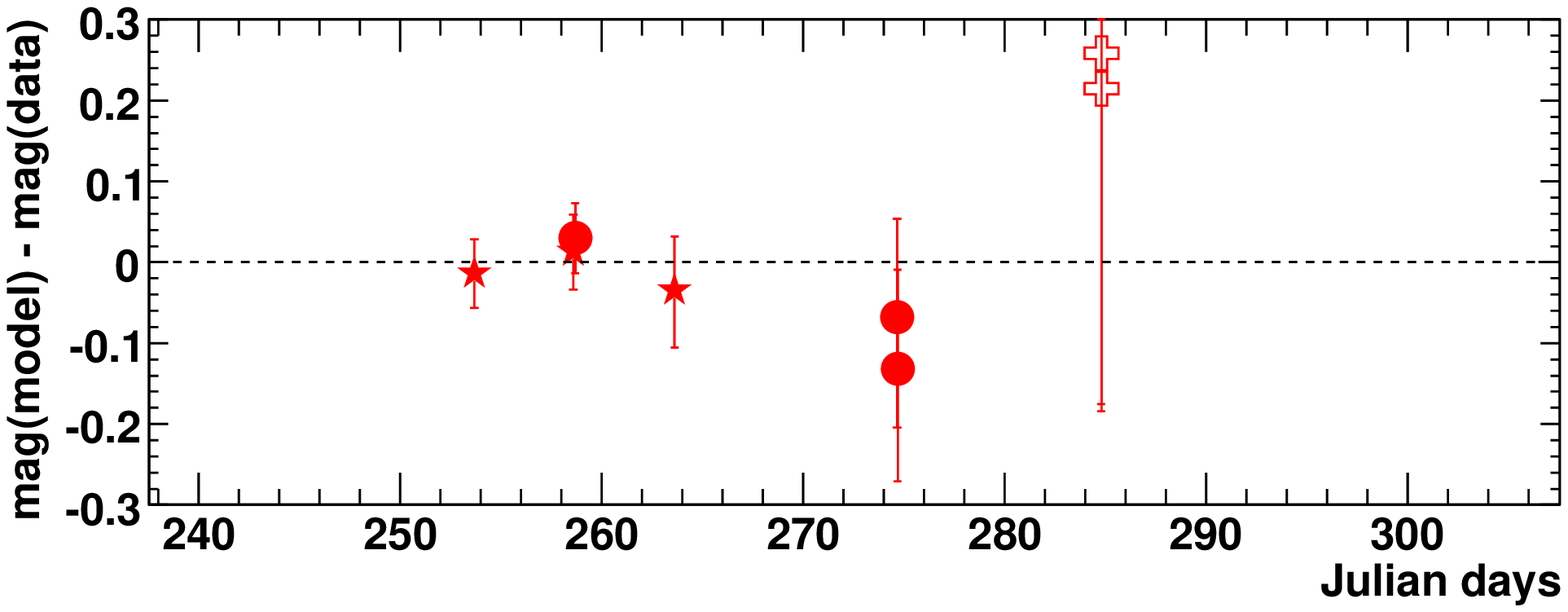}
\end{center}
\end{minipage}

\end{figure*}

\begin{figure*}
\begin{minipage}{0.5\linewidth} 
\begin{center}
\includegraphics[width=\twidth\textwidth]{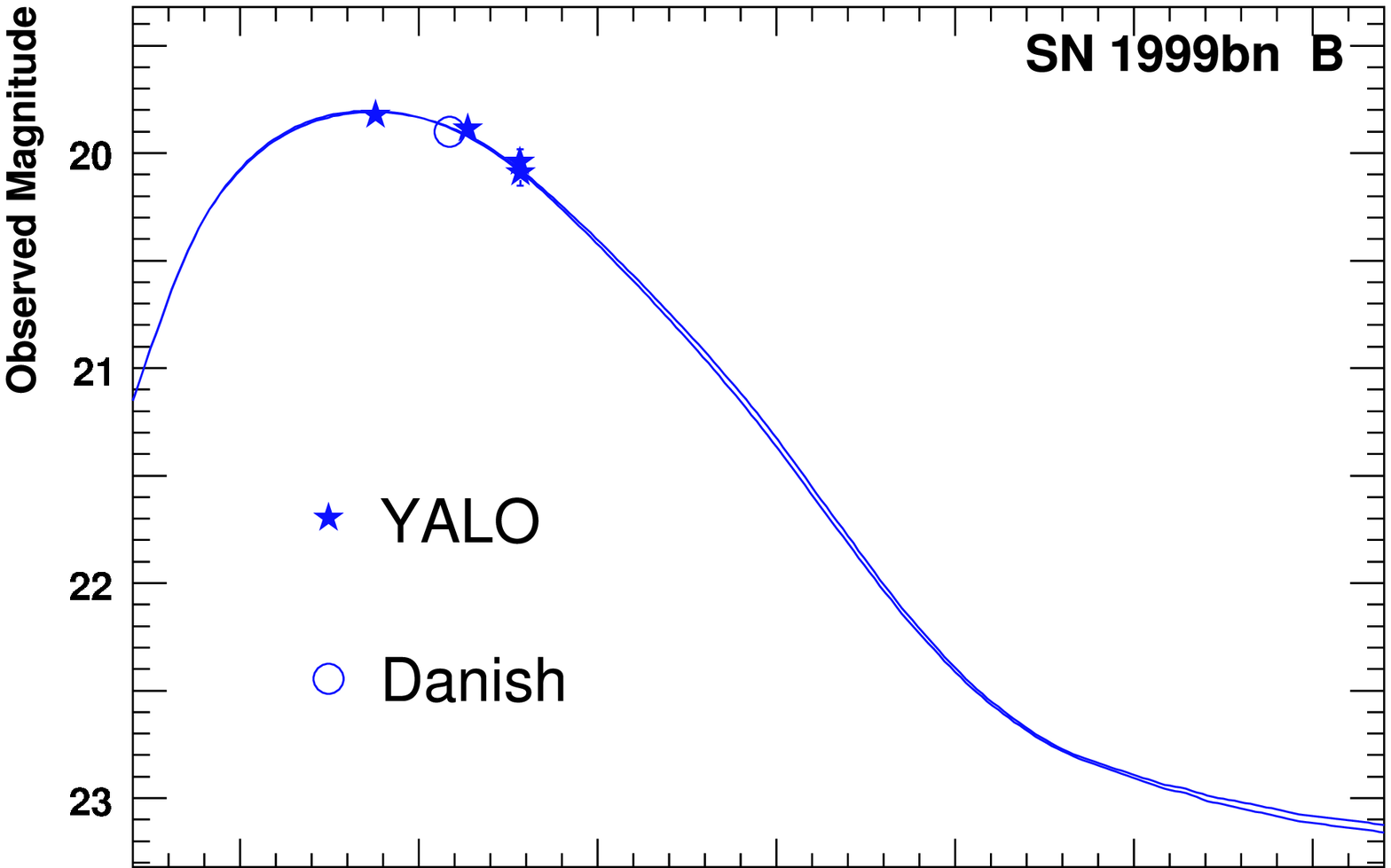}
\end{center}
\end{minipage}
\begin{minipage}{0.5\linewidth} 
\begin{center}
\includegraphics[width=\twidth\textwidth]{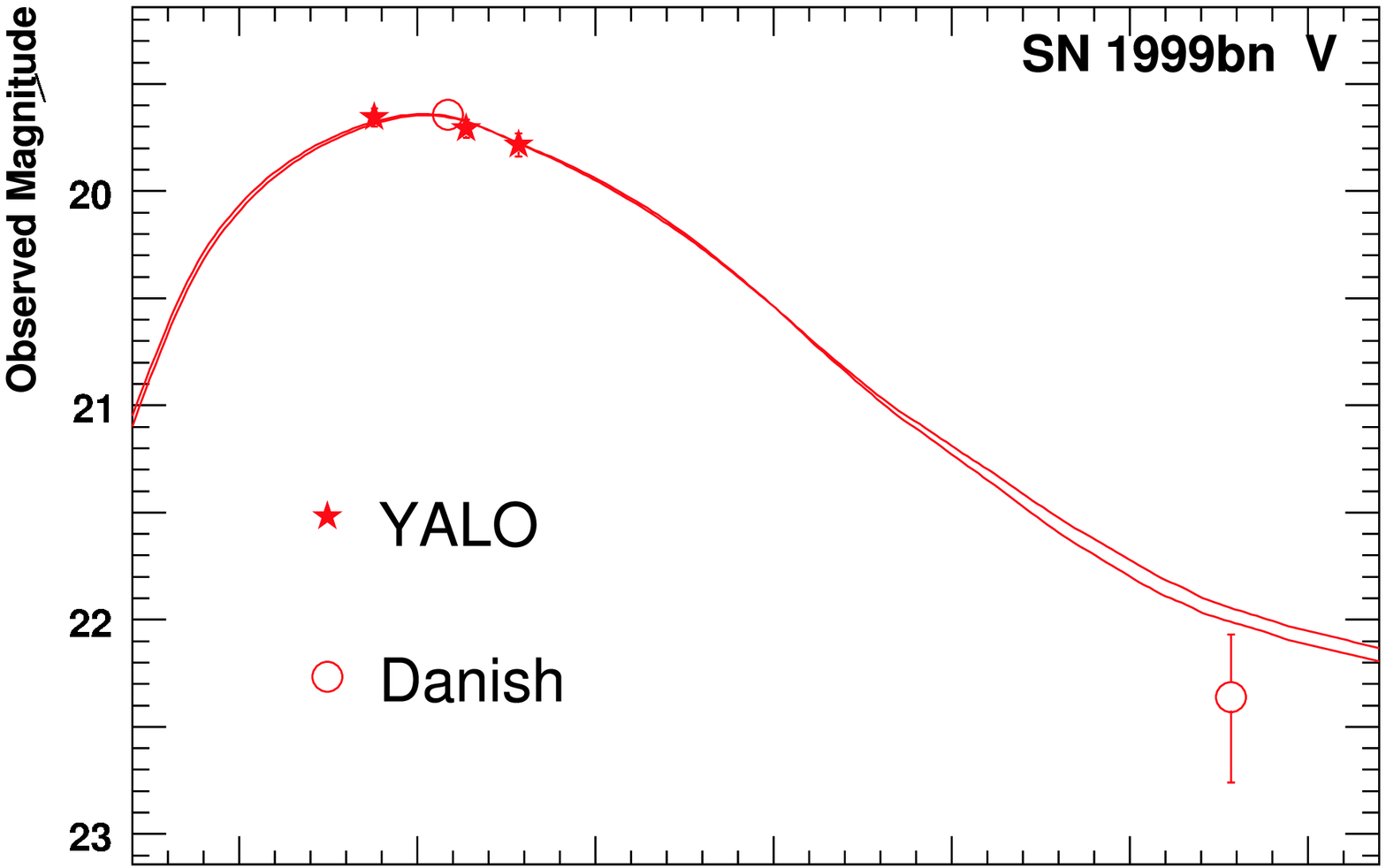}
\end{center}
\end{minipage}

\begin{minipage}{0.5\linewidth} 
\begin{center}
\includegraphics[width=\twidth\textwidth]{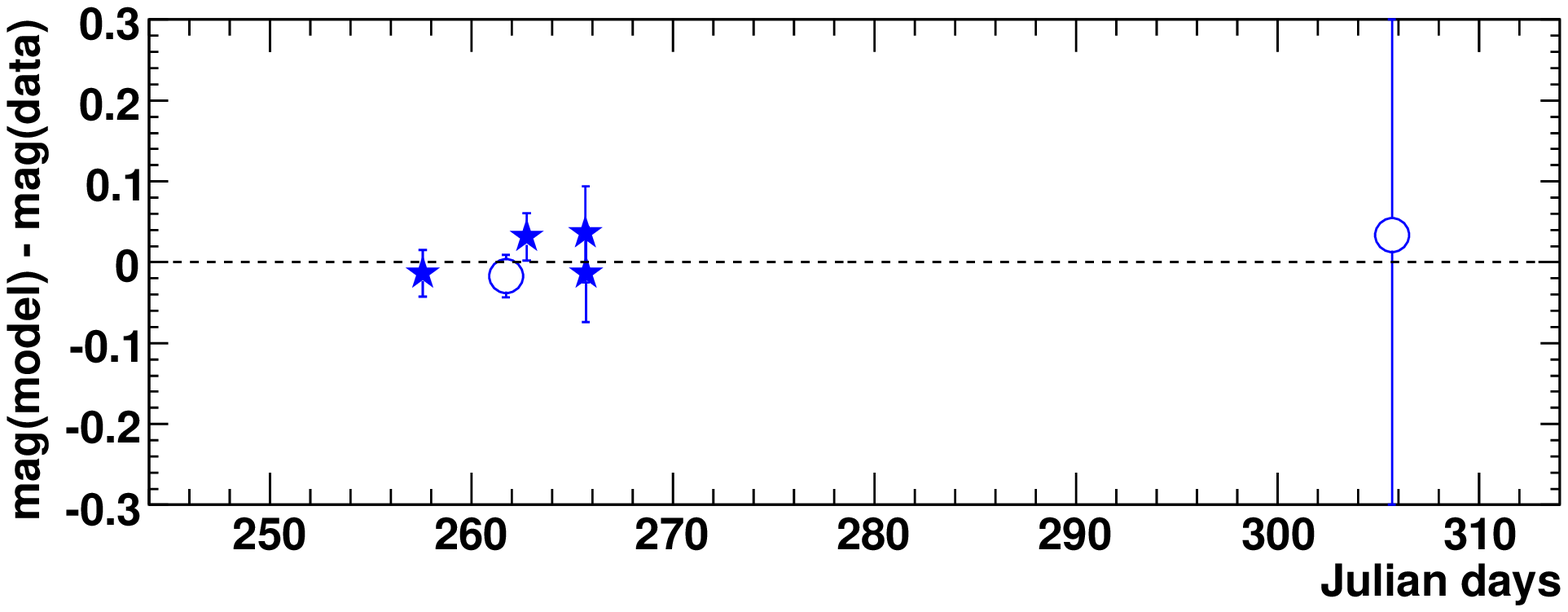}
\end{center}
\end{minipage}
\begin{minipage}{0.5\linewidth} 
\begin{center}
\includegraphics[width=\twidth\textwidth]{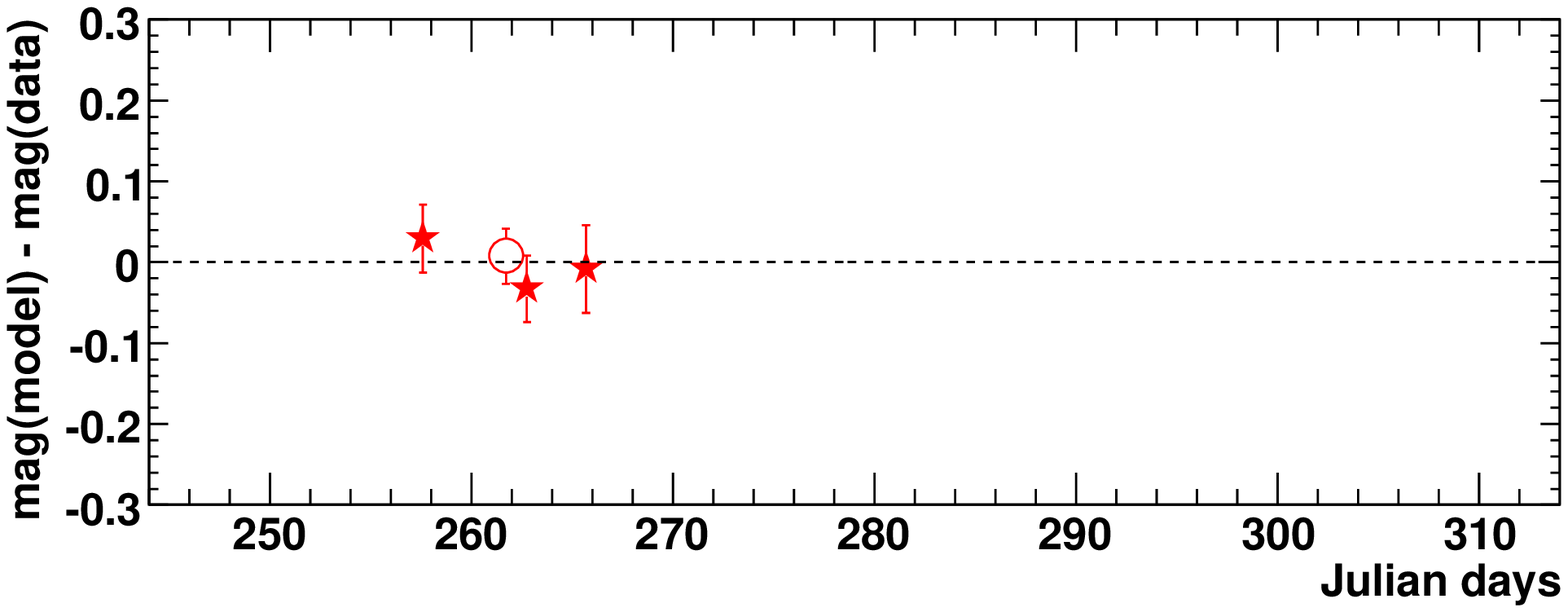}
\end{center}
\end{minipage}
\addtocounter{figure}{-1}
\caption{Continued}
\end{figure*}


\begin{figure*}
\begin{minipage}{\xdiv\linewidth} 
\begin{center}
\includegraphics[width=\twidth\textwidth]{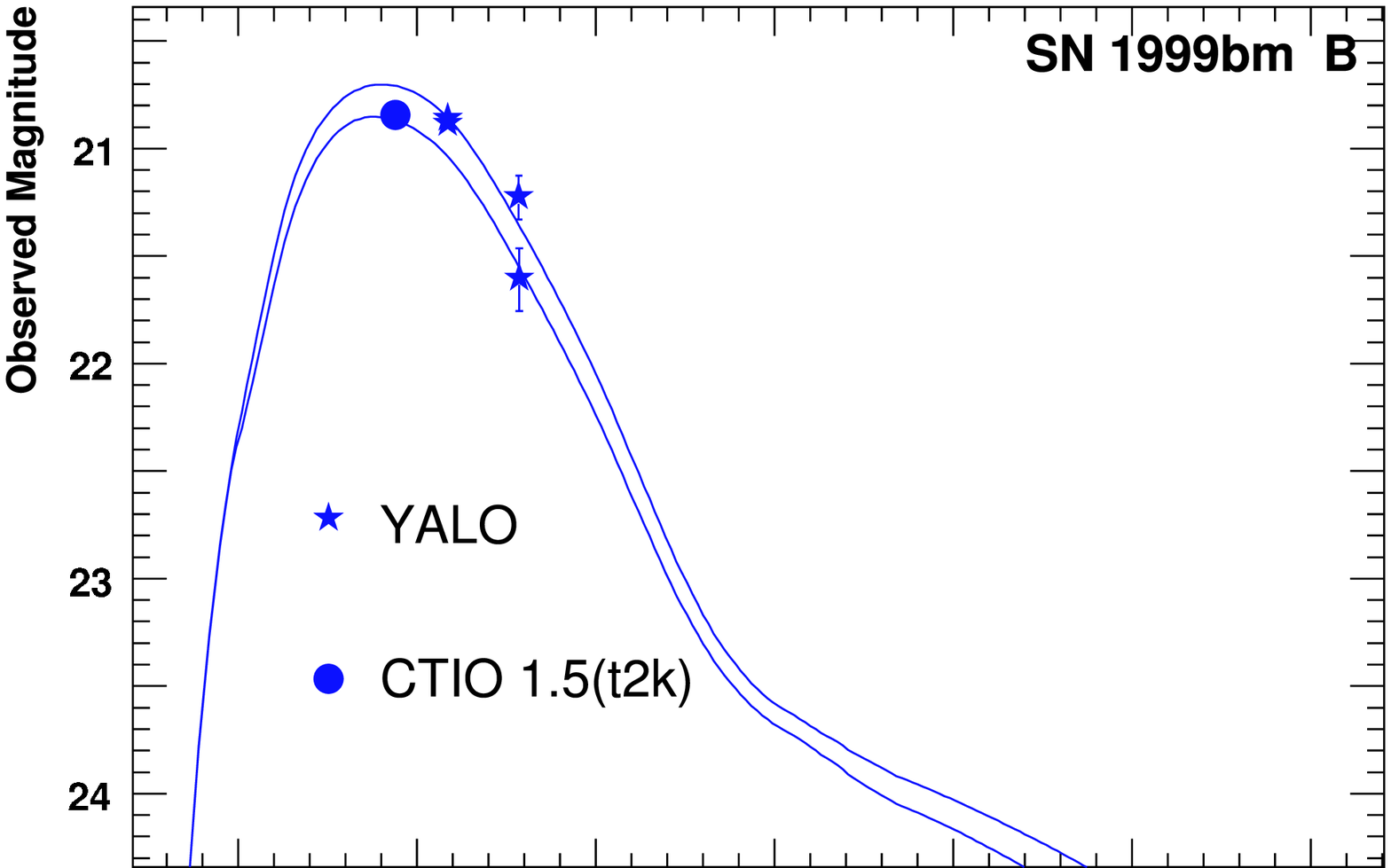}
\end{center}
\end{minipage}
\begin{minipage}{\xdiv\linewidth} 
\begin{center}
\includegraphics[width=\twidth\textwidth]{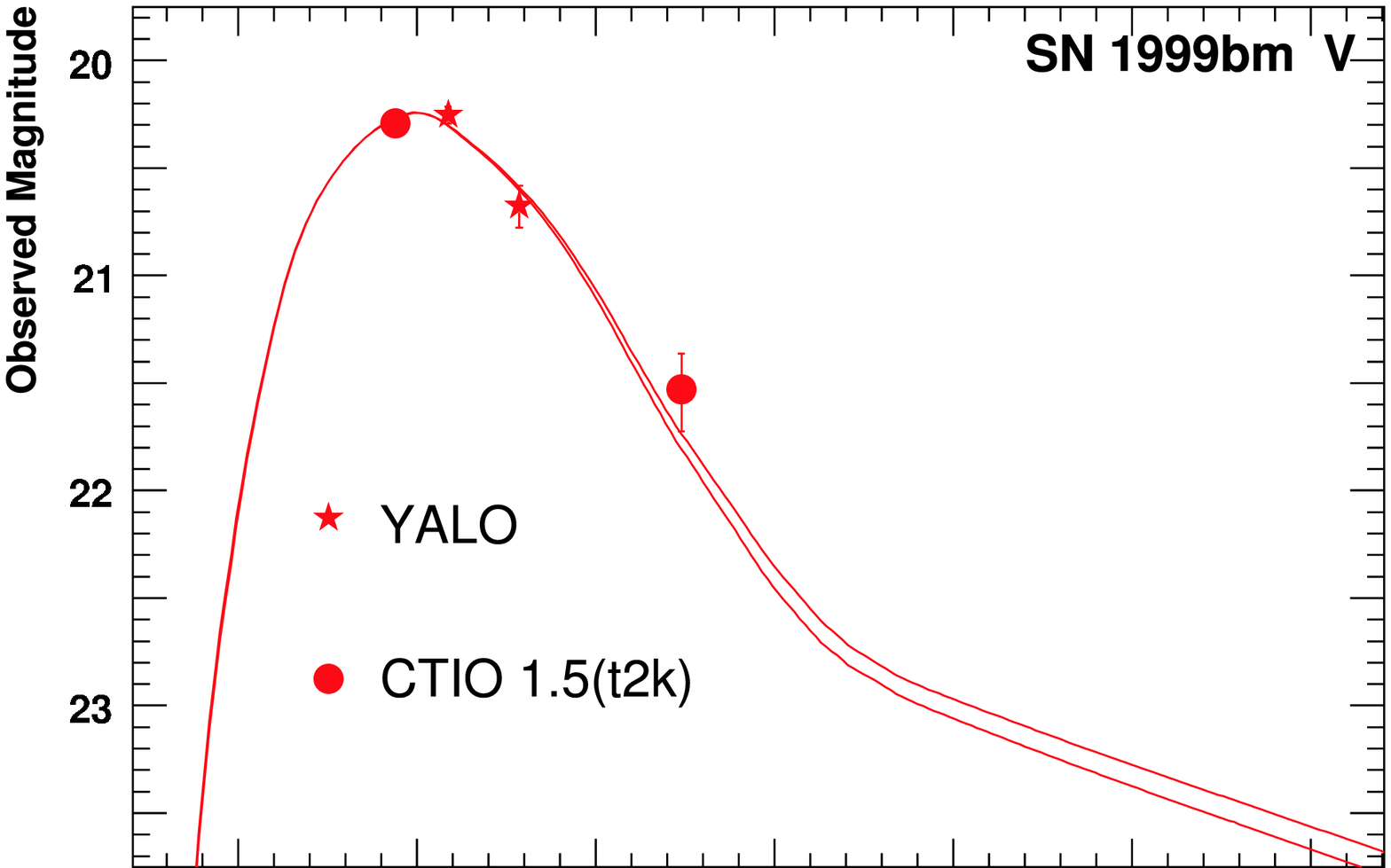}
\end{center}
\end{minipage}

\begin{minipage}{0.5\linewidth} 
\begin{center}
\includegraphics[width=\twidth\textwidth]{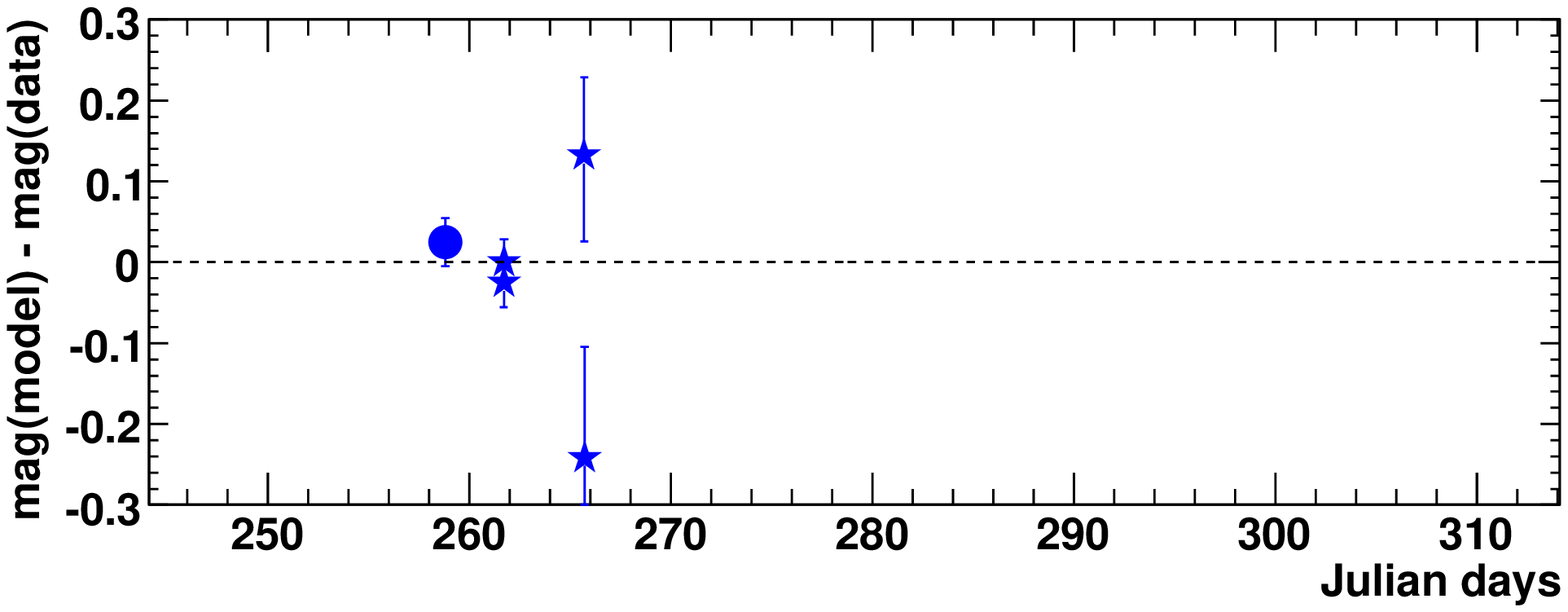}
\end{center}
\end{minipage}
\begin{minipage}{0.5\linewidth} 
\begin{center}
\includegraphics[width=\twidth\textwidth]{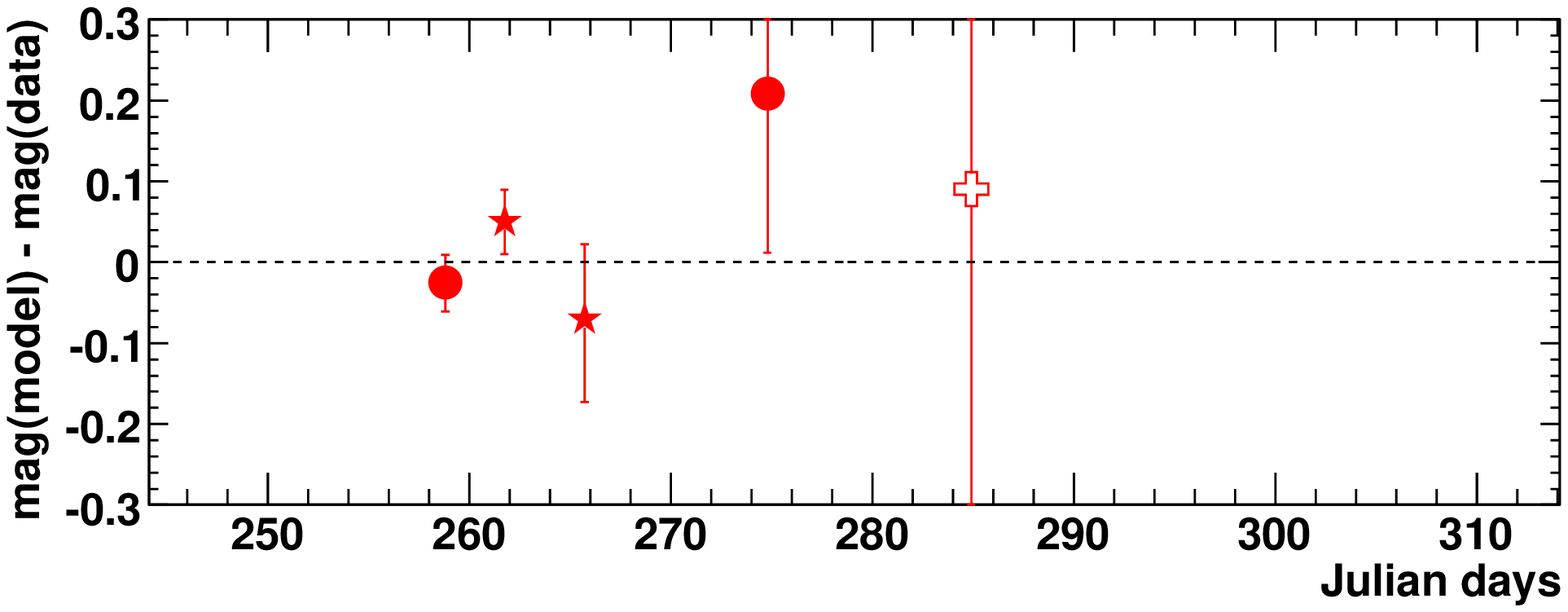}
\end{center}
\end{minipage}
\end{figure*}


\begin{figure*}
\begin{minipage}{\xdiv\linewidth} 
\begin{center}
\includegraphics[width=\twidth\textwidth]{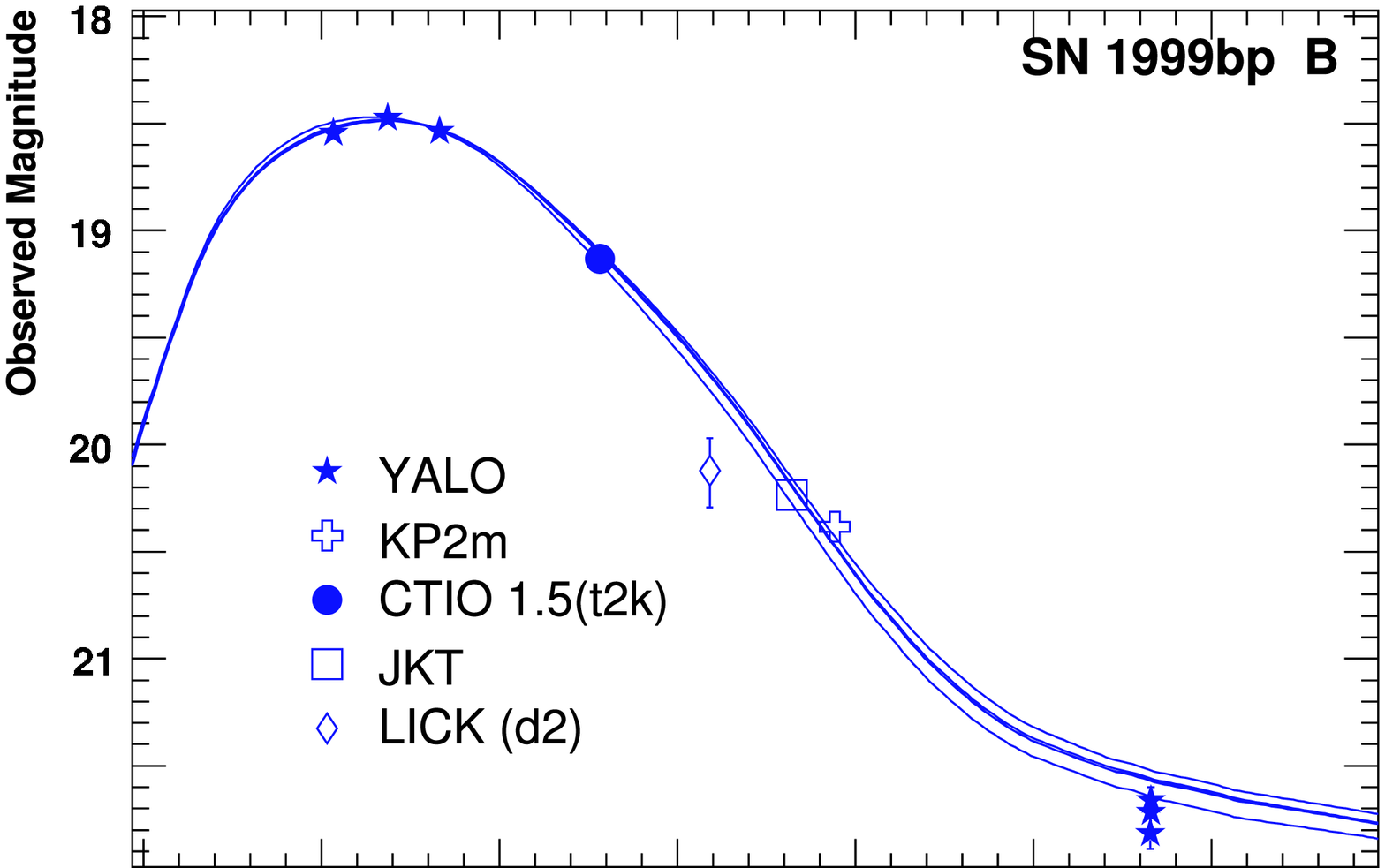}
\end{center}
\end{minipage}
\begin{minipage}{\xdiv\linewidth} 
\begin{center}
\includegraphics[width=\twidth\textwidth]{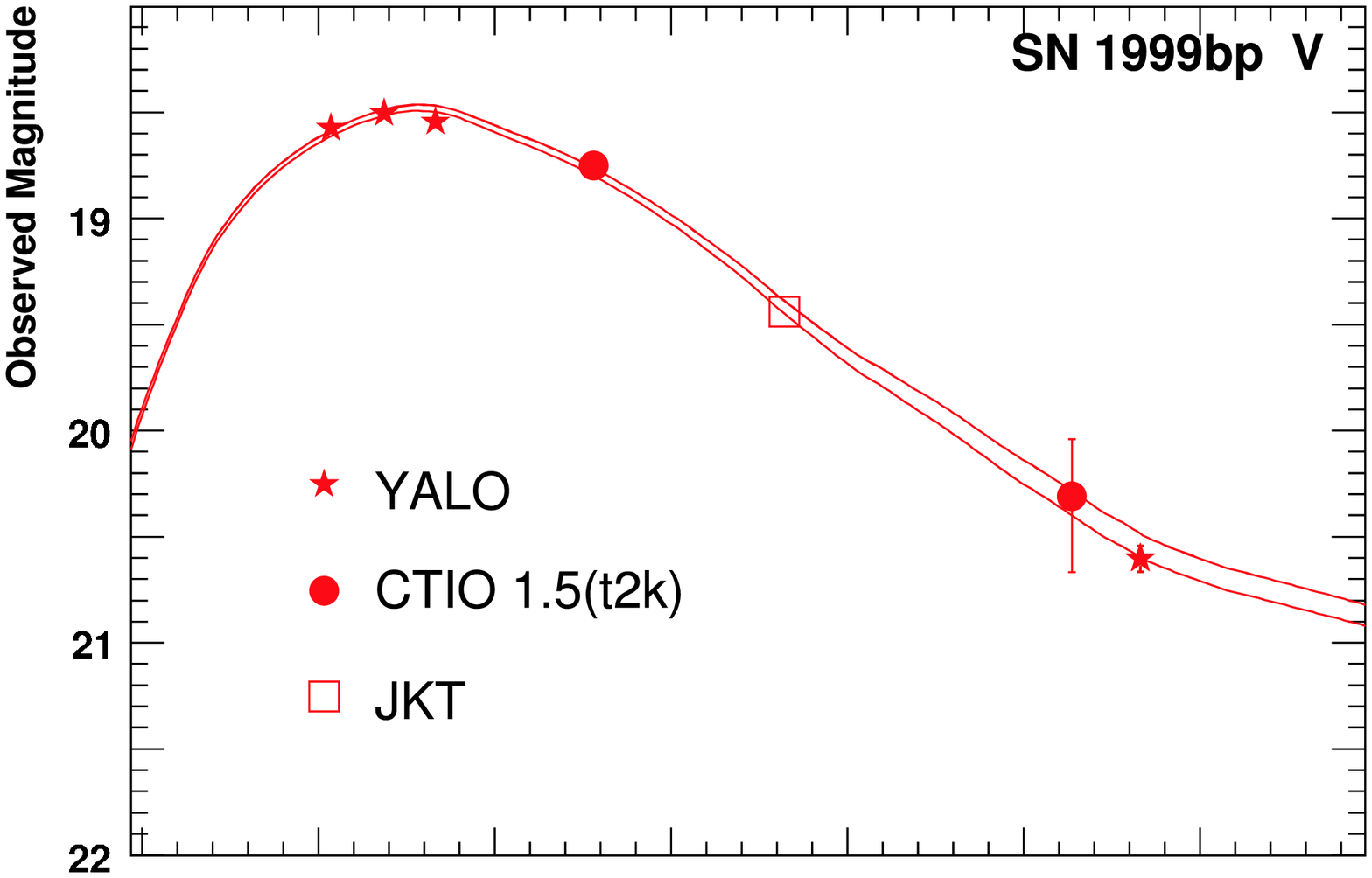}
\end{center}
\end{minipage}

\begin{minipage}{0.5\linewidth} 
\begin{center}
\includegraphics[width=\twidth\textwidth]{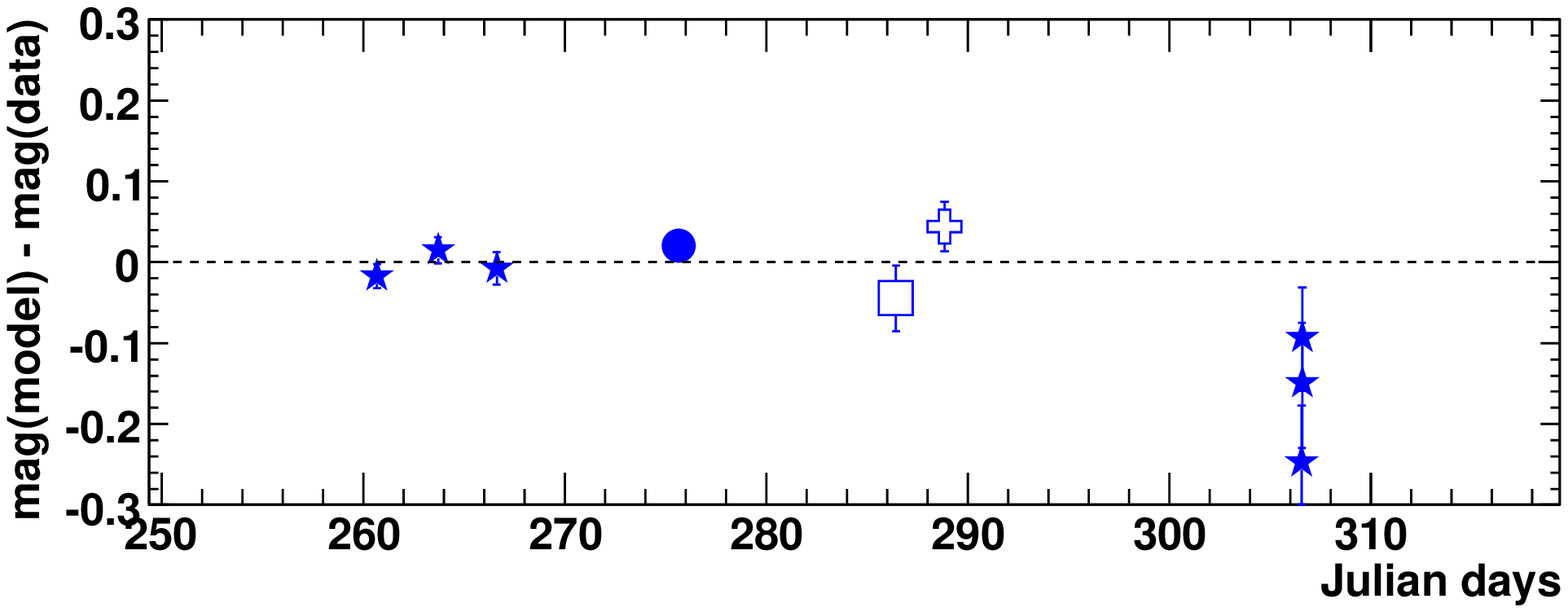}
\end{center}
\end{minipage}
\begin{minipage}{0.5\linewidth} 
\begin{center}
\includegraphics[width=\twidth\textwidth]{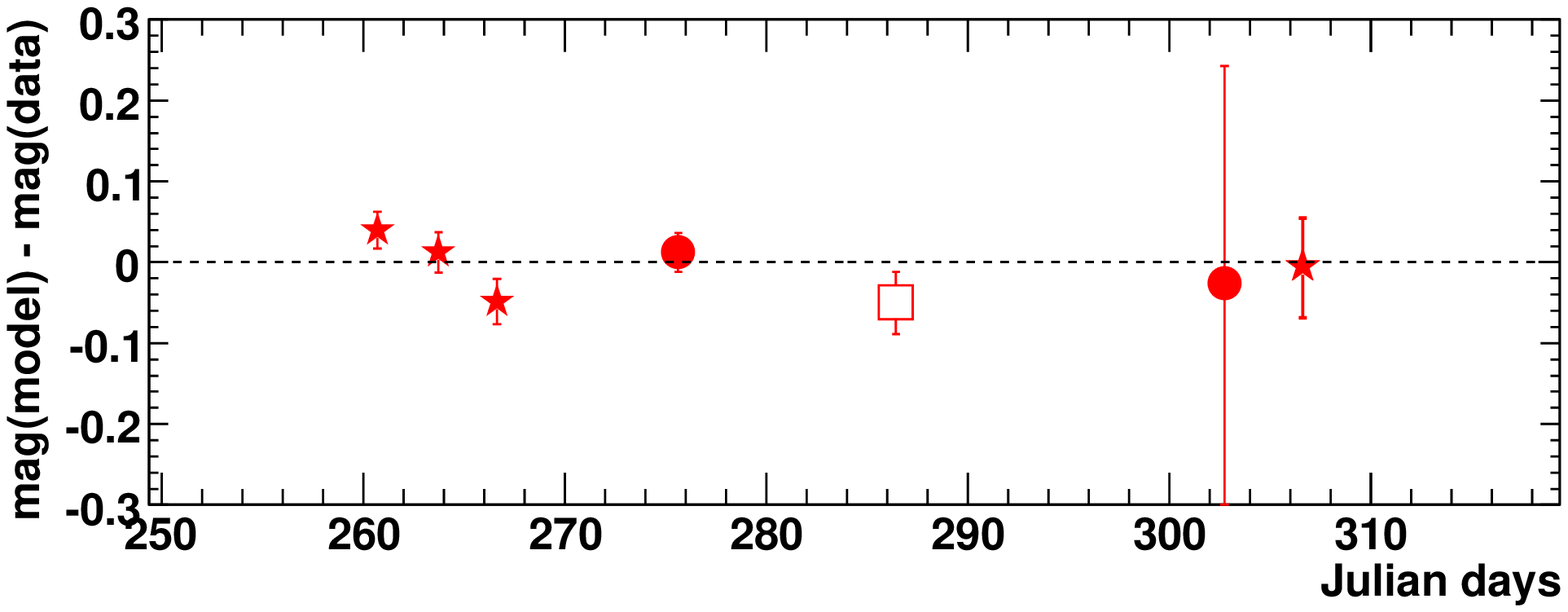}
\end{center}
\end{minipage}
\addtocounter{figure}{-1}
\caption{Continued}
\end{figure*}
\begin{figure*}[htp]
\vspace{1cm}
\begin{center}
\includegraphics[width=0.32\textwidth]{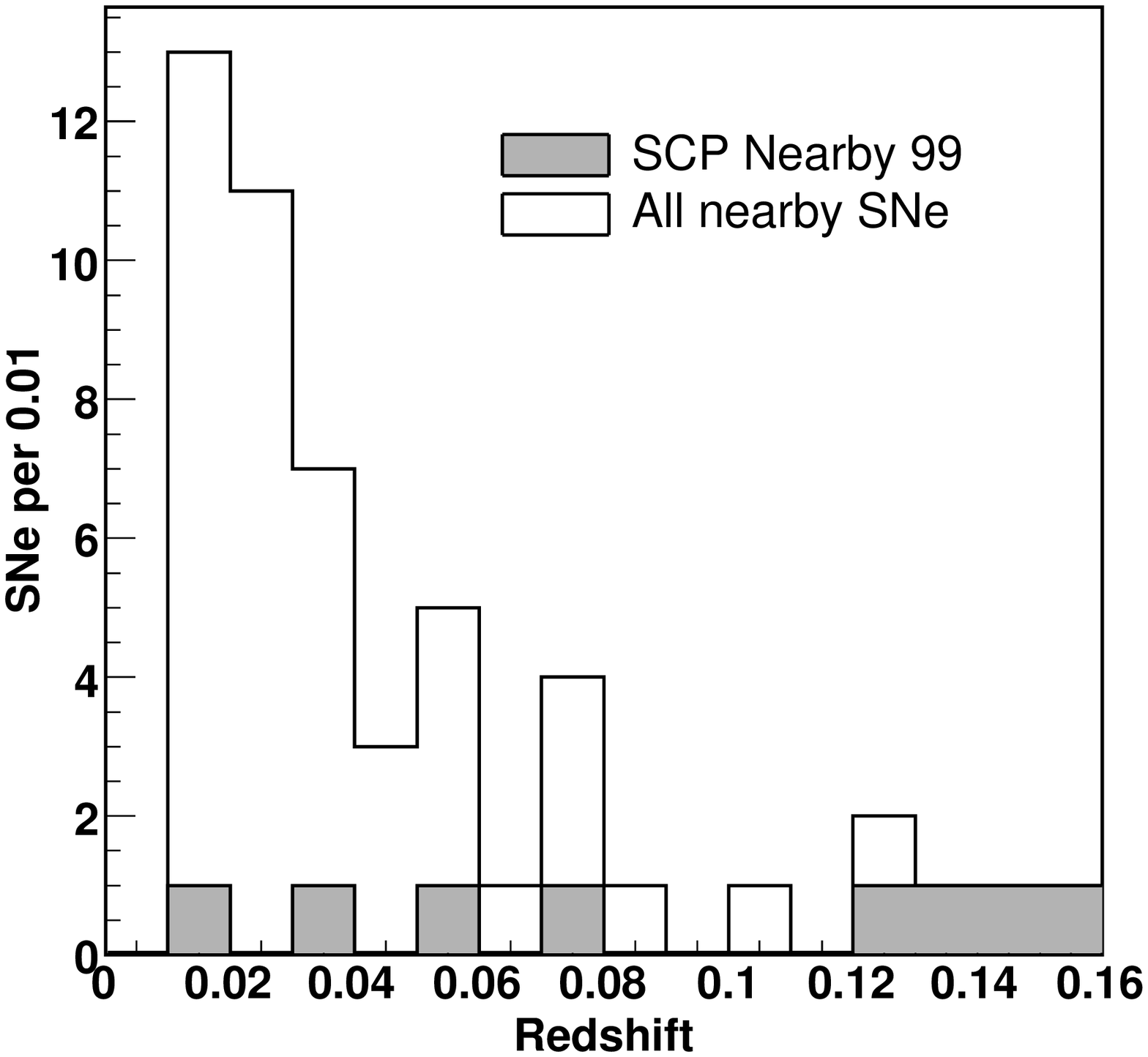}
\includegraphics[width=0.32\textwidth]{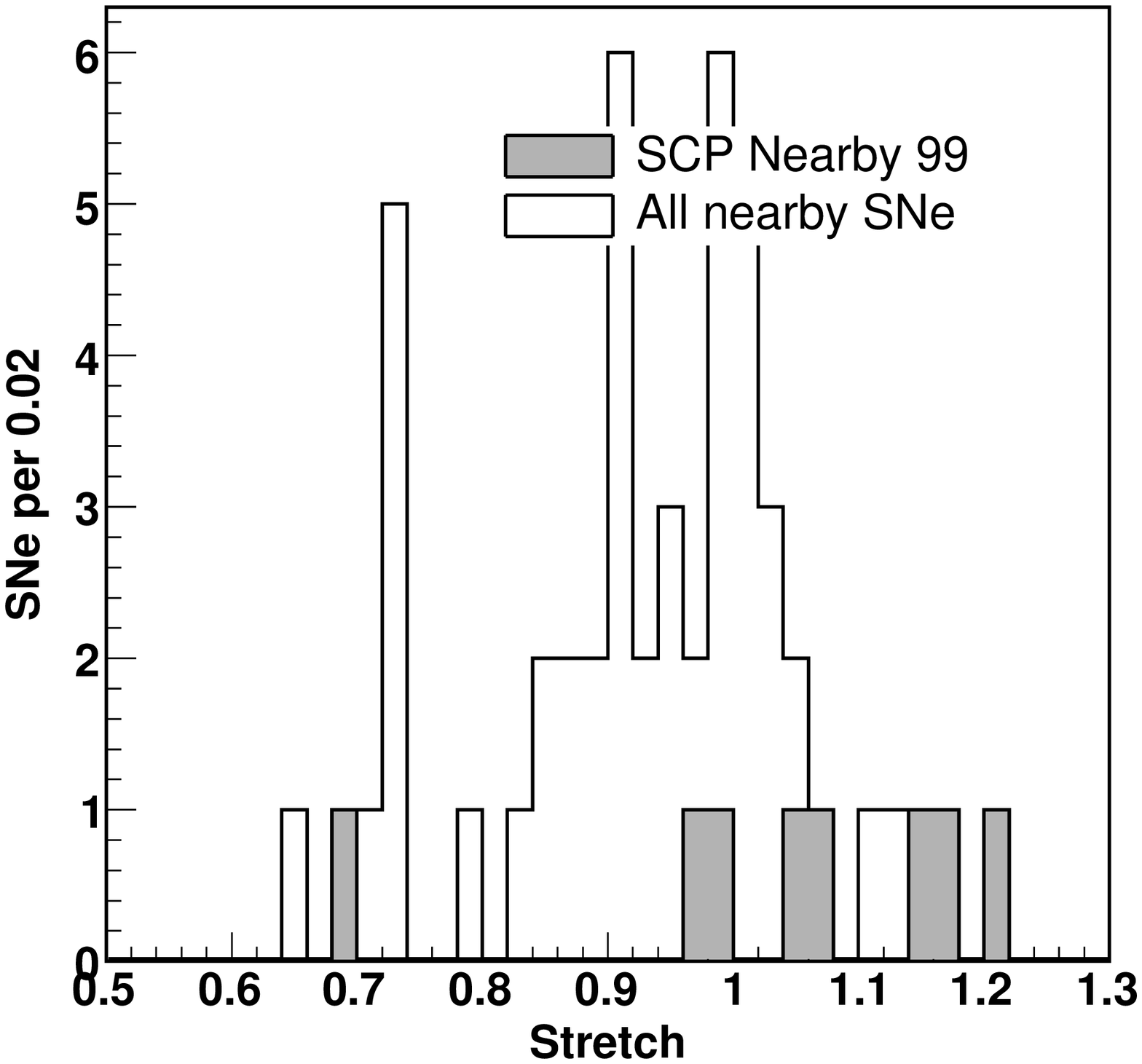}
\includegraphics[width=0.32\textwidth]{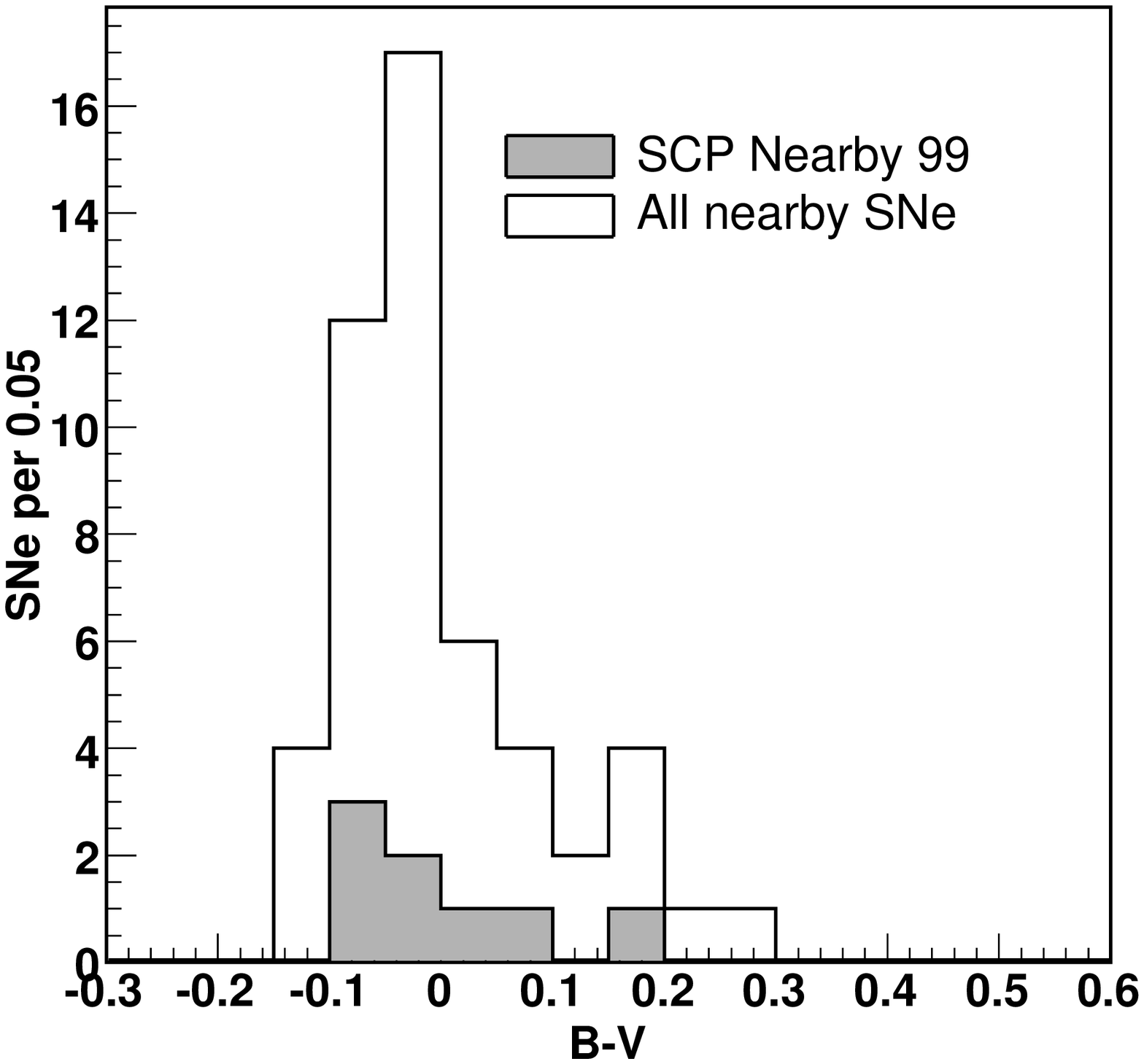}
\caption{Left panel: redshift distribution; middle panel: stretch distribution; right panel: $B-V|_{t=B_{\rm max}}$ distribution.}
\label{fig:zsc}
\end{center}
\end{figure*}
\section{Lightcurves}
\label{sec:ltcvs}
\subsection{Lightcurves from the SCP Nearby 1999 Supernova Campaign}
Figure \ref{fig:ltcv} shows the $BVRI$ lightcurves from the SCP Nearby 1999 campaign (the data are provided in Table \ref{tb:nb99_ltcv}).  
Different telescopes are marked by different symbols. Empty symbols 
represent uncorrected photometric data and filled symbols represent data 
corrected 
for non-standard band passes, the so called S-corrections \citep{suntzeff}. 
The S-corrections represent the magnitude shift needed to bring the 
data obtained with different band passes to a common standard system (in our case the Bessell system).
The S-corrections are obtained from a synthetic photometry calculation using the ``instrument dependent'' band pass functions described above and a spectrophotometric lightcurve model. The  spectrophotometric SN lightcurve model was adjusted using spline-functions to match the colors of the lightcurve models of our SNe.
The lightcurve models are 
shown in Fig.\ \ref{fig:ltcv} to guide the eye only. They are obtained
in two different ways  depending on the quality of the available 
photometric data. 
For the four SNe with $z<0.1$, we have used the fit method explained in 
\citet{wang06} which has six parameters per band.
This method allows
effective fitting of R and I band data, which exhibits a second ``bump'' of
variable strength appearing approximately 30 days after the maximum.
  However, since this fit method has six free parameters per 
fitted band one can use it only for lightcurves with dense temporal
sampling with high signal-to-noise.
For the more distant SNe 1999ar, 1999bi, 1999bm, 1999bn we use a more 
constrained lightcurve
fitting method based on template matching. A library of template 
lightcurves obtained from well observed supernovae is 
K-corrected to the observed redshift. The best matching lightcurve is chosen 
as a model for the supernova. The lightcurve models along with the 
S-corrections shown in Figure \ref{fig:ltcv} are meant to guide the eye and 
will not be used in the remainder of the paper; we continue with the concept 
of using instrumental magnitudes along with instrument-dependent passbands
when fitting the lightcurve parameters.  

The lightcurve parameters such as peak 
magnitude, stretch and color at maximum are obtained using the spectral 
template method of \citet{salt} which is 
described in 
more detail in section \ref{sec:ltcvfit}. The method 
is well suited to this task since it uses telescope-specific band pass functions 
for modeling the observer-frame lightcurves. The $B$- (left) 
and $V$-band (right)   
observer-frame lightcurves are shown in Figure~\ref{fig:saltlc},
along with the lightcurves predicted by the spectral template for 
 the corresponding band pass. In the bottom part of 
the plots we show the residuals from the model prediction. In most cases 
the model describes the data reasonably well, with $\chi^2/{\rm DOF}\sim1$. 
Systematic deviations, such as observed in the late-time behavior of the $B$-band 
lightcurve of SN~1999aw, are likely to be attributable to the limitations 
of the two-parameter spectral template model in 
capturing the full diversity of Type Ia supernovae lightcurves.
 
Figure \ref{fig:zsc} (right and middle) shows the fitted $B-V$ color at 
maximum, as well as the stretch distribution. The 
stretch distribution has one 
low stretch supernova (SN~1999bm), but is otherwise dominated by supernovae 
with larger stretches. Two lower stretch SNe~Ia were found in these searches,
but are not presented here because of their faintness --- in one case combined
with proximity
to the cuspy core of an elliptical host --- prevented an analysis using the
techniques described here. In any case, the larger number of high 
stretch supernovae is not very significant (a K-S test resulted in a  
20\% probability that the two distributions are consistent with each other). 

For two of the eight supernovae, lightcurve data have previously been
published.  \citet{jha06,krisc00,altavilla04} present 
independent data 
on SN~1999aa. When comparing the fit results for SN~1999aa we find agreement to 
within $1\%$ in maximum $B$-band luminosity, color and stretch.  
Spectroscopic and photometric data on SN~1999aw are previously reported by \citet{strolger_aw}. While the raw data of \citet{strolger_aw} are largely the same 
as that presented here, the reduction pipelines used are independent. A main difference is 
the treatment of non-standard band passes. We report the original
 magnitudes and correct for non-standard band passes during the fit of the 
lightcurve, while in \citet{strolger_aw} corrections based on  the color 
coefficient have been applied to the data. When fitting for 
peak $B$-band magnitude, color and stretch we obtain differences of 
$\Delta B = 0.04,\Delta (B-V)= 0.02$ and $\Delta s= 0.005$. 

Figure \ref{fig:zsc} (left) shows the redshift distribution relative to the sample of other nearby supernovae (see \ref{sec:lc} for a definition of that sample).  As 
can be seen, the distribution extends to redshifts $z\sim0.15$, an
 underpopulated region in the Hubble diagram.

\subsection{Literature supernovae}

\label{sec:lc}

Here we discuss the  
set of previously published nearby and distant supernovae included in the 
analysis. Not all SN lightcurves are of sufficiently good quality to allow their use
in the following cosmological analysis. 
For all supernovae in the sample, we require that 
 data from at least two bands with rest-frame central wavelength between 
3470~\AA~ ($U$-band)
and 6600~\AA~ ($R$-band) exist and that there are in total at least five data points available. 

Further, we require that there is at least one 
observation existing between 15 days before and 6 days after the date of 
maximal $B$-band brightness, as obtained from an initial fit to the 
lightcurves 
(see section \ref{sec:ltcvfit}).  The 6 day cut is scaled by stretch for
 consistency. In addition, we observed that for a smaller number of poorer 
lightcurves, 
the uncertainties resulting from the fits are unphysically small compared 
to what is expected from the photometric data. 
In these cases, we  randomly perturb each data point by a tenth (or if necessary by a fifth) of its 
photometric error  and refit the lightcurves. 
The remaining 16  SNe, where convergence can  not be obtained 
even after perturbation of the 
data, are excluded from further analysis (note that these SNe are generally poorly measured and would have low weight in any cosmological analysis).

For the nearby SN sample, we use only supernovae with CMB-centric
redshifts $z>0.015$, in order to reduce  the impact of uncertainty due to 
host galaxy peculiar velocities. We checked that our results do not 
depend significantly on the value of the redshift cut-off 
(tested for a range $z=0.01-0.03$). 

The number of SNe passing these cuts are summarized in Table \ref{tb:cuts}. Each individual supernova is listed in Table~11, and the last column indicates any cuts that the supernova failed.

The list contains 17 supernovae from \citet{ct}, 
11 from \citet{riess99}, 16 from \citet{jha06}, and 6 from 
\citet{krisc04a,krisc04b,krisc01}.   
 Our lightcurve data for SN~1999aa are 
merged with that of \citet{jha06}.
To this list of nearby supernovae from the literature we add the new nearby 
supernovae presented here. For SN~1999aw, we use only the lightcurve 
data presented in this paper. Hence the sample contains 58 nearby supernovae.

The sample of high redshift supernovae is comparably heterogeneous.
We use all of the 11 SNe from \citet{knop03} that have 
lightcurves obtained with HST.
 Of the 42 supernovae from \citet{perl99}, 30 satisfy the selection 
cuts described above (as can be seen in the photometry data of 
Table \ref{p99_ltcv}). Of the 16 SNe used by the High-Z Team (HZT)
 \citep{riess98, garnavich98,schmidt98}, two are already included in the \citet{perl99} 
sample and of the remaining 14, 12 pass our cuts.

Included also are 22 SNe from \citet{barris04}, and 
the 8 SNe from \citet{tonry03} that are typed to be secure or likely SNe Ia. 
We do not use 
SN~1999fv and SN~1999fh, as the number of available data points does not 
exceed the number of lightcurve fit parameters.

We add the 73 SNe Ia from the first year of SNLS  \citep{astier05}, of which
one does not pass the first phase cut (03D3cc). 
Note that, in \citet{astier05}, 2 of the 73 
supernovae were excluded from their cosmological parameter fits 
because they were significant outliers 
(see discussion in section \ref{sec:outlier}). 
\citet{riess04,riess06} have published 37 supernovae which were discovered 
and followed using HST. Of these, 29 passed our lightcurve quality cuts.  
This sample contains the highest redshift supernovae in our compilation. 
Finally, we use the 84 SNe from the ESSENCE survey \citep{essence_m_07,essence_wv_07}, of which 75 pass our cuts. 

\begin{table}[htbp]
\begin{center}
\begin{tabular}{ l c  }
\tableline
\tableline
Requirement & $N_{\rm SN}$\\
\tableline
all & 414\\
$z>0.015$     & 382 \\
Fit successful  & 366 \\
Color available & 351 \\
First phase $< 6~$d& 320 \\
$\geq 5$ data points & 315 \\
Outlier rejection & 307 \\

\tableline
\tableline
\end{tabular}
\end{center}
\caption{Number of SNe after consecutive application of cuts. See \ref{sec:outlier} for a discussion of the outlier rejection cut.}
\label{tb:cuts}
\end{table}

\subsection{Lightcurve fitting}
\label{sec:ltcvfit}

 The spectral-template-based fit method 
of \citet{salt} (also known as SALT) is used to fit consistently 
both new and literature lightcurve data. 
This method is based on a spectral template \citep{nugent02}  which has been adapted in an 
iterative 
procedure to reproduce a training set of nearby SNe $UBVR$ lightcurve data. The training set consists of mostly $z<0.015$ SNe and hence does not overlap with the sample we use for determination of cosmological parameters. To obtain 
an expected magnitude for a supernova at a certain phase,
 the model spectrum is first 
redshifted to the corresponding redshift followed by an integration of   
the product of  spectrum and band pass transmission.  
The spectral-template based fit  method 
has the advantage that it consistently allows the simultaneous fit of 
multi-band light 
curves with arbitrary (but known) band pass transmission functions. 
In view of the large number of filters and instruments used for the new 
nearby SN samples as well as the very 
diverse lightcurve data found in the literature, this is particularly important here.
In addition, frequent practical problems associated with K-corrections---such as the 
propagation of photometric errors---are handled naturally.

The spectral template based fit method of \citet{salt} fits for the time of 
maximum, the flux normalization as well as rest-frame 
color at maximum defined as $c$=\BV{} and time-scale stretch $s$. 
It is worth noting that by construction, the stretch in SALT has a related 
meaning to the conventional time-axis 
stretch \citep{stretch, goldhaber01}. However, as a parameter of the lightcurve model it also 
absorbs other, less pronounced, stretch dependent lightcurve dependencies.
The same is true for the color $c$.

Recently, direct comparisons between alternative fitters, such as SALT, its update \citep{salt2} 
as well as MLCS2k2 \citep{mlcs2k2} show good consistency between the fit results, e.g. the amount of 
reddening \citep{hubbub_conley}. Our own tests have shown that for well observed supernovae, the method produces very consistent results (peak magnitude, stretch) when compared to the more traditional method of using light-curve templates \citep{stretch}. However, we noticed that fits of poorly observed 
lightcurves in some cases do not converge properly. Part of the explanation 
is that in the case of the spectral template based fit method, the data before 
$t<-15$~days is not used as an additional constraint. More typically, the SALT
fitter can fall into an apparent false minimum and we then found it 
necessary to restart it repeatedly to obtain convergence.
Note that the small differences between the lightcurve fit parameters of 
Table~11 and the values
shown in Table~10 of \citet{essence_wv_07} are primarily cases where the 
\cite{essence_wv_07} SALT fit
did not converge (some of which are noted in \citet{essence_wv_07}) and a few 
cases where we found it necessary to remove an extreme outlier photometry 
point from the lightcurve.

The lightcurves from Barris et al. (2003) and the I-band lightcurves of 4 
supernovae of P99 
(SNe 1997O, 1997Q, 1997R, and 1997am, see also Knop et al 2003) need a different analysis procedure, since in these cases the light of the host galaxy was not fully subtracted 
during the image reduction. We hence allow for a constant contribution of light from the host galaxy in the lightcurve fits. The supernovae were fit with additional parameters: the zero-level of the I-band lightcurve in case of the four SNe from the P99 set and the zero-level of all the bands in case of the  \citet{tonry03} data. The additional uncertainties due to these unknown zero-levels have been propagated into the resulting
lightcurve fit parameters.

The fitted lightcurve parameters of all SNe 
 can be found in Table~11 which is also available in electronic form\footnote{http://supernova.lbl.gov/Union}.

\section{Hubble diagram construction and cosmological parameter fitting}
\label{sec:par_estimate}
The full set of lightcurves as described in section \ref{sec:lc} have been 
fitted, yielding B-band maximum magnitude $m^{\rm max}_B$, stretch $s$, and color $c$=\BV{}. 
In this section, these are input to the determination of the distance modulus.
The analysis method is chosen to minimize  
 bias in the estimated parameters (see section \ref{sec:ubias}).   
An outlier rejection based on truncation 
is performed which is further described in section \ref{sec:outlier}, before constraints on the cosmological parameters are computed. 

\subsection{Blind analysis}
\label{sec:blind}
Following \citet{conley06}, we adopt a blind analysis strategy. The basic 
aim of pursuing a blind analysis is to remove potential 
bias introduced by the analyst. In particular, there is a documented tendency 
(see for example \citet{pdg})
 for an analysis to be checked for errors in the procedure 
(even as trivial as bugs in the code)
 up until the expected results are found but not much beyond.  
The idea of a blind analysis is 
to hide the experimental outcome until the analysis strategy is finalized and 
debugged.
However, one does not want to blind oneself 
entirely to the data, as the analysis 
strategy will be partially determined by the properties of the data. 
The following blindness strategy is used, which is 
similar to the one invented in \citet{conley06}. The data is fit 
assuming a \lcdm{} cosmology, with the resulting fit for $\om$ stored without being reported. 
The flux of each supernova  data point is then rescaled according to the ratio of luminosity distances obtained from the fitted parameters and arbitrarily chosen dummy  parameters (in this case $\om=0.25, \ola=0.75$). This procedure preserves the stretch and color distribution, and as long as the fitted parameters are not too different from the target parameters approximately preserves the residuals from the Hubble diagram. In developing the analysis, one is only exposed to 
data blinded by the procedure described above. Only after the analysis is 
finalized and the procedure frozen, is the blinding  turned off. 
				   
Note that this prescription  allows --- in a consistent way --- the inclusion of 
future data samples. A new data sample would be first investigated in a blind 
manner following the tests outlined in section \ref{sec:diversity}, and if no 
 anomalies are observed, one would combine it with the other data sets.

\subsection{Unbiased parameter estimation}
\label{sec:ubias}

Type Ia supernova obey a
redder-dimmer relation and a wider-brighter relation \citep{phillips93}. The redder-dimmer relation in principle can be explained by dust extinction; however, the total to selective extinction ratios generally 
obtained empirically are  smaller than expected from Milky-Way-like dust \citep{tripp98,tripp99,parodi00,salt,wang06}.  
At the same time, the exact slope of the stretch-magnitude relation is not 
(yet) 
predicted by theory. The absence of a strong theoretical prediction motivates
an empirical treatment of stretch and color corrections. Here we adopt the
 corrections of \citet{tripp98} (see also \citet{tripp99,wang06,salt} and \citet{astier05}): 

\begin{equation} 
\mu_B = m^{\rm max}_B - M + \alpha(s-1)-\beta c,
\label{eq:mub}
\end{equation}

 Since the $\beta$-color correction term must account for both dust and any intrinsic color-magnitude relation, it is clearly an empirical approximation.
The validity of $\beta$-color correction relies on only one assumption, 
that is, nearby supernovae and distant supernovae have an identical 
magnitude-color relation. 
If either the intrinsic SNe properties or the dust extinction properties of the supernovae are evolving with redshift, these assumptions may be violated. Observational selection effects may also introduce biases which invalidate equation \ref{eq:mub}. These potential sources of systematic error will be 
evaluated in section \ref{sec:alpha_beta_sys}.

The $\chi^2$ corresponding to that of Eq.\ \ref{eq:mub} is given as:

\begin{equation}
\chi^2=\sum_{\mathrm{SNe}}{\frac{(\mu_B - \mu(z,\om,\ola,w))^2}{
\sigma_{\mathrm{tot}}^2 + \sigma_{\mathrm{sys}}^2+ \sum_{ij}c_{i}c_jC_{ij}}.}
\label{eq:biased_chi}
\end{equation}
The sum in the dominator represents the statistical uncertainty as obtained from the light-curve fit with  $C_{ij}$ representing the covariance matrix of fit 
parameters: peak magnitudes, color and stretch (i.e. $C_{11}=\sigma_{m_B}^2$) 
and 
$c_i=\{1,\alpha,-\beta\}$ are the corresponding correction parameters. 
$\sigma_{\rm tot}$ represents an 
astrophysical dispersion obtained by adding in quadrature the dispersion due to lensing, $\sigma_{\rm lens}=0.093z$ (see Section \ref{sec:lensing}), the 
uncertainty in the Milky-Way dust extinction correction 
(see Section \ref{sec:mw_sys_ext}) and  a term reflecting the uncertainty due 
to host galaxy peculiar velocities of 300~km/s.
The  dispersion term $\sigma_{\mathrm{sys}}$ contains  an 
observed sample-dependent dispersion due to possible 
unaccounted-for systematic errors. In section \ref{sec:outlier} we discuss 
the contribution  $\sigma_{\rm sys}$ further.

Note that  Eq.\ \ref{eq:biased_chi} can be derived using minimization of a generalized $\chi^2$.
  Defining a residual vector for a supernova $\textbf{R} = \left(\mu_{B} - \mu_{\mathrm{model}}, s - s', c - c' \right)$ and supposing that the light-curve fit returns covariance matrix $\textbf{C}$, we can write 
\begin{equation}
\chi^2 = \sum_{\mathrm{SNe}} \textbf{R}^T \textbf{C}^{-1} \textbf{R}.
\end{equation}
Here, $s'$ and $c'$ take the role of the true stretch and color, which have to be estimated from the measured ones. 
 Minimizing this equation over all possible values of $s'$ and $c'$ gives the $\chi^2$ in Eq.\ \ref{eq:biased_chi}.  
The $\chi^2$ is minimized, not marginalized, over $\alpha$ and $\beta$; marginalization would yield a biased result due to the asymmetry of the $\chi^2$ about the minimum.  

\begin{figure}[htp]
\begin{center}

\includegraphics[width=0.23\textwidth]{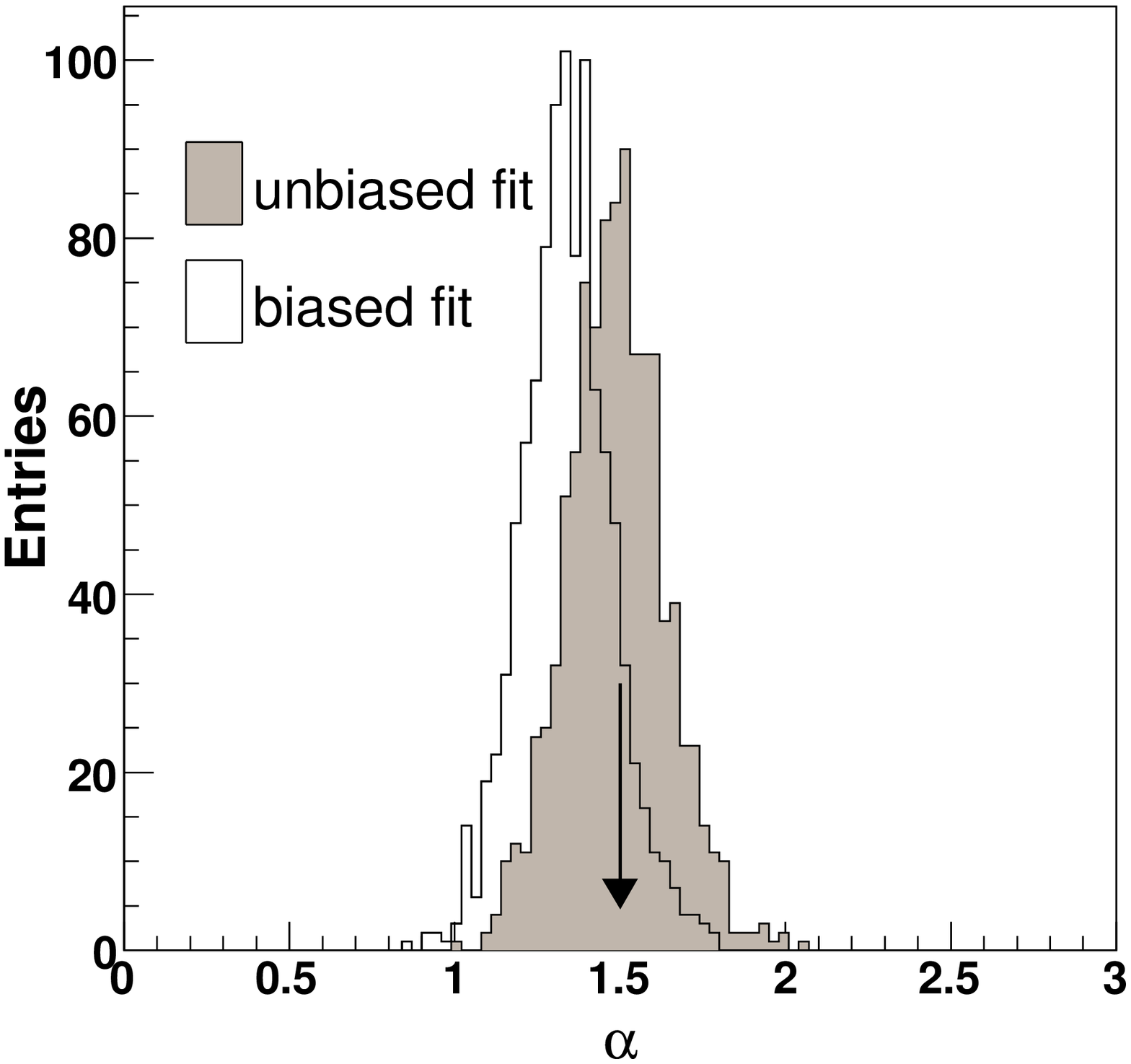}
\hfill
\includegraphics[width=0.23\textwidth]{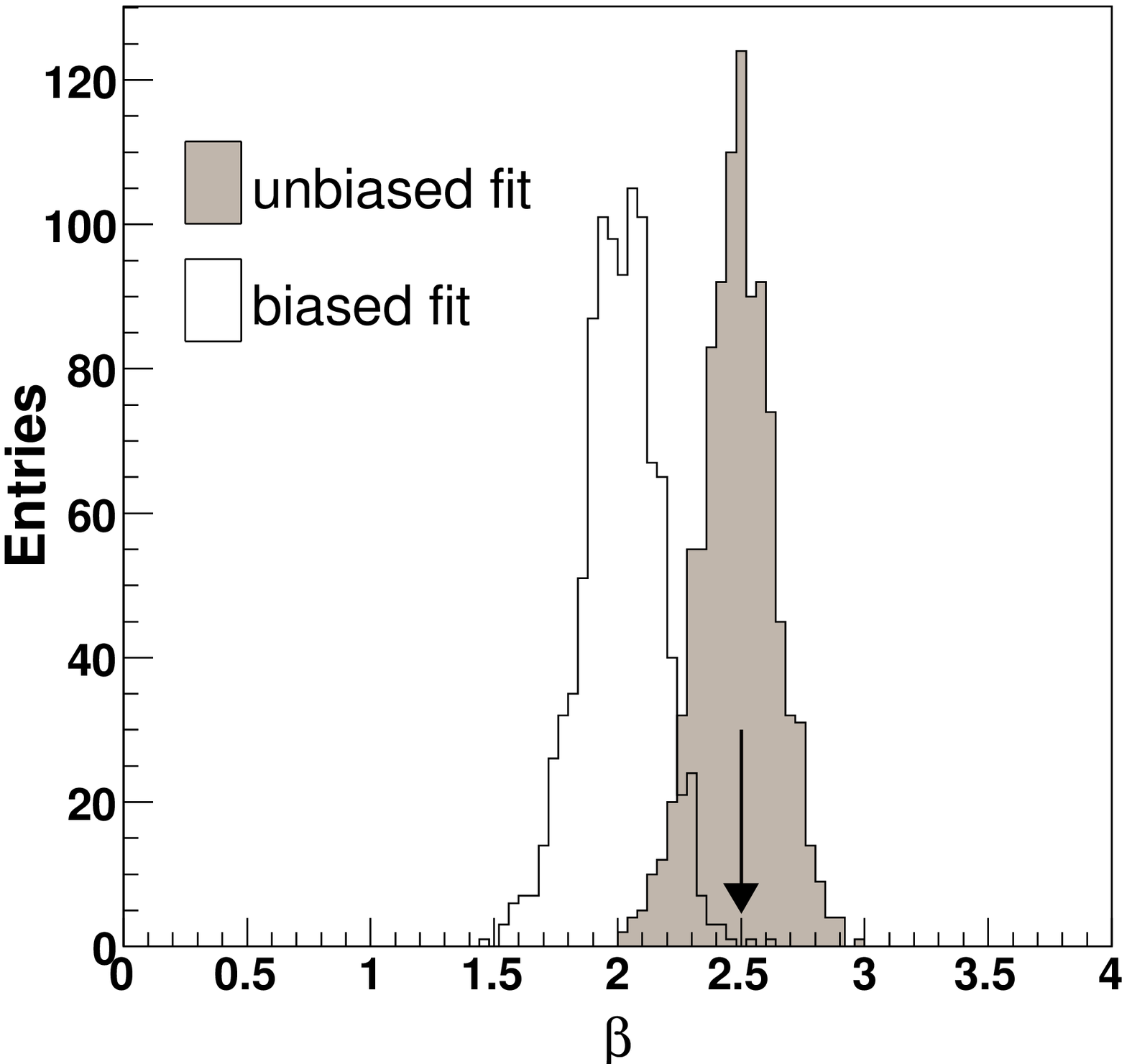}
\caption{Monte Carlo simulation of the resulting $\alpha$ (left) and $\beta$ (right) distributions as fitted with the unbiased and biased method. The true values $\alpha=1.5$ and $\beta=2.5$ are represented by the arrows. }
\label{fig:betabias}
\end{center}
\end{figure}

Frequently, Eq. \ref{eq:biased_chi} is minimized by updating the denominator 
iteratively, i.e. only between minimizations (see for example 
\cite{astier05}). 
As shown in Fig. \ref{fig:betabias} and discussed next, 
this method produces biased fit results, an artifact previously noted by 
\citet{wang06}.

We use a Monte Carlo simulation to estimate any biases from the 
fitting procedure. Random supernova samples resembling the observed 
one are generated and then fitted.
The true stretch and color are sampled from 
a normal distribution of width 0.1 and for the 
peak magnitude an intrinsic dispersion of 0.15 magnitudes is assumed. A 
further dispersion
 corresponding to the measurement errors is added. By construction, 
the SN 
samples have the same 
redshift and stretch, color and 
peak magnitude uncertainties as the real sample. The test values 
for $\alpha$ and $\beta$ were chosen as 1.5 and 2.5.
This bias on  $\alpha$ and  $\beta$, as would be obtained from the iterative method's 
fits to the simulated 
data sets, is visible in Figure \ref{fig:betabias} as the unshaded histogram. 
The large potential bias on $\beta$ ($\Delta\beta \sim -0.5$), if the $\chi^2$
 had been chosen  according to Equation  \ref{eq:biased_chi} with the iteratively updated 
denominator, is a result of the fact that the measurement error on $c$ for 
high redshift SNe is
 similar to and often even exceeds the width of the color distribution 
itself. 

We have investigated other sources of bias in the fitted parameters.
A measurement bias 
will be introduced because overall, brighter SNe will have smaller 
photometric errors, and hence larger weights, than dimmer ones. 
If the photometric error bars are small enough that the intrinsic dispersion 
dominates the uncertainty, this bias will 
be small. Hence low-redshift, well observed SNe are biased less than   
high-redshift, poorly observed SNe, resulting in biased cosmological 
parameters. This bias was studied using the Monte Carlo simulation described above. For 
the sample under investigation it was found to introduce a bias $\delta M =0.01$. In principle this bias can be corrected; however, 
since it is roughly a factor of three 
smaller than the statistical or systematic uncertainties, we choose not to carry out this step.

\subsection{Robust statistics}
\begin{figure}[ht]
\begin{center}

\includegraphics[angle=90,width=0.23\textwidth]{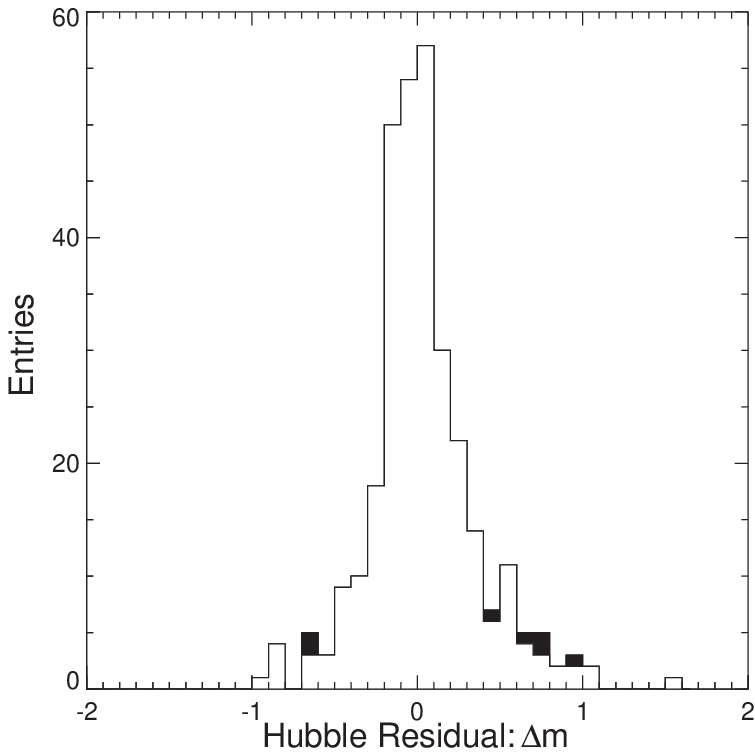}
\hfill
\includegraphics[angle=90,width=0.23\textwidth]{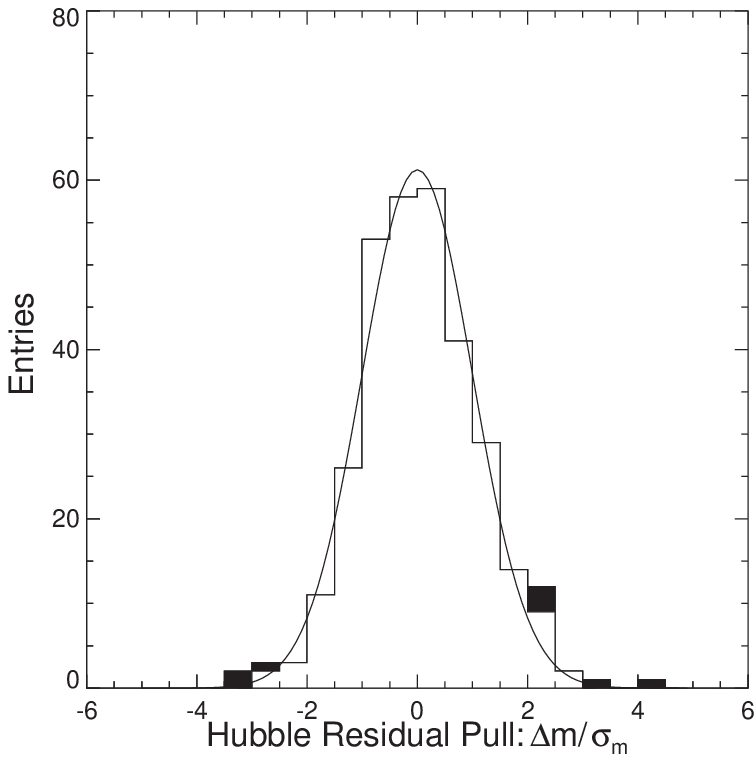}
\caption{Residual of restframe, stretch and color corrected,  B-band magnitude (left) and pull distribution  (right) from the best fitting cosmology. The filled histogram shows the rejected outliers. The pull distribution is overlayed with a normal distribution of unit width.}
\label{fig:hpull}
\end{center}
\end{figure}

Figure \ref{fig:hpull} shows the distribution of rest-frame B-band
corrected magnitude residuals (left)  from the 
best fit as obtained with the full data set. The right plot shows the pull distribution, where  the pull is  defined as the corrected B magnitude residual 
divided by its 
uncertainty. The 
distributions have outliers which, if interpreted as statistical 
fluctuations, are highly improbable. Hence these 
outliers point to non-Gaussian behavior of the underlying data, due to either 
systematic errors in the observations, contamination or 
intrinsic variations in 
Type Ia SNe. The fact that an outlier is present even in the high 
quality  SNLS supernova set (see Table \ref{tb:cut_number}) suggests that 
 contamination or unmodeled intrinsic variations might be 
present. However, other samples that typically were observed with a more heterogeneous set of telescopes and instruments 
show larger fractions of outliers, 
indicating additional potential observation-related problems.

\label{sec:outlier}

In order to limit the 
influence of outliers, we use 
a robust analysis technique. First,  the SN samples are fit for $M$, the absolute magnitude of the SNe, using median 
statistics (see \citet{Gott:2000mv} for a discussion of median statistics in the context of SN cosmology). The quantity  minimized is  $\chi = \sum_{\rm SNe} \frac{|\mu_B - \mu_{\mathrm{model}}|}{\sigma}$, where the uncertainty $\sigma$ in the denominator includes the covariance terms in the denominator of the right hand side of Equation \ref{eq:biased_chi}.  We then proceed to fit each sample by itself using the $\alpha$, $\beta$, and $\om$ from the combined fit, as $\chi$ is not a well-behaved quantity for small numbers of SNe.

For each sample, we remove SNe with a  pull exceeding a certain value $\sigma_{\rm cut}$ relative 
to the median fit of the sample. 
Currently available algorithms, which correct the peak magnitude using,  e.g.,
stretch or $\Delta m_{15}$,  are capable of standardizing SNe Ia to a level of
 $\sim0.10-0.15$ magnitudes. To reflect this we add in quadrature 
a systematic dispersion
  to 
the known uncertainties. The list of known uncertainties 
include observational errors, 
distance modulus uncertainties due to peculiar velocities 
(with $\Delta v=$300~km/s) and gravitational lensing 
(relevant only for the highest redshift SNe; see section 
\ref{sec:lensing} for a discussion). 
The additional systematic dispersion has two components: a common irreducible 
one, possibly associated with intrinsic variations in the SN
explosion mechanism, as well as an observer-dependent component.
To obtain 
self-consistency the systematic dispersion is 
recalculated during the analysis. 
One starts by assuming 
a systematic dispersion of $\sigma_{\rm sys}=0.15$~magnitudes, then computes the best fitting cosmology for the particular sample using median statistics, 
removes the outlier SNe with residuals larger than a cut value $\sigma_{\rm cut}$, iterates $\sigma_{\rm sys}$ such that the total $\chi^2$ per degree of freedom is unity, and in a final step redetermines the best fitting cosmology using regular $\chi^2$ statistics to obtain an updated $\sigma_{\rm sys}$. From that point in the analysis, after outliers are rejected and  $\sigma_{\rm sys}$ determined, only regular $\chi^2$ statistics are applied.

\begin{figure}[htp]
\begin{center}
\includegraphics[width=0.235\textwidth]{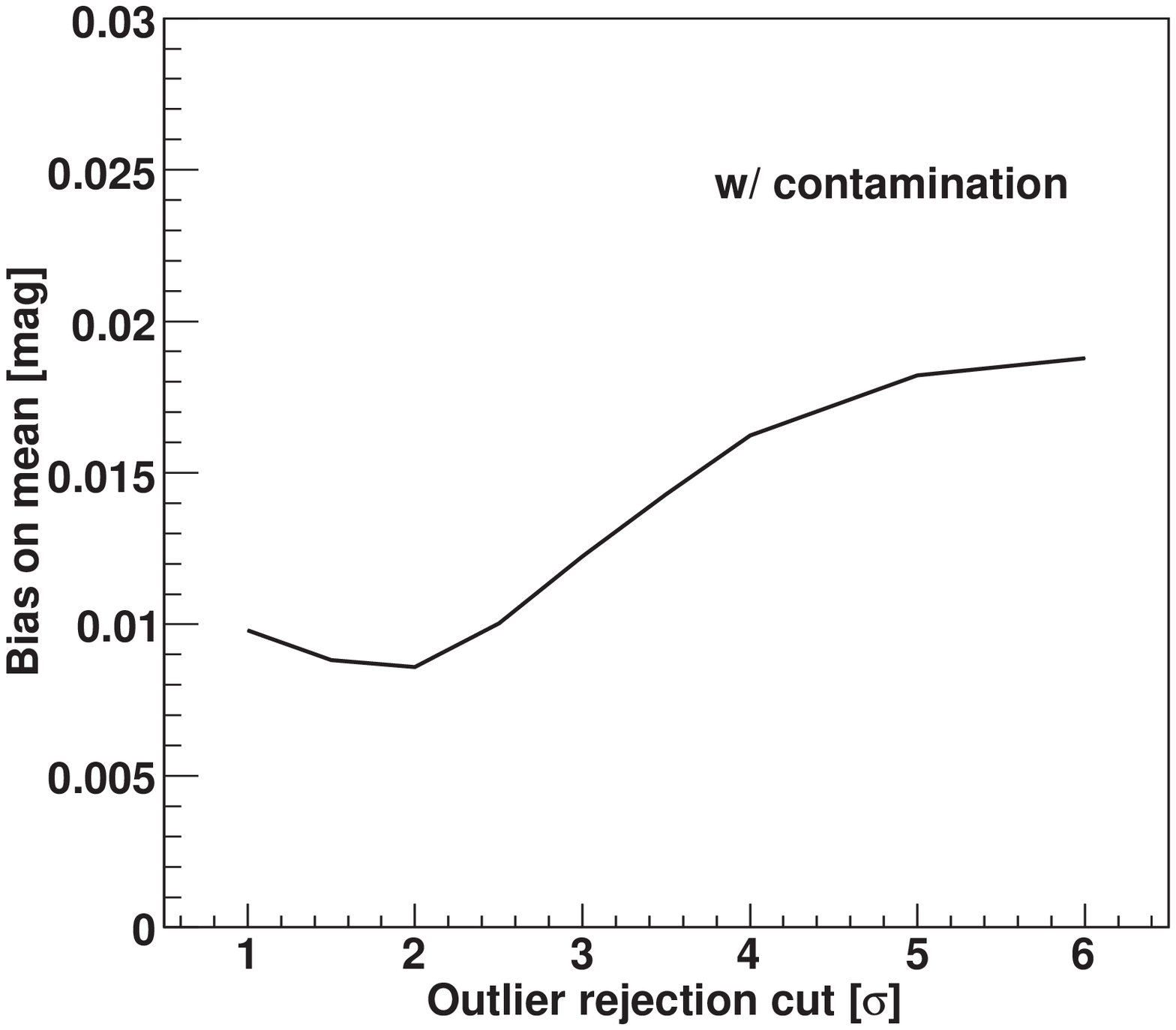}
\includegraphics[width=0.235\textwidth]{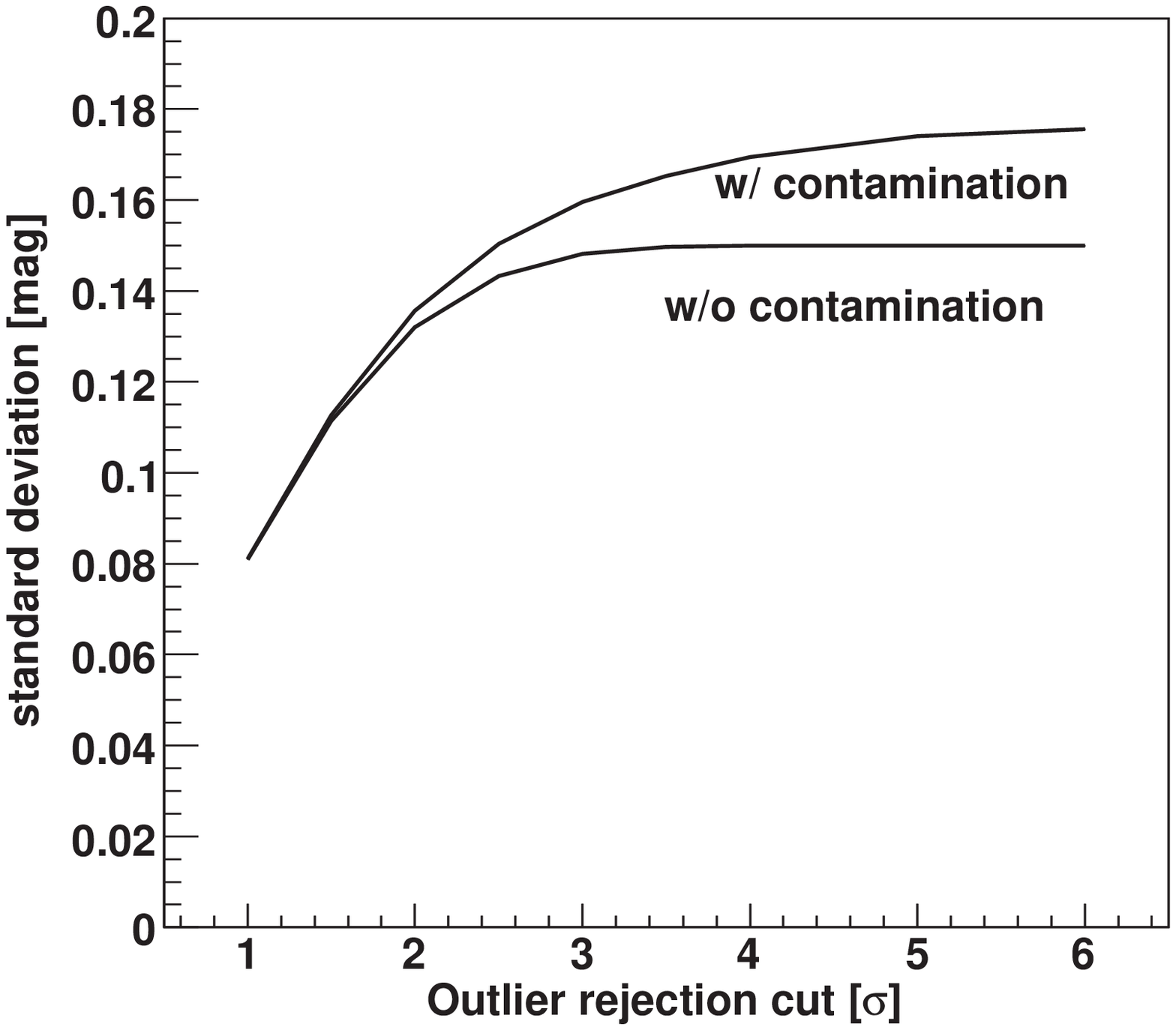}
\caption{Mock simulation of bias (left panel) and standard deviation (right) 
of the mean magnitude as a function of the outlier rejection cut. 
The simulated SN set consists of one population of 
270 SNe with intrinsic dispersion of 0.15 magnitudes and 
 zero mean and a second population of 50 SNe with intrinsic 
 dispersion of 0.26 mag and mean 0.13 mag.
The effect of outlier rejection on a single population without contamination 
is shown as a reference curve.}
\label{fig:bias_std}
\end{center}
\end{figure}

When using a robust analysis, it is necessary to check that $a)$ in the 
absence of contamination the results are not altered from the Gaussian case 
and $b)$ in presence of a contaminating contribution, the impact of it 
is indeed reduced.
In order to investigate this, we begin with a model for the contamination. 
We assume the data sample to be  
composed of two  types of objects, 
one representing the desired SNe Ia and a second contribution 
 characterizing the contamination.  We then use a maximum-likelihood
analysis of the observed pull distribution shown in Figure \ref{fig:hpull} (right) to determine the 
normalization, width and mean of the 
contaminating distribution. The uncontaminated pull distribution is
assumed to be a Gaussian distribution of unit width 
and zero mean.
The observed pull distribution
is best fitted by an additional contaminating contribution that is 50\% wider ($\sigma_{m}=0.23$ mag) and which has a mean
shifted by $\Delta m = 0.3 \sigma_m$, normalized to 18~\% of the area.
A mock simulation that is based on this
superposition of two normal distributions illustrates 
the benefits of using the 
robust analysis. 
Figure \ref{fig:bias_std} (right) shows the bias of the mean 
relative to the center of the main component as a function of the outlier 
rejection cut value.  Outlier rejection can reduce the bias by 
a factor of three with a remaining bias of less than 0.01 magnitude. Even for 
 a wide range  of contaminant parameters 
($\sigma_m=0.15-2;\Delta m=0-2$ magnitudes) the 
bias obtained for the robust analysis 
remains below 0.015 magnitudes. Only in cases where the contamination is  
larger than 30\% does the outlier rejection algorithm become 
unstable.

Besides reducing the potential bias due to contamination, 
robust statistics can also lead to tighter parameter constraints through 
reduction of the intrinsic dispersion.
The right panel of Figure  \ref{fig:bias_std}  shows 
for the simulated data the average standard 
deviation as a function of the outlier rejection cut for the 16~\% 
contamination case described above. As a reference, the case
 of a  
single uncontaminated population of SNe is shown as well. Note 
that a cut at $3\sigma$ reduces the dispersion noticeably in the case of a 
contaminated sample, while the uncontaminated single population is affected 
negligibly (the standard deviation is reduced by 1.3~\%, e.g. from 0.15 to 0.148 magnitudes).

For the real data, we consider two values $\sigma_{\rm cut}=2,3$ as well as 
the case in which all SNe are kept.
We chose as our main cut value $\sigma_{\rm cut}=3$ since, after application 
of the outlier rejection, 
standard $\chi^2$ statistics is still a good approximation while 
at the same time a potential bias introduced by contamination is 
significantly reduced. Note also that the impact of individual SNe 
that have residuals close to 
$\sigma_{\rm cut}$ is small for large statistics: an additional SN 
will shift the mean distance modulus of $N_{\rm SNe}$ by at 
most $\sigma_{\rm cut}/\sqrt{N_{\rm SNe}}$ standard deviations.  Hence for 
$N_{\rm SNe}\gsim 10$ and $\sigma_{\rm cut}=3$ the algorithm 
can be considered stable relative to fluctuations of individual SNe.   
\newpage 
\subsection{Sample characteristics, dispersion and pull}
\label{sec:diversity}
\begin{figure*}
\vspace{1cm}
\begin{center}
\includegraphics[width=1.2\textwidth,angle=90]{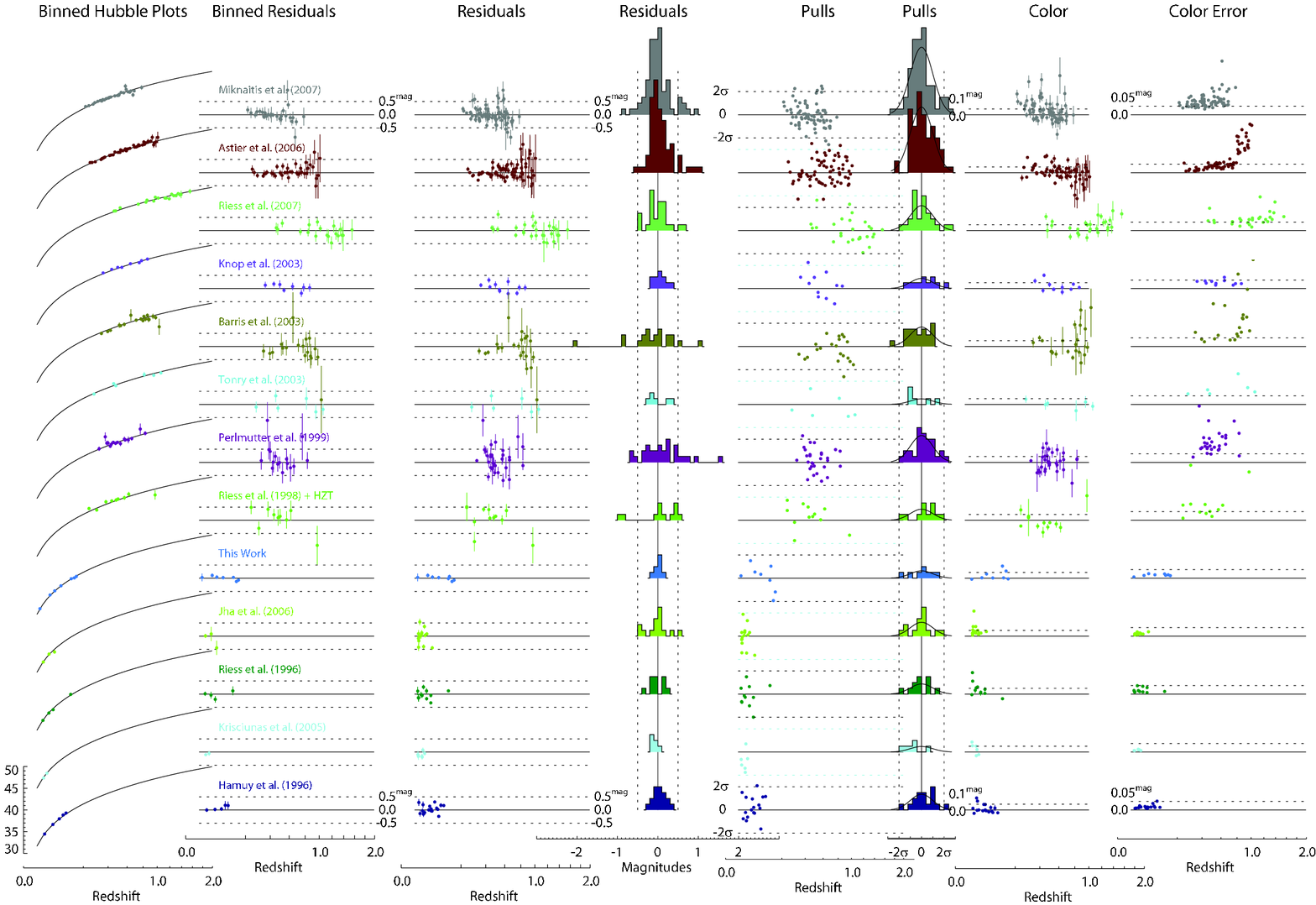}
\caption{From left to right: a) Hubble diagrams for the various samples; b) binned magnitude residuals from the best fit (bin-width: $\Delta z =0.01)$; c) unbinned magnitude residuals from the best fit; d) histogram of the residuals from the best fit; e) pull of individual SNe as a function of redshift; f) histogram of pulls; g) SN color as a function of redshift; h) uncertainty of the color measurement as an illustration of the photometric quality of the data.}
\label{fig:hubbleind}
\end{center}
\end{figure*}

Figure \ref{fig:hubbleind}  illustrates the heterogeneous character of the samples. It shows the Hubble and residual diagrams for the various samples, as well as the histogram of the SN residuals and pulls from the best fit. 
 The difference in photometric quality is 
 illustrated in the right-most column of Fig.\ \ref{fig:hubbleind},  
by showing the error on the color 
measurement. As can be seen, some samples show a significant redshift 
dependent gradient in the errors, while others have small, nearly 
constant errors (most notably the sample of \citet{knop03}). The sample of 
\citet{astier05} shows a small color uncertainty up to $z\leq 0.8$, and 
degrades significantly once the color measurement relies on the poorer 
$z$-band data (c.f. SALT2 \citep{salt2}, which is capable of incorporating lightcurve data bluer than 
rest-frame U).

\begin{table*}[htp]
\caption{Shown is the number of SNe passing the different outlier rejection cuts, as well as the sample dependent systematic dispersion ($\sigma_{\rm sys}$) and the RMS around the best fit model. The compilation obtained with the $\sigma_{\mathrm{cut}}= 3$ cut will be referred to as the Union robust set. \label{tb:cut_number}}
\begin{tabular}{ c | c c c | c c c | c c c }
\tableline
\tableline
 & 
\multicolumn{3}{c}{No Outlier Cut} & 
\multicolumn{3}{c}{\boldmath$ \sigma_{\mathrm{cut}}= 3$} & 
\multicolumn{3}{c}{$\sigma_{\mathrm{cut}}=2$} \\
Set & SNe & $\sigma_{\mathrm{sys}} (68\%)$ & RMS $ (68\%)$ &  SNe & $ \sigma_{\mathrm{sys}} (68\%) $ & RMS \boldmath$(68\%)$ & SNe & $\sigma_{\mathrm{sys}} (68\%)$ & RMS $ (68\%)$ \\
\tableline

Hamuy et al. (1996) & 17 & $ 0.14 ^{+ 0.04}_{- 0.03} $  & $0.16 ^{+ 0.03}_{- 0.03}$ & {\bf 17} & $ {\bf 0.14 ^{+ 0.04}_{- 0.03}}$ & \boldmath$ 0.16 ^{+ 0.03}_{- 0.03}$ &
 16 & $0.12 ^{+ 0.05}_{- 0.03}$ & $0.15 ^{+ 0.02}_{- 0.03}$ \\
Krisciunas et al. (2005) & 6 & $ 0.06 ^{+ 0.11}_{- 0.05} $  & $0.10 ^{+ 0.03}_{- 0.04}$ & {\bf 6} & $ {\bf 0.05 ^{+ 0.11}_{- 0.05}}$ & \boldmath$ 0.10 ^{+ 0.03}_{- 0.04}
$ & 6 & $0.08 ^{+ 0.12}_{- 0.07}$ & $0.12 ^{+ 0.03}_{- 0.04}$ \\
Riess et al. (1996) & 11 & $ 0.16 ^{+ 0.07}_{- 0.04} $  & $0.18 ^{+ 0.03}_{- 0.04}$ & {\bf 11} & $ {\bf 0.16 ^{+ 0.07}_{- 0.03}}$ & \boldmath$ 0.17 ^{+ 0.03}_{- 0.04}$ &
 11 & $0.18 ^{+ 0.08}_{- 0.04}$ & $0.20 ^{+ 0.04}_{- 0.05}$ \\
Jha et al. (2006) & 16 & $ 0.30 ^{+ 0.09}_{- 0.05} $  & $0.31 ^{+ 0.05}_{- 0.06}$ & {\bf 15} & $ {\bf 0.26 ^{+ 0.08}_{- 0.05}}$ & \boldmath$ 0.27 ^{+ 0.05}_{- 0.06}$ & 1
1 & $0.10 ^{+ 0.08}_{- 0.06}$ & $0.15 ^{+ 0.03}_{- 0.04}$ \\
This Work & 8 & $ 0.01 ^{+ 0.06}_{- 0.01} $  & $0.09 ^{+ 0.02}_{- 0.03}$ & {\bf 8} & $ {\bf 0.00 ^{+ 0.05}_{- 0.00}}$ & \boldmath$ 0.07 ^{+ 0.02}_{- 0.02}$ & 8 & $0.07 ^
{+ 0.06}_{- 0.03}$ & $0.12 ^{+ 0.03}_{- 0.04}$ \\
\tableline
Riess et al. (1998) + HZT & 12 & $ 0.29 ^{+ 0.20}_{- 0.11} $  & $0.50 ^{+ 0.09}_{- 0.12}$ & {\bf 12} & $ {\bf 0.28 ^{+ 0.19}_{- 0.10}}$ & \boldmath$ 0.48 ^{+ 0.09}_{- 0.
11}$ & 10 & $0.16 ^{+ 0.19}_{- 0.10}$ & $0.49 ^{+ 0.10}_{- 0.13}$ \\
Perlmutter et al. (1999) & 30 & $ 0.43 ^{+ 0.13}_{- 0.09} $  & $0.65 ^{+ 0.08}_{- 0.09}$ & {\bf 29} & $ {\bf 0.33 ^{+ 0.10}_{- 0.07}}$ & \boldmath$ 0.50 ^{+ 0.06}_{- 0.0
7}$ & 24 & $0.19 ^{+ 0.11}_{- 0.09}$ & $0.43 ^{+ 0.06}_{- 0.07}$ \\
Tonry et al. (2003) & 6 & $ 0.00 ^{+ 0.33}_{- 0.00} $  & $0.24 ^{+ 0.06}_{- 0.09}$ & {\bf 6} & $ {\bf 0.06 ^{+ 0.28}_{- 0.06}}$ & \boldmath$ 0.24 ^{+ 0.06}_{- 0.09}$ & 6
 & $0.00 ^{+ 0.32}_{- 0.00}$ & $0.26 ^{+ 0.07}_{- 0.09}$ \\
Barris et al. (2003) & 22 & $ 0.31 ^{+ 0.12}_{- 0.07} $  & $0.64 ^{+ 0.09}_{- 0.11}$ & {\bf 21} & $ {\bf 0.23 ^{+ 0.12}_{- 0.08}}$ & \boldmath$ 0.62 ^{+ 0.09}_{- 0.10}$ 
& 19 & $0.11 ^{+ 0.16}_{- 0.11}$ & $0.71 ^{+ 0.11}_{- 0.13}$ \\
Knop et al. (2003) & 11 & $ 0.10 ^{+ 0.08}_{- 0.04} $  & $0.17 ^{+ 0.03}_{- 0.04}$ & {\bf 11} & $ {\bf 0.10 ^{+ 0.07}_{- 0.04}}$ & \boldmath$ 0.17 ^{+ 0.03}_{- 0.04}$ & 
11 & $0.11 ^{+ 0.08}_{- 0.05}$ & $0.18 ^{+ 0.04}_{- 0.04}$ \\
Riess et al. (2007) & 29 & $ 0.22 ^{+ 0.05}_{- 0.04} $  & $0.31 ^{+ 0.04}_{- 0.04}$ & {\bf 27} & $ {\bf 0.16 ^{+ 0.05}_{- 0.04}}$ & \boldmath$ 0.26 ^{+ 0.03}_{- 0.04}$ &
 24 & $0.08 ^{+ 0.05}_{- 0.06}$ & $0.22 ^{+ 0.03}_{- 0.03}$ \\
Astier et al. (2006) & 72 & $ 0.14 ^{+ 0.03}_{- 0.02} $  & $0.31 ^{+ 0.03}_{- 0.03}$ & {\bf 71} & $ {\bf 0.12 ^{+ 0.03}_{- 0.02}}$ & \boldmath$ 0.29 ^{+ 0.02}_{- 0.03}$ 
& 70 & $0.12 ^{+ 0.03}_{- 0.02}$ & $0.30 ^{+ 0.02}_{- 0.03}$ \\
Miknaitis et al. (2007) & 75 & $ 0.21 ^{+ 0.04}_{- 0.03} $  & $0.32 ^{+ 0.02}_{- 0.03}$ & {\bf 73} & $ {\bf 0.18 ^{+ 0.04}_{- 0.03}}$ & \boldmath$ 0.30 ^{+ 0.02}_{- 0.03
}$ & 66 & $0.00 ^{+ 0.05}_{- 0.00}$ & $0.23 ^{+ 0.02}_{- 0.02}$ \\
\tableline
Union & 315 &  &  & {\bf 307} &  &  & 282 & \\
\tableline
\end{tabular}
\end{table*}

Our analysis is optimized for large, multi-color samples such as that of Astier et al. (2006), since these now dominate the total sample.  There is often a better analysis approach for any given specific sample that would emphasise the strengths of that sample's measurements and yield a tighter dispersion and more statistical weight.   However, for this combined analysis of many samples it was more important to use a single uniform analysis for every sample, at the cost of degrading the results for some of the smaller samples.  This particularly affects some of the very earliest samples, such as \citet{riess98}, \citet{perl99}, and \citet{barris04}, where the color measurements had originally been used with different priors concerning the dust distribution.   Treating these samples with the current analysis thus gives significantly larger dispersions (and hence less weight) to these samples than their original analyses.   As a check, we have verified that by repeating the analysis according to \citet{perl99} we reproduce the original dispersions.

\begin{figure}
\begin{center}
\includegraphics[width=.42\textwidth,angle=180]{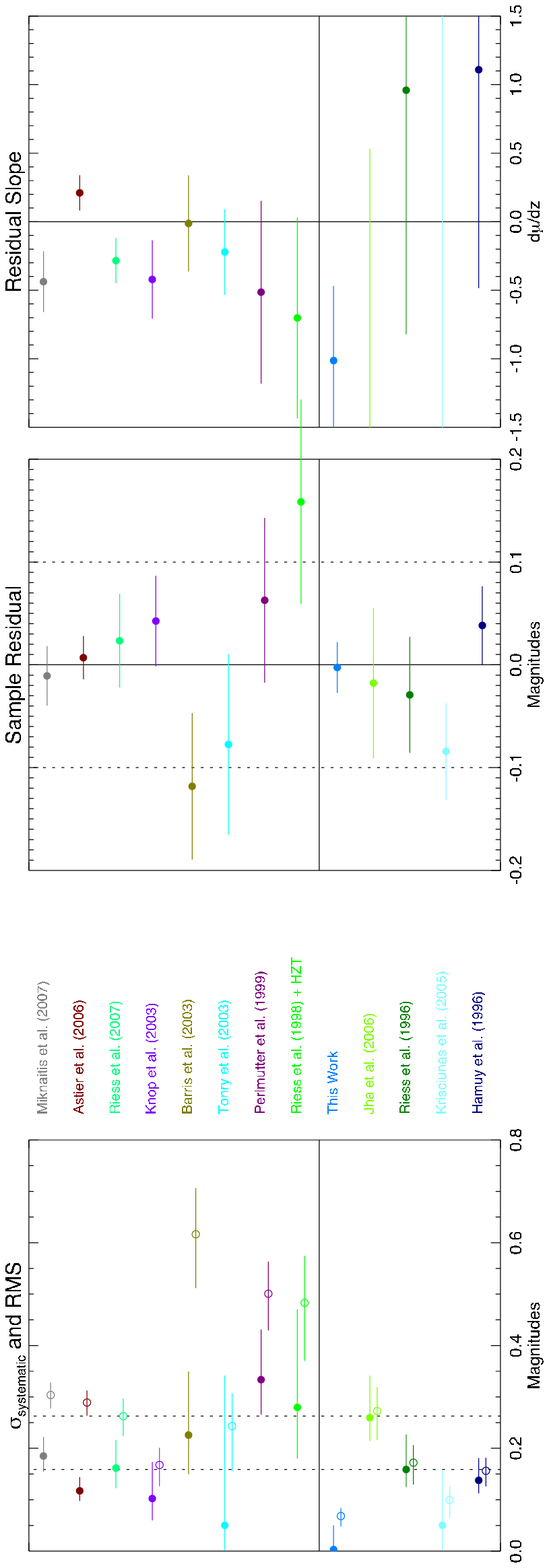}
\caption{From left to right: Systematic dispersion (filled circles) and RMS around the best fit model (empty circles); The mean, sample averaged, deviation from the best fit model; The slope of the Hubble-residual (in magnitudes) 
versus redshift, $d\mu_{\rm residual}/dz$. The parameters characterizing 
the different samples are used to uncover potential systematic problems. }
\label{fig:diagnostics}
\end{center}
\end{figure}

Figure \ref{fig:diagnostics} shows diagnostic variables used to 
test for consistency between the various samples.
The leftmost plot shows the systematic dispersion and RMS around the best fit model. 
One expects that there 
is an intrinsic dispersion associated with all SNe, which provides a lower 
limit to the sample dependent systematic dispersion. To estimate the intrinsic 
dispersion one can look at various quantities, as for example the smallest 
$\sigma_{\rm sys}$ or, perhaps more appropriate, by the median of 
$\sigma_{\rm sys}$. The median of $\sigma_{\rm sys}$, which is about 0.15 mag (shown as the leftmost dashed vertical line), 
is a robust measure of the intrinsic dispersion, as long as the 
majority of samples are not dominated by observer dependent, unaccounted-for
uncertainties. 

As a test for tension between the data sets, we compare for each sample the average residual from the best 
fit cosmology. This is shown in the middle panel of 
Fig.\ \ref{fig:diagnostics}. As can be seen, 
most samples fall within $1\sigma$ and none deviate by more then $2\sigma$. 
The larger samples show no 
indication of inconsistency. This changes 
if one considers, instead of the mean, the slope, $d\mu_{\rm residual}/dz$, 
of the residuals as a function of the redshift. The right panel of Fig.\  
\ref{fig:diagnostics}  
shows a large fraction of significant outliers in the slope. The largest slope 
outlier is found for the \citet{essence_m_07} sample (see also the middle panel of 
Fig.\ \ref{fig:hubbleind}). The sign of the slope is consistent with the presence of 
a Malmquist bias (see \citet{essence_wv_07} for a discussion). The uncertainties associated with such a Malmquist bias are 
discussed in Section \ref{sec:malmquist}.
While in general there is no clear trend in the 
sign of the slope deviations, it is clear that any results that depend on the 
detailed slope, such as a changing equation of state, should be treated with caution.

\begin{figure}[htp!]
\begin{center}
\includegraphics[width=0.35\textwidth,angle=90]{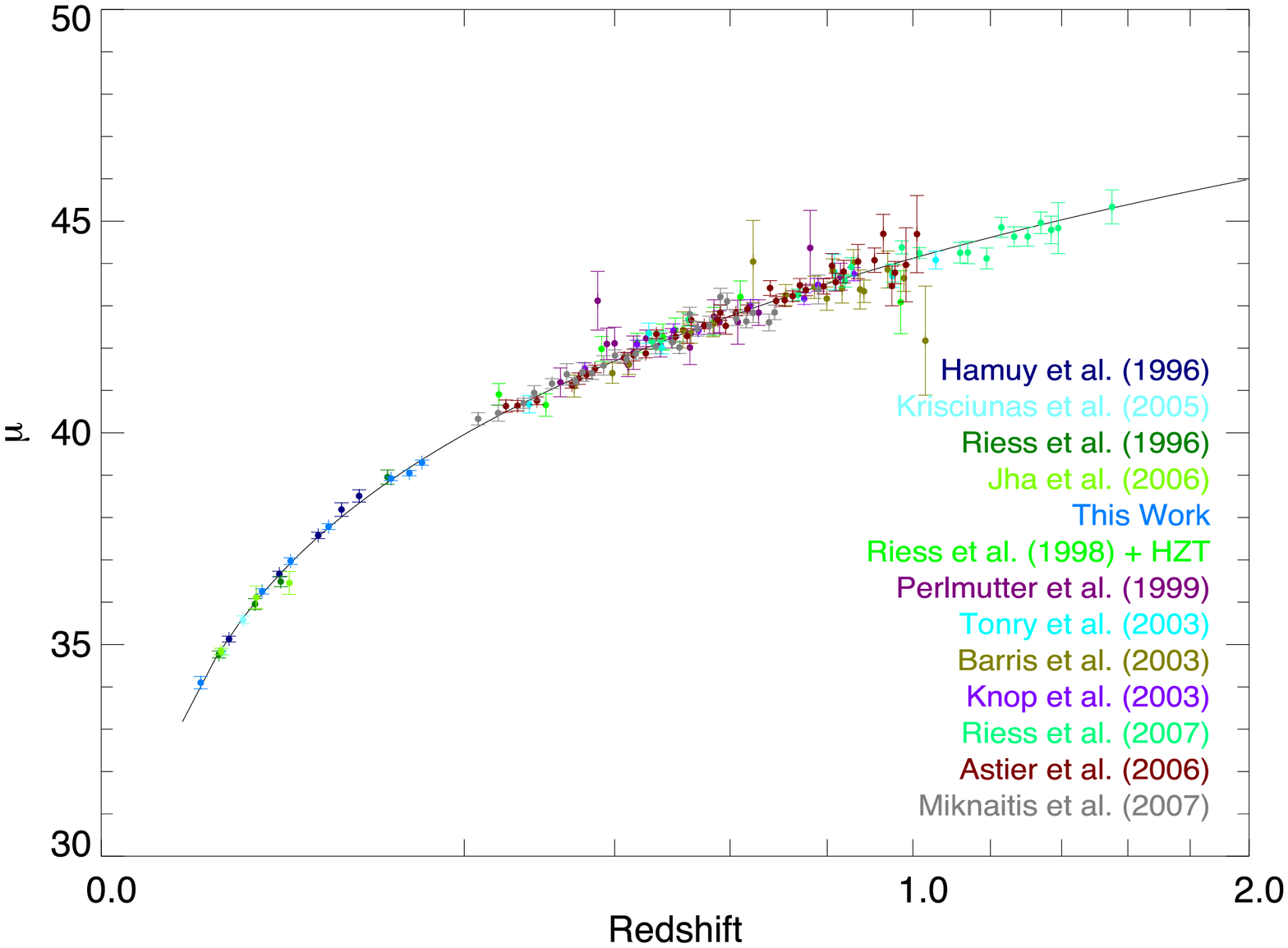}
\includegraphics[width=0.35\textwidth,angle=90]{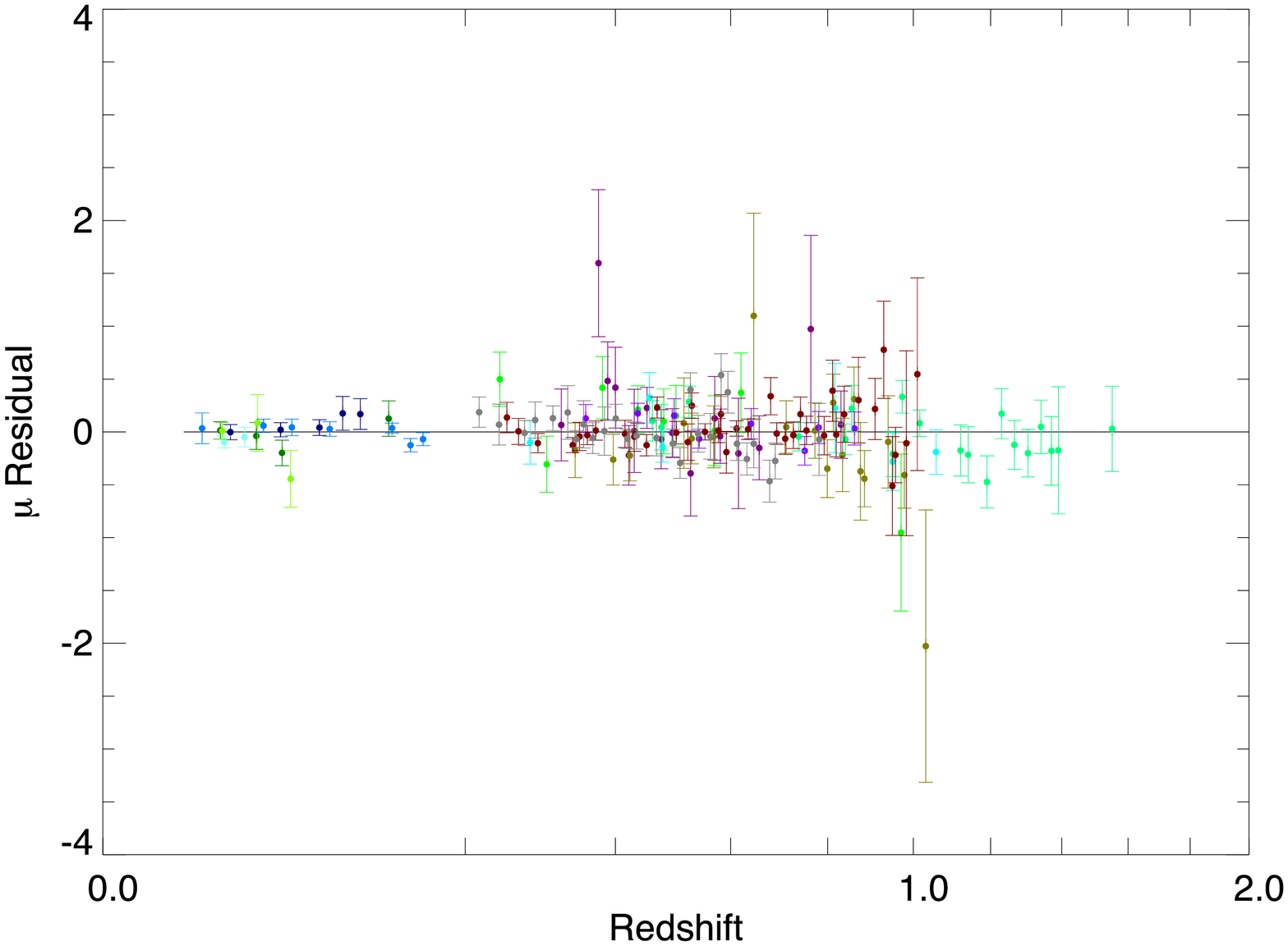}
\caption{Top: Binned Hubble diagram (bin-size $\Delta z=0.01$). 
Bottom: Binned residuals 
from the best fitting cosmology.}
\label{fig:hubble_diagram}
\end{center}
\end{figure}


\section{Systematic errors}
\label{sec:systematics}
Detailed studies of the systematic effects have been published as part of the 
analysis of individual data sets. The list includes 
photometric zero points, Vega spectrum, lightcurve fitting, contamination, evolution, 
Malmquist bias, K-corrections and gravitational lensing, which have also 
been discussed in earlier work by authors of this paper \citep{stretch,perl99,knop03,astier05, pilar07, essence_wv_07}. 

Some sources of systematic errors are common to all surveys and will be 
specifically addressed for the full sample. 
Other sources of 
systematic errors are controlled by the individual observers. 
The degree with which 
this  has been done for the various data samples entering the analysis is very 
different. 
The SNLS---which is using a single telescope and 
instrument for the search and followup, and which has detailed multi-band 
photometry for nearly all its SNe---has a strong handle on 
a subset of the observation-dependent  systematics uncertainties. With the exception of the ESSENCE SN data sample, other high redshift samples  
are smaller and will contribute less to the final results.

We handle the two types of systematic errors separately: systematic errors that can be associated with a sample (e.g. due to observational effects), and those that are common to all the samples (e.g. due to astrophysical or fundamental calibration effects). 
To first order, the measurement of cosmological parameters depends on the 
relative brightness of nearby SNe 
($z\sim0.05$) compared to their high redshift counterparts ($z\sim 0.5$). If low and high redshift SNe are different, this can be absorbed in different 
absolute magnitudes $M$. 
We hence cast the common systematic uncertainties into an uncertainty in the 
difference $\Delta M = M_{\rm low-z}-M_{\rm high-z}$.
`We have chosen 
$z_{\rm div}=0.2$ as the dividing redshift as it conveniently splits the 
samples according nearby and distant SN searches.  Note, however, 
that our resulting systematic errors change by less than 25\% of its value for 
$z_{\rm div}$ in the range $0.1-0.5$.
In addition we allow 
for a set of extra parameters, $\Delta M_i$, one for each sample $i$.

Systematic uncertainties are then propagated by adding these nuisance 
parameters to $\mu_B$: 
\begin{equation} 
\mu_B^\prime = 
\left\{ 
\begin{array}{ll}
\mu_B  + \Delta M_i & \mbox{for~~} z_{\rm div}<0.2\\
\mu_B  + \Delta M_i + \Delta M & \mbox{for~~} z_{\rm div}\ge 0.2\\
\end{array}
\right.
\label{eq:nuisance}
\end{equation}
with the term ${\Delta M^2}/{\sigma_{\Delta M}^2}+\sum_{i=1}^{N_{\rm samples}}{\Delta M_i^2}/{\sigma_{\Delta M_i}^2}$ being added to the $\chi^2$ 
as defined through Eq. \ref{eq:biased_chi}. 
We have checked that this treatment 
of systematic errors is consistent (in our case to better than 5~\% of its 
value) with 
the  more common procedure, applicable to one-dimensional constraints, 
in which part of the input data is offset by $\pm\sigma_{\Delta M}$ to obtain 
the systematic variations in the resulting  parameter (e.g. $w$ or $\om$).

In the following we discuss the different contributions to 
$\sigma_{\Delta M}$, and summarize them in section \ref{sec:sys_summ}.
 The resulting systematic errors on the 
cosmological parameters are discussed in section \ref{sec:cosmo}.

\subsection{Stretch \& evolution}
\label{sec:alpha_beta_sys}
With the large statistics at hand one can test the errors associated with the empirical stretch and color corrections.
These corrections would become sources of systematic error
if $a)$ different SN populations were to require 
different corrections and $b)$ if the SN populations were to 
show differences 
between nearby and distant objects (either due to selection effects or 
due to evolution of the SN environment). 

A potential 
redshift dependence of the correction parameters can be 
tested by separately fitting low redshift and high redshift objects. 
For this test, a \lcdm{} cosmology was assumed with $\om=0.28$ and $\om=0.72$ 
(the values we obtain from the fit of the full sample);
however, the results 
are rather insensitive to the assumed cosmological parameters. 
The obtained fit parameters  $\alpha$ and  $\beta$ are presented in 
table \ref{tb:system}. 

The values of $\beta$ at high and low redshift agree very well, providing strong constraints on evolution of the color-correction. Such evolution effects could arise, for example, due to a different mix of dust reddening and intrinsic color at different redshifts. The fact that $\beta$ agrees so well supports the choice of the empirical color correction\footnote{
Note that if $\beta$ were not obtained by fitting but instead was 
fixed, e.g. $\beta=R_{\rm B}=4.1$, a bias can be expected (and might have already been observed, see \citet{hubbub_conley}) if the average reddening changes as a function of redshift.}.

\label{sec:colorstretch}
\begin{table}
\begin{center}
\begin{tabular}{ c c c c c c }
\tableline
\tableline
subset & $N_{\rm SN}$ & $\alpha$ & $\beta$ & $\om^{a}$ & $w^{b}$  \\ 
\tableline
all & 307 & 1.24(0.10) & 2.28(0.11) & 0.29(0.03) & -0.97(0.06) \\ 
$z>0.2$ & 250 & 1.46(0.16) & 2.26(0.14) &  - & -  \\ 
$z\le 0.2$ & 57 & 1.07(0.12) & 2.23(0.21) &  - & -  \\ 
$s < 0.96$ & 155 & 1.56(0.27) & 2.18(0.18) & 0.27(0.05) & -0.98(0.09) \\ 
$s \ge 0.96$ & 152 & 1.51(0.37) & 2.34(0.17) & 0.30(0.04) & -0.93(0.07) \\ 
\tableline

\end{tabular}
\end{center}
\caption{Fit parameters as obtained for different SN subsamples. $(^a)$A flat Universe was assumed in the constraints on $\om$.  $(^b)$Constraints on $w$ were obtained from combining SNe
with  CMB and BAO measurements. A flat Universe was also assumed.
(see section \ref{sec:cosmo} for more details).}
\label{tb:system}
\end{table}
The $\alpha$ at low-redshift and high-redshift are only 
marginally 
 consistent with each other. We will take the difference at face value and 
estimate the impact it would have on the final result.  The average stretch  
is $\left<s\right>\approx0.96$ and 
hence the difference in the average stretch correction is 
$\left<1-s\right> \Delta \alpha \approx 0.015$. If $\alpha$ indeed is 
redshift dependent and this was 
not accounted for, one would obtain a bias of $\Delta M =  0.015$ mags.

Effects of potentially different SN populations should be considered  as well. 
It has recently been argued by \citet{sb05} and \citet {mannucci06} that one needs to allow 
for two types of SN-progenitor timescales to explain the observed rates in 
different galaxy types. One class of objects traces the 
star formation rate directly, while the second class has a  
 delay time trailing the star formation rate by a few billion years.
If indeed two populations are present, they might evolve  
differently with 
redshift. It is therefore important to check that the empirical corrections 
suit both populations.
To test the effect of different SN populations one can subdivide the 
sample according to  SN sub-types or host environments 
\citep{sullivan03, Howell:2007rb}. 
\citet{sullivan06} have found using well observed 
SNe and hosts from SNLS 
that the stretch of a light curve is  correlated with its 
host environment as well as with the two classes of SN-progenitor systems postulated by \citet{sb05,mannucci06}.
Therefore, we divide the SN sample into two approximate 
equally large samples with $s<0.96$ and $s \ge 0.96$. The two independent samples
are then fitted, with the results shown in Table 
\ref{tb:system}. The resulting parameters $\om$ (for a flat Universe) 
and $w$ (for a flat Universe together with BAO+CMB) for the two samples are 
less than $1\sigma$ apart and hence there is no evidence for an underlying 
systematic effect. 
Nevertheless, this will be a very important number to watch, once future 
high quality  SN data sets will be added. (Note that, while the resulting values of $\alpha$s for the two samples are 
consistent with each other, they appear inconsistent with the value obtained for the complete sample. This apparent inconsistency 
arises in part due to a bias introduced by dividing the stretch distribution in the middle. 
Larger stretch SNe, misclassified due to measurement errors as belonging to the low stretch SNe sample, 
as well as lower stretch SNe, misclassified as belonging  to the large stretch SNe sample, will for both samples 
result in a $\alpha$ biased to larger values.)

We have also investigated whether  the sample can be sub-divided according to the 
color of the SNe. We found that the results of such a test can be very misleading. 
While in principle one would expect to find that the 
best fitted cosmological parameters do not depend on color selection 
criteria (e.g. $c<c_{\rm cut}$ and $c>c_{\rm cut}$), we find by means of Monte Carlo simulation 
described in Section \ref{sec:ubias} that a 
significant bias is introduced into the measurements. This bias is also observed in the data.
For example, by choosing $c_{\rm cut}=0.02$ we find that for our sample of SNe  
$\om$ changes by $\pm0.1$. The bias arises from truncating an asymmetric 
distribution. In the case of color, the asymmetry in the distribution  
is introduced by the fact that extinction by dust leads only to reddening.
Hence the number of objects which would belong to 
$c_{\rm true}<c_{\rm cut}$ but, due to a large measurement error, 
are fitted with  $c_{\rm observed}>c_{\rm cut}$,  are 
not compensated by objects misclassified in the opposite way. The number of 
 misclassified objects is a function of the measurement errors, and hence is 
larger towards higher redshift. The simulated data sets result in  
a significant bias 
both in $\om$ as well as $\beta$. The size of the bias, however, depends on 
assumptions made for the underlying color distribution. Hence, for the current data sample, splitting the data set in two color bins introduces a bias so large and difficult to control, that the results of the test become meaningless. 
Note that if one had very small error bars on the color measurement 
over the full 
redshift range (as obtained from a dedicated space based survey \citep{aldering2005}),  the bias
can be kept small. This would allow for additional  tests of systematic uncertainties due to reddening corrections.

\subsection{Sample contamination}

As discussed in Section \ref{sec:outlier}, the method of robust statistics was applied to limit the effect of outliers, which could be present if the data is contaminated by  non Type Ia SNe, or by other events which do not have the standard candle properties of regular SN~Ia. It was shown in Section  \ref{sec:outlier} that the bias due to contamination can be limited for this analysis 
to $\Delta M = 0.015$ mag, which we hence use as the uncertainty due to contamination.

In  previous compilations, such as that of \citet{riess04,riess06}, no formal outlier criteria 
were applied. Instead, with some exceptions, the original classifications made by 
the authors of 
the data sample were used. Spurious candidates are sometimes 
removed 
from the data samples by hand (see for example \citet{astier05}), 
making it extremely difficult to estimate the effect of the remaining 
contamination. Our method of outlier rejection provides a simple and objective 
alternative. 

\subsection{Lightcurve model \& $K$-corrections} 

The lightcurve model \citep{salt} is a parametric description with 
two free parameters. As such it has limitations in capturing the full diversity 
of Type Ia SNe. By visual inspection we find, 
for example, that the fitted maximum magnitude can differ from the data by
 a few hundredths of a magnitude. A particular problem could arise if the 
observation strategies for  nearby and distant SNe differ. In fact, the 
high-redshift data sets have on average earlier observations of the lightcurve,
which is a result 
of the rolling-search techniques frequently used to find and follow-up 
SNe. Hence, when comparing low-z to high-z SNe, the 
fitted lightcurve parameters are obtained from slightly different 
parts of the lightcurve. The mismatch between template and the data lightcurve might thus  be more pronounced in one sample than the other. 
To quantify the effect, we have performed an 
extensive Monte Carlo simulation. A set of BVR lightcurve templates are 
obtained from a quartic spline fit  
to data including the 
well observed SNe 1990N, 1994D, 1998aq, 2001el, 2002bo, 2003du, 2004eo, and 2005cf \citep{mark07}. The templates are then used to sample random realizations of the lightcurves with cadence, signal-to-noise and date of the first detection 
of the nearby and distant 
SN sample. These simulated lightcurves are then fitted. 
The difference in the 
stretch and color corrected peak magnitude between the nearby and distant 
sample can be used to estimate the systematic uncertainty. For the nine 
templates we obtain the average difference between nearby and distant SNe of $-0.02$ magnitudes with an RMS 
scatter of 0.015. 
We adopt an associated systematic uncertainty of $\Delta M=0.02$ magnitudes due to this.

There is another source of uncertainty arising from the diversity of 
SNe Ia lightcurves.  
If a certain class of SNe is misrepresented  (for example 
if they are brighter than average for their typically 
fitted stretch value) and if the fraction of such SNe changes as a 
function of redshift, it will lead to a systematic bias in the cosmological 
parameters. Section \ref{sec:colorstretch} has addressed this issue by 
subdividing the sample according to stretch and redshift. If a significant 
lightcurve misrepresentation were present, one would expect to see 
differences in the fitted lightcurve-correction parameters. No statistically 
significant differences have been observed and we assign no additional
contribution to the uncertainties from such an effect. 
 
The lightcurve model is based on a spectral template series. 
It thereby eliminates the need for a separate $K$-correction (see Section \ref{sec:ltcvfit}). The model has been trained with nearby SNe data and  hence 
will be affected by systematic errors associated with that training data. 
These are largest for the U-band, which suffers from low 
training statistics and difficult flux calibration.
However, the validity of the model in  the U-band has been verified 
with the SNLS data set to better than 0.02 magnitudes \citep{astier05}. 
Here we adopt their 
assessment of the resulting systematic error of $\Delta M = 0.02$. 

\subsection{Photometric zero points}
\label{zb_uncertainty}
With present methods, ground based photometric zero point calibration is generally limited to an accuracy of $\gsim1\%$  \citep{stubbstonry}. The largest contribution to the photometric error of the peak magnitude arises from the color correction  $\Delta M \sim \beta\Delta c$.  
The color measurement is based on the measurement of the relative 
flux in two (or more bands), and as a result some of the uncertainties cancel. 
Nevertheless, since the color of SNe at different redshifs are 
obtained from different spectral 
regions, the uncertainty in the reference Vega spectrum 
limits the achievable accuracy to $\Delta c \approx 0.01-0.015$ mag \citep{stritz05,bohlin04}. 

Here we assume an uncertainty of $\Delta M=0.03$ for the photometric peak 
magnitude due to zero point calibration. Part of this uncertainty is common to all samples (as the same set of calibration stars is being used), 
while the other part is sample 
dependent (e.g. tied to the calibration procedure) and we divide the error 
equally among the two categories.

\subsection{Malmquist bias}
\label{sec:malmquist}
Malmquist bias arises in flux limited surveys, when SNe are detected 
because they are overly bright.  What 
matters for cosmology is whether the bias is
different for the low-z and high-z samples.
\citet{perl99}, \citet{knop03} and \citet{astier05} have evaluated the 
effects of Malmquist bias for the SCP and SNLS 
SN samples as well as 
the nearby SN sample and found that they nearly cancel. 
Since an individual estimate of Malmquist bias for all the different samples is beyond the 
scope of this work, we attribute a conservative 
systematic uncertainty of $\Delta M=0.02$ \citep{astier05} for all samples, which is 
consistent with previous estimates. 

In addition, we investigated whether the significant redshift dependence of the Hubble residuals observed for the \citet{essence_m_07} sample (see section \ref{sec:diversity}), if interpreted as due to Malmquist bias,  exceeds our claimed uncertainty.  A simulation was performed in which we introduced a magnitude cut-off such that the resulting slope, $d\mu/dz$, matches 
the observed slope of $-0.6$. The associated Malmquist bias with that sample is 
then $\sim0.05$ mags. If this is compared to the average Malmquist bias obtained for magnitude limited searches, the extra bias is only 0.03 mags larger-- not much larger than the systematic uncertainty we have adopted.  While we do not treat the ESSENCE data sample differently from the others, we note that \citet{essence_wv_07} made their extinction prior redshift-dependent to account for the fact that at higher redshifts an increasingly larger fraction of the 
reddened SNe was not detected. The  linear color correction employed in our analysis is independent of a prior and therefore unaffected by a 
redshift dependent reddening distribution. 

\subsection{Gravitational lensing}
\label{sec:lensing} 
Gravitational lensing decreases the mode of the 
brightness distribution and causes increased 
dispersion in the Hubble diagram at high redshift. 
The effect has been discussed in detail in the literature \citep{sasaki87,linder88,bergstroem00,holz05}. We treat lensing as a statistical phenomenon only, 
although with the detailed optical and NIR data available for the
 GOODS field, the 
mass-distribution in the 
line-of-sight and hence the lensing (de)magnification may be estimated for 
individual 
SNe \citep{Jonsson06}. Important for this work is that they find 
 no evidence for selection effects (i.e. Malmquist bias) due to lensing of 
the high redshift SNe.

Considering both strong and weak lensing, 
\citet{holz05} found that lensing will add a 
dispersion  of $0.093 \times z$ mag, which if the statistics of SNe is 
large enough, can be approximated as an additional Gaussian error. 
Here, we added 
the additional dispersion from gravitational lensing in quadrature
to the ``constant'' systematic dispersion and observational error. 
This effectively deweights the
high redshift SNe. However, only for the highest redshift SNe is 
the additional uncertainty comparable to that of the intrinsic dispersion.

\begin{table}[htbp]
\begin{center}
\begin{tabular}{ l| c | c}
\tableline
\tableline
Source & common & sample-dependent  \\
 & (mag) & (mag) \\
\tableline
$\alpha \& \beta $ correction\ & 0.015 & -   \\
Contamination             & -    & 0.015    \\
Lightcurve model\ & 0.028 & -   \\
Zero point             & 0.021   & 0.021    \\

Malmquist bias\ & - & 0.020    \\
Gravitational lensing & - & 0.009$^*$    \\
Galactic extinction  & 0.013 & - \\

\tableline
\tableline
Total in mag &$\Delta M = 0.040$ & $\Delta M_i=0.033$ \\
\tableline
\end{tabular}
\end{center}
\caption{Most relevant common and sample dependent systematic errors of 
this analysis (in magnitudes).}
\label{tb:syserror}
\end{table}

Flux magnification and demagnification effects 
due to over- or under-densities of  matter near the line of sight cancel.
But one 
obtains a bias if magnitudes instead of fluxes are used. However, 
the bias is $0.004\times z$ mag and therefore still much smaller than the 
statistical error on the luminosity distance obtained from the ensemble of
high redshift SNe. While not yet relevant for this analysis,  
future high-statistics samples  will have to take this effect into account. 

Another potential bias is introduced by the $3\sigma$ outlier rejection, 
since the lensing PDF is asymmetric. Using the PDFs of \citet{holz05} 
we have checked that the bias is never larger than $0.006\times z$ mag. We take the worst case value of  $0.009$ magnitude (i.e. for a SNe at $z\approx1.5$) as a conservative systematic uncertainty for gravitational lensing, since this is still an almost
 negligible value.
\subsection{Gray intergalactic dust} 
The possibility that SNe are dimmed 
due to hypothetical gray intergalactic dust, as suggested by \citet{Aguirre99}, was 
constrained by \citet{oestman05, moertsell03} by studying the colors of 
high-redshift quasars. Applying their constraints on intergalactic dust, 
we find that the cosmological parameters are shifted by about 
one statistical standard deviation, i.e. for a flat Universe 
$\Delta\om=-0.03$. This should not be considered a systematic uncertainty, 
but rather an upper
limit on the effect of hypothetical large grains of
cosmic dust in the line of sight.

\subsection{Galactic Extinction}
\label{sec:mw_sys_ext}
All lightcurve data were corrected for Galactic extinction using the extinction law of \citet{ccm} using an $R_V$ of 3.1.  The $E(B - V)$ values were derived 
from the sky map of \citet{schlegel98} and have a typical 
statistical error of $10\%$.  For nearby SNe we hence obtain an additional uncertainty of \begin{equation}
\Delta \mu_B \approx (R_B-\beta)\cdot \sigma\left(E(B - V)\right) \approx 0.2\cdot E(B - V), 
\end{equation}
where $\beta$ is the color correction coefficient from Eq.\ \ref{eq:mub}. 
We add this statistical error in quadrature to each nearby SNe.  
High redshift SNe are measured in redder bands and, since $R_R\approx\beta$, are less affected by Galactic extinction. 

There is also a common systematic error of $10\%$ in the overall reddening normalization.  The average Galactic $E(B - V)$ for the low redshift sample 
is 0.063 and we add $0.063\cdot 0.2=0.013$ mag systematic uncertainty to $\Delta M$.
\begin{figure}[htp]
\begin{center}
\includegraphics[width=0.35\textwidth]{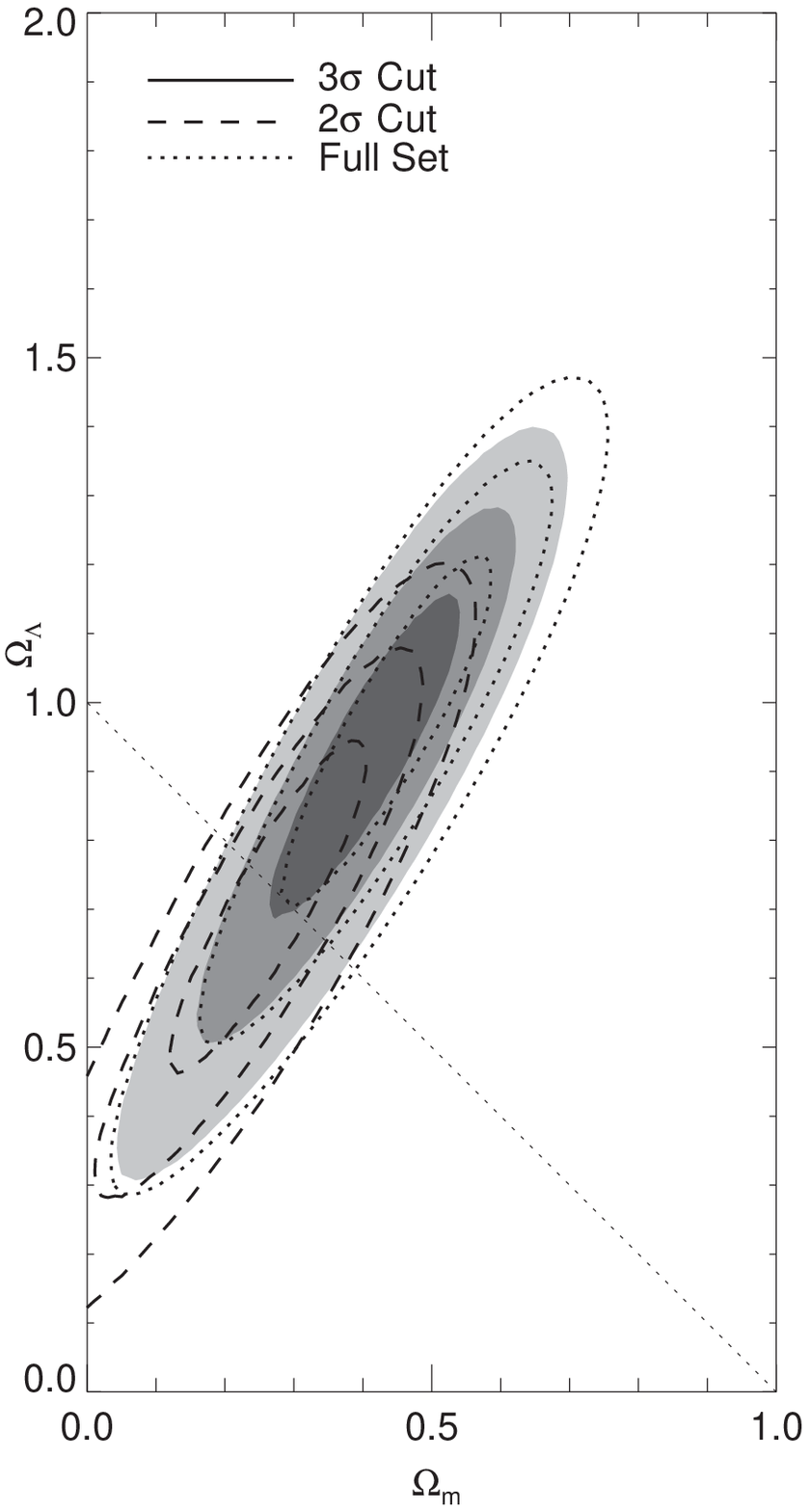}
\caption{68.3~\%, 95.4~\% and 99.7\% confidence level contours on 
$\ola$ and $\om$ plane from the Union SNe set. The result from the 
robustified set, obtained with a $\sigma_{\rm cut}=3$ outlier cut, is  shown as filled contours. 
The empty contours are obtained with
the full data set (dotted line) and   $\sigma_{\rm cut}=2$ outlier rejected data set (dashed line). As can be seen, outlier rejection shifts the contours along the degenerate axis by as much as $0.5\sigma$ towards a flat Universe. 
In the remaining figures, we refer to the $\sigma_{\rm cut}=3$ outlier rejected set as the Union set. 
}
\label{fig:omol_fit3}
\end{center}
\end{figure}

\subsection{Summary of systematic errors}
\label{sec:sys_summ}

In our treatment of the above systematic errors we distinguish between systematic errors common between datasets, which are largely of astrophysical nature, 
and the more observer dependent ones associated with individual samples. 
Table \ref{tb:syserror} summarizes what are considered the  relevant 
contributions to the 
systematic uncertainties in this analysis. They are propagated into the final 
result 
through Eq. \ref{eq:nuisance}.  

\section{Cosmological fit results}
\label{sec:cosmo}
Our analysis of cosmological model fits includes both statistical and 
systematic errors. The individual contributions to the systematic error 
identified in Table \ref{tb:syserror} are of very different nature and hence are assumed uncorrelated. We hence obtain the combined systematic error by adding 
in quadrature the 
individual contributions. The resulting error was propagated according to the 
prescription described in Section 
\ref{sec:systematics}. Our constraints on the matter 
density $\om$, assuming a flat Universe, are summarized in Table \ref{tb:results}. Both statistical (68~\% CL) and systematic errors  
are quoted.

\begin{figure}[htp]
\begin{center}
\includegraphics[width=0.33\textwidth]{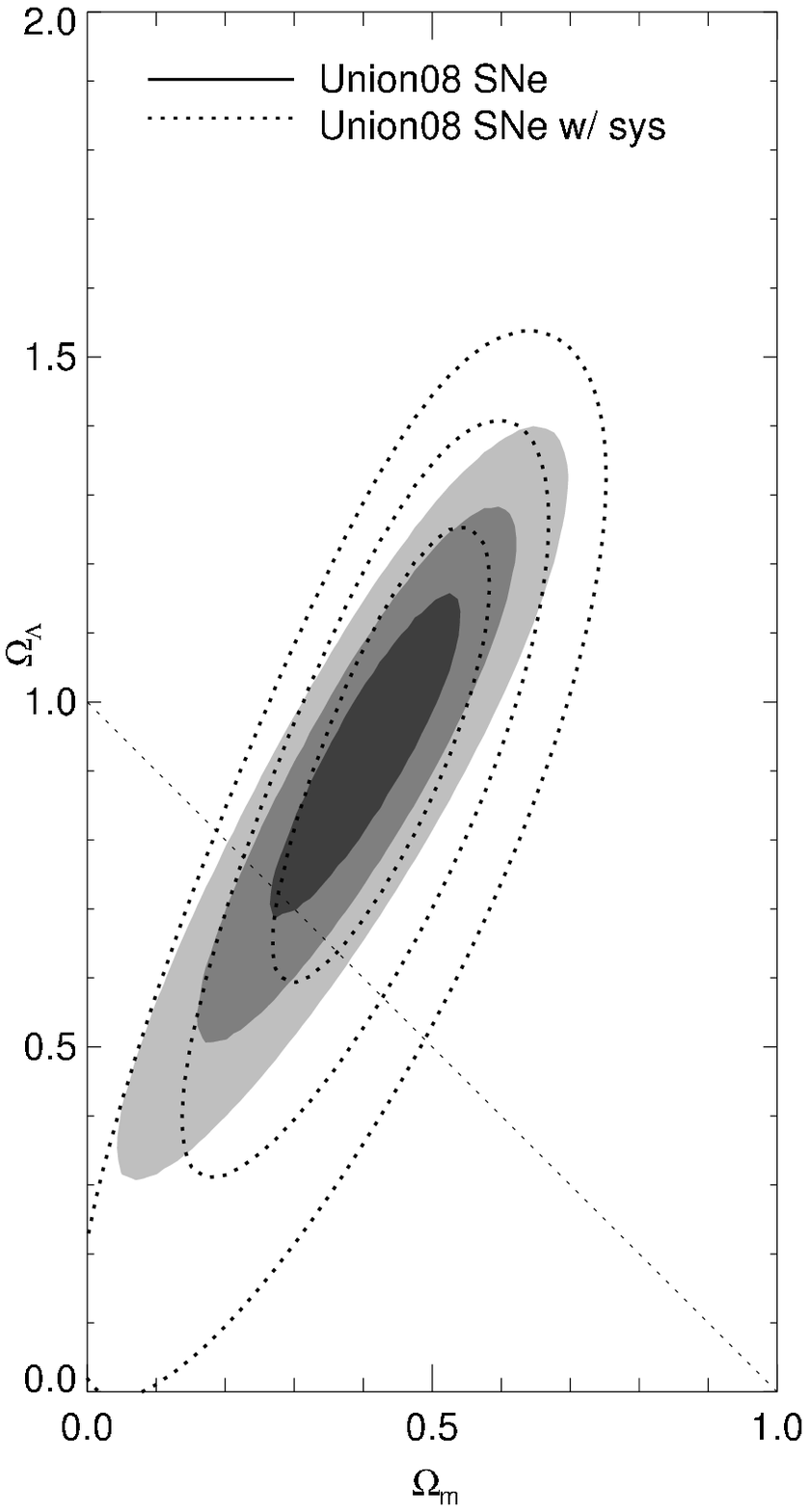}
\hfill
\includegraphics[width=0.33\textwidth]{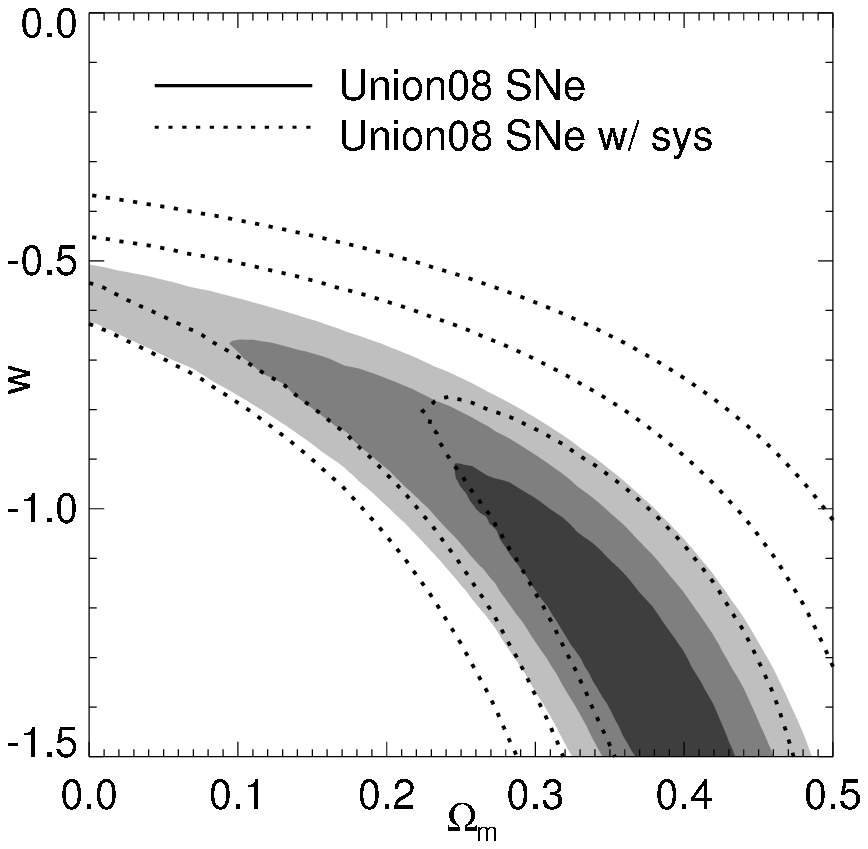}
\caption{Top: 68.3~\%, 95.4~\% and 99.7\% confidence level contours on $\ola$ and $\om$ obtained with the Union set, without (filled contours) and with inclusion of systematic errors (empty contours).
Bottom: The corresponding confidence 
level contours on the equation of state parameter $w$ and $\om$, assuming a constant $w$.}
\label{fig:cosmo_sys}
\end{center}
\end{figure}

\begin{figure*}[htp]
\begin{center}
\includegraphics[width=0.7\textwidth]{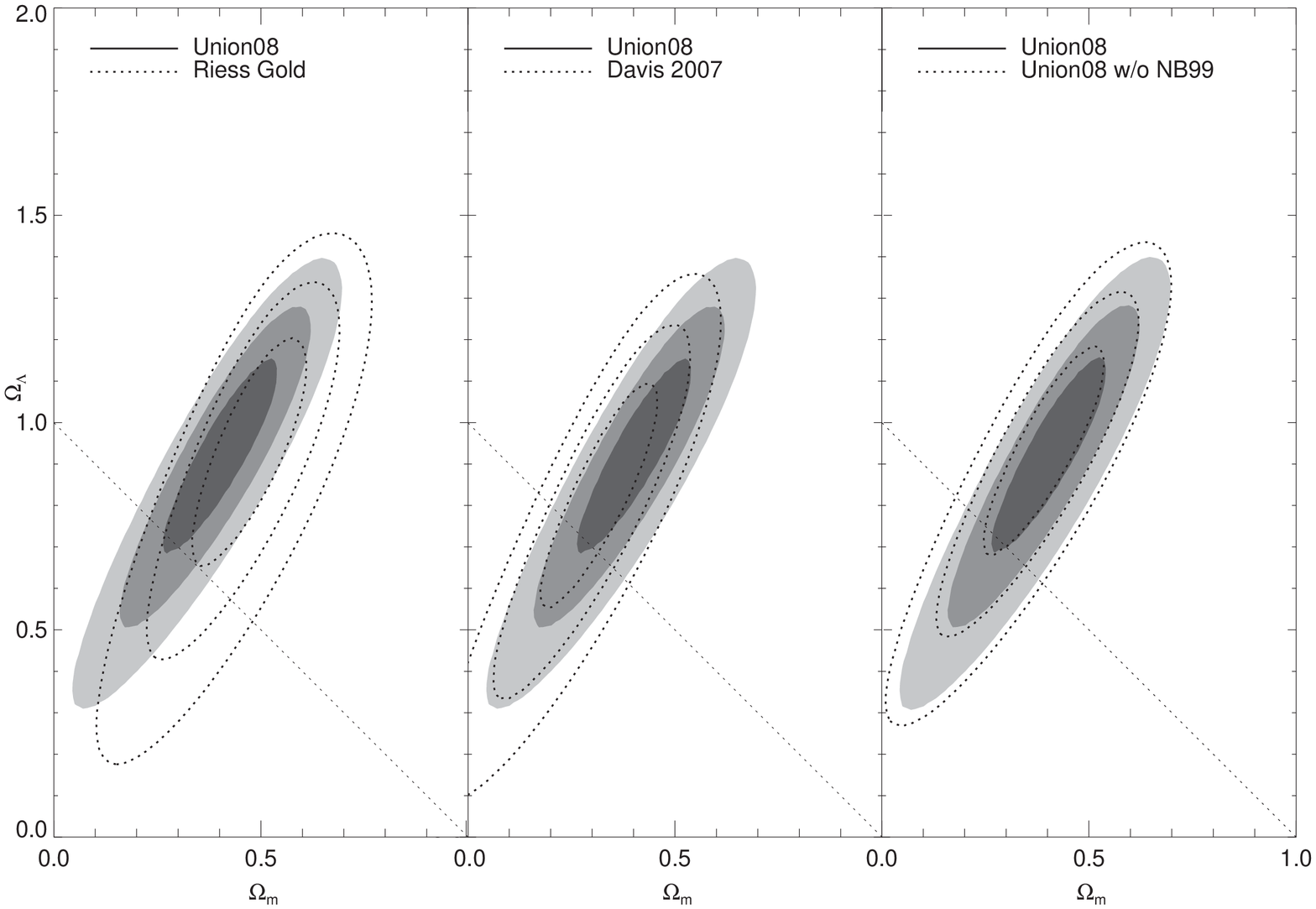}
\vspace{0cm}
\includegraphics[width=0.7\textwidth]{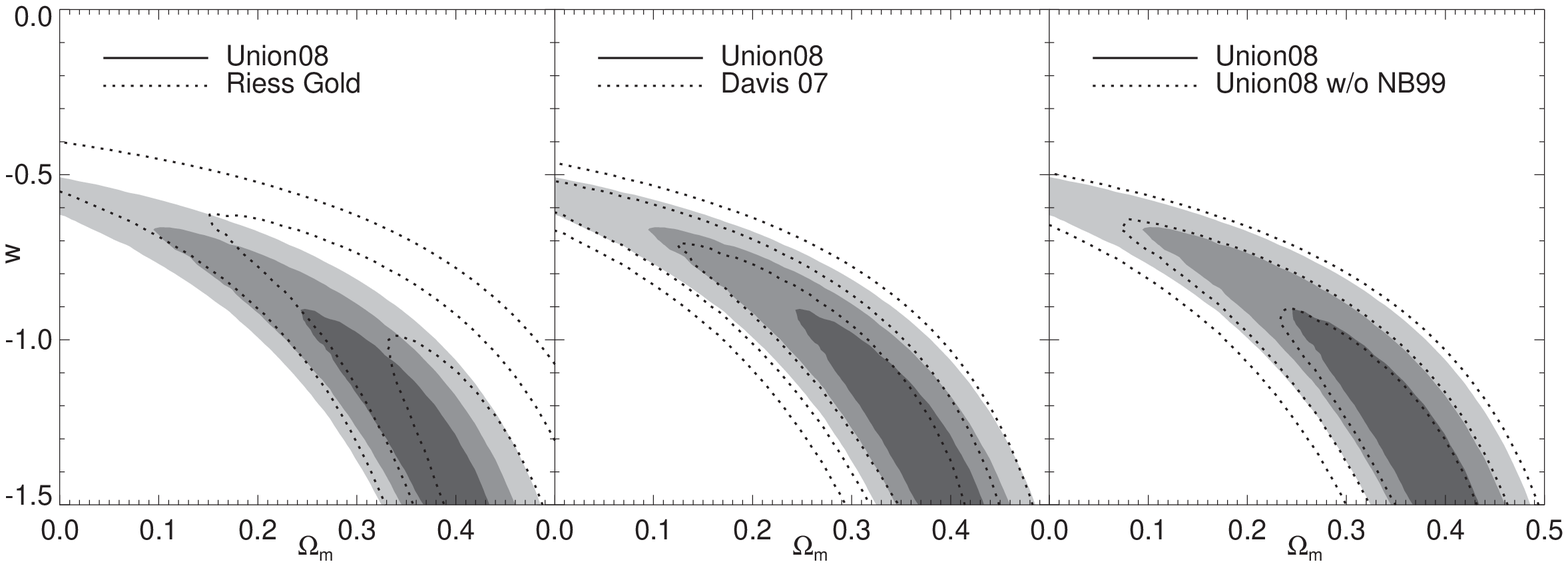}
\caption{68.3~\%, 95.4~\% and 99.7\% confidence level contours on 
$\ola$ and $\om$ (top row) and $\om$ and $w$ (bottom row). The results from the 
Union 
set are  shown as filled contours. The empty contours in the left column represent the 
Gold sample \citep{riess04,riess06} and  the middle column the constraints from 
\citet{davis07}. While our results are statistically consistent with the previous work, 
the improvements in the constraints on the cosmological parameters are evident. The 
right column shows the impact of the SCP Nearby 1999 data.}
\label{fig:omol_comp}
\label{fig:omw_comp} 
\end{center}
\end{figure*}

Figure \ref{fig:omol_fit3} plots our results for the joint fit to the matter 
density and 
cosmological constant energy density, $\om$ and $\ola$, 
and the effect of varying the outlier cut, while Fig.\ \ref{fig:cosmo_sys} 
illustrates the effects of systematics.  For comparison with previous work, 
Figure \ref{fig:omol_comp} shows our joint constraints on $\om$ 
and $\ola$ (statistical error only) and the \citet{riess06} 
constraints obtained from the Gold compilation of data primarily from the 
HZT, SCP and SNLS \citep{riess06} and a recent compilation of \citet{davis07}, which is based 
on lightcurve fits from \citet{riess06} and \citet{essence_wv_07}.
The results obtained in this work are consistent with those of previous studies; however, compared to the recent SN fit results of \citet{astier05, riess06,
 essence_wv_07, davis07}, we obtain a 15-30~\% reduction in the statistical 
error.

\begin{figure}[hbt]
\begin{center}
\includegraphics[width=0.49\textwidth]{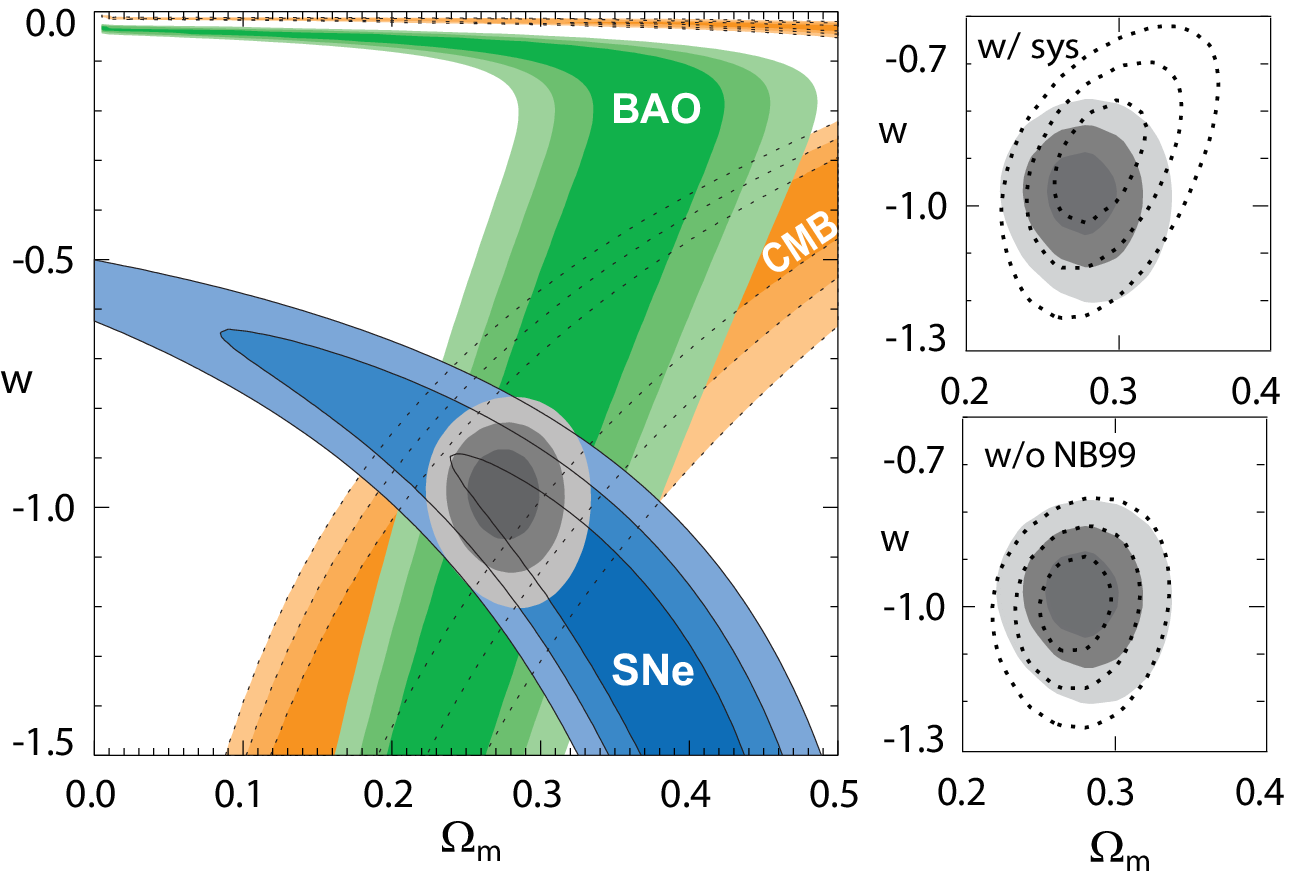}
\caption{68.3~\%, 95.4~\% and 99.7\% confidence level contours on 
$w$ and $\om$, for a flat Universe. The top plot shows the individual 
constraints 
from CMB, BAO and the Union SN set, as well as the combined
constraints (filled gray contours, statistical errors only). 
The upper right plot shows the effect of including systematic errors. The 
lower right plot illustrates the impact of the SCP Nearby 1999 data.}
\label{fig:omegam_w}
\end{center}
\end{figure}

About half the improvement can be attributed to the new SCP 
Nearby 1999 SNe. Their 
impact is evident in the 
rightmost column of Fig.\  \ref{fig:omol_comp} (as well as in 
Fig.\ \ref{fig:omegam_w}). The impact of these SNe is somewhat larger 
because the sample has a best-fit systematic uncertainty of zero. If instead one would introduce the requirement that $\sigma_{\rm sys}\geq0.1$, there would be an increase of about 10\% in the uncertainties of the cosmological parameters.

\begin{figure}[t]
\includegraphics[width=0.49\textwidth]{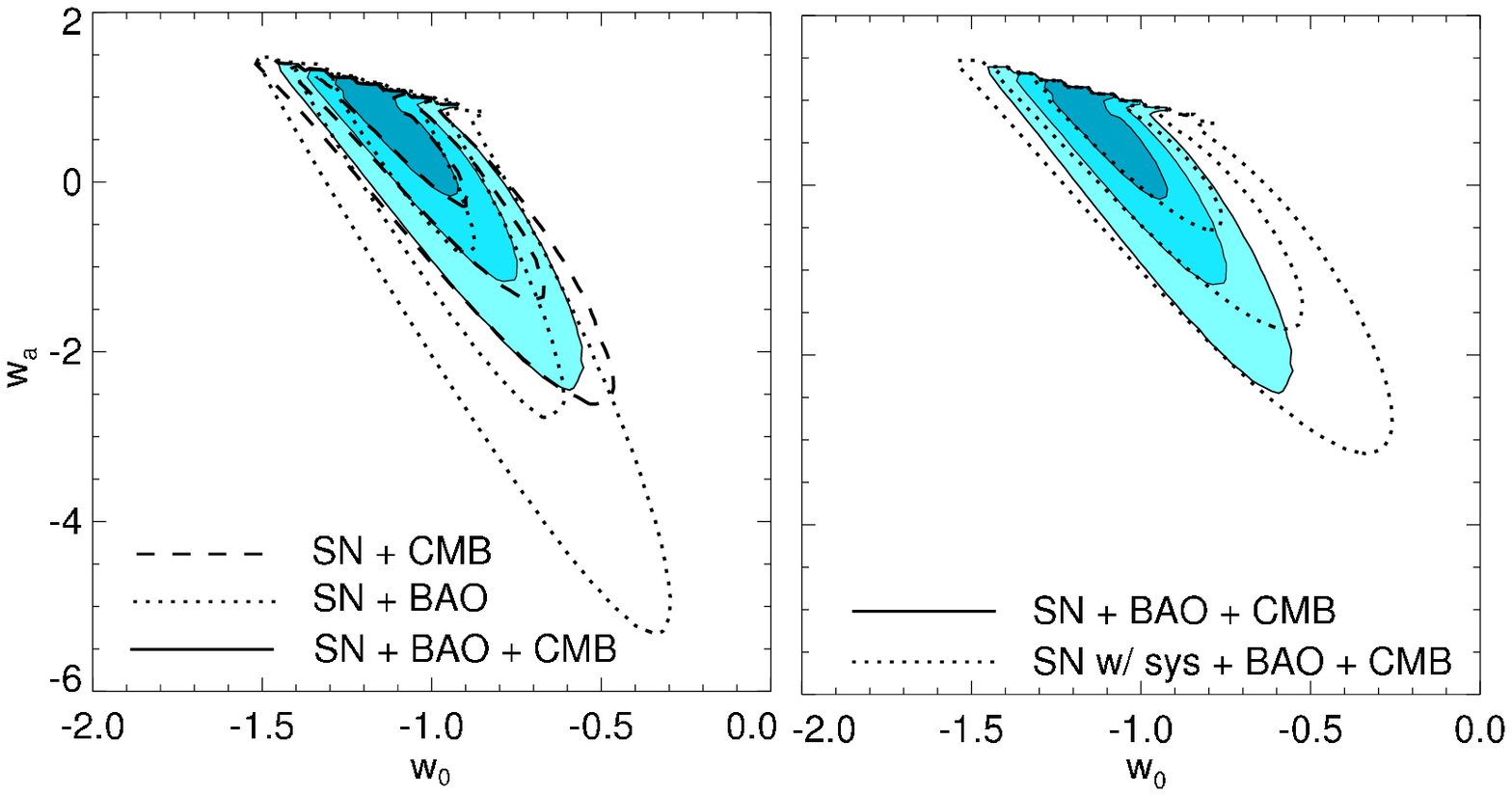}
\caption{68.3~\%, 95.4~\% and 99.7\% confidence level contours on 
$w_a$ and $w_0$ for a flat Universe. 
Left: The Union SN set was combined with CMB or BAO constraints. Right: Combination of SNe, CMB and BAO data, with and and without systematic uncertainties included. The diagonal line represents $w_0+w_a=0$; note 
  how the likelihoods based on observational data remain below it, 
  favoring matter domination at $z\gg1$.  \label{fig:waw}}
\end{figure}

Figure \ref{fig:omw_comp} shows the constraints on the equation of state parameter $w$ (assumed constant) and $\om$. A flat Universe was assumed. Again, the constraints are 
consistent with, but stronger than, those from \citet{riess06} and \citet{davis07}.
The current SN data do not provide strong
constraints on the equation of state parameter $w$ by itself, 
since it is to a large extent degenerate with $\om$. However, the degeneracy can be broken by combining with other measurements involving $\om$. Figure  \ref{fig:omegam_w} shows the constraints obtained from the detection of baryon acoustic oscillations (BAO) \citep{eisenstein05} and from the five year data release of the Wilkinson Microwave Anisotropy Probe (CMB) \citep{dunkley08}. The constraints from the
CMB data follow from the reduced distance to the surface 
of last scattering at $z=1089$  (or shift parameter).  It is important to realize that for parameter 
values far from the concordance model, the shift in the sound 
horizon must also be taken into account.  The reduced distance $R$ 
is often written as 
\begin{equation} 
R_{\rm conc}=(\om H_0^2)^{1/2}\int_0^{1089} dz/H(z),  
\label{eq:shiftp}
\end{equation} 
where the Hubble parameter is \newline $H(z)=H_0 \left[\om(1+z)^3+(1-\om)(1+z)^{3(1+w)} \right]^{1/2}$. 
The WMAP-5 year CMB data alone yields $R_0=1.715\pm0.021$ for a fit assuming a constant $w$ \citep{Komatsu:2008hk,Lambda_website}.
Defining the corresponding $\chi^2$ as $\chi^2 =[(R_{\rm conc}-R_0)/\sigma_{R_0}]^2$ one can then 
deduce constraints on $\om$ and $w$.
However, this assumes a standard matter (and radiation) dominated epoch for calculating 
the sound horizon.  The more proper expression for the shift parameter 
accounts for deviation in the sound horizon: 
\begin{eqnarray}
R&=&(\Omega_m H_0^2)^{1/2}\int_0^{1089} dz/H(z)\\
& \times &  \bigg[\int_{1089}^\infty \nonumber
dz/\sqrt{\Omega_m (1+z)^3}
{\displaystyle \bigg/}\int_{1089}^\infty dz/(H(z)/H_0)\bigg]. 
\label{eq:rcmb}
\end{eqnarray} 
Since dark energy is generally negligible at high redshift, the factor 
in square brackets is usually unity (for example, it deviates from unity 
by less than 1\% even for $w_0=-1$, $w_a=0.9$, i.e.\ $w(z=1089)=-0.1$).  
However, for extreme models that upset 
the matter dominated behavior at high redshifts, the correction will be 
important in calculating whether the geometric shift parameter accords 
with CMB observations (apart from any issue of fitting other observations). Violation of early matter domination causes the ``wall'' in likelihood apparent in Fig.\ \ref{fig:waw}.  Also see, for example, \citet{linmiq,wright07}.

BAO measurements from the SDSS data \citep{eisenstein05} provide a distance constraint at a redshift $z=0.35$. 
\citet{percival2007} have derived BAO distances  
for $z=0.2$, in addition to the $z=035$ SDSS-data point,  using the combined 
data from SDSS   
and 2dFGRS. However, some points of tension were noted between the data sets (Percival et al 2007, see also \citet{2008MNRAS.tmp..243S}), especially evident for 
$\Lambda$CDM models.
We confirm this observation 
and found that the $z=0.2$ data 
point, if combined with SN and CMB data according to the prescription in Appendix A of Percival (2007)
leads to an 2.5 sigma inconsistency. Neither the $z=0.35$ BAO data point 
from \citet{percival2007} nor the slightly weaker constraint 
from \citet{eisenstein05}  
shows such kind of tension. 
Given the differences between the two data sets, we use the $z=0.35$ SDSS data point of \citet{eisenstein05}, but with the caveat that BAO constraints need further clarification.
\citet{eisenstein05} provides a constraint
on the distance parameter $A$: 
\begin{eqnarray} 
A(z)&=&(\om H_0^2)^{1/2} H(z)^{-1/3} z^{-2/3} \left[\int_0^z dz'/H(z')\right] ^{2/3} \\ 
& \times & \left[\int_{1089}^\infty 
dz/\sqrt{\Omega_m (1+z)^3}\bigg/\int_{1089}^\infty dz/(H(z)/H_0)\right] \nonumber,
\label{eq:bao}
\end{eqnarray} 
to be $A(z=0.35)=0.469\pm0.17$.  Note that BAO also depend on accurate accounting of the sound horizon and receive the same correction factor shown in brackets in Eq.~\ref{eq:rcmb}.  This results in a similar wall to the acceptable confidence contour reflecting violation of early matter domination.  To see that such violation has severe implications, note that most models above the wall have a total linear growth factor a factor ten below the concordance cosmology.

The joint constraints from SN data, BAO, and CMB are shown 
in Fig.\ \ref{fig:omegam_w} and the corresponding numbers are given in Table 
\ref{tb:results}. 
As can be seen, the constraints obtained from combining 
 either BAO or CMB with SNe data give consistent results and comparable 
error bars, while the combination of all three measurements improves only the
 statistical error. 
The impact of including systematic errors (only from SNe, from Eq.~\ref{eq:nuisance}) 
is shown in the upper right panel of Fig. \ref{fig:omegam_w}. 


\begin{table*}[htp]
\begin{center}
\begin{tabular}{ c |  c c c  }
\tableline
\tableline
Fit & $\om$ & $\Omega_k$ & $w$ \\
\tableline
SNe & $0.287^{+0.029+0.039}_{-0.027-0.036}$ & 0 (fixed) & -1 (fixed)  \\ 
SNe+BAO & $0.285^{+0.020+0.011}_{-0.020-0.009}$ & 0 (fixed) & $-1.011^{+0.076+0.083}_{-0.082-0.087}$  \\ 
SNe+CMB & $0.265^{+0.022+0.018}_{-0.021-0.016}$ & 0 (fixed) & $-0.955^{+0.060+0.059}_{-0.066-0.060}$  \\ 
SNe+BAO+CMB & $0.274^{+0.016+0.013}_{-0.016-0.012}$ & 0 (fixed) & $-0.969^{+0.059+0.063}_{-0.063-0.066}$  \\ 
SNe+BAO+CMB & $0.285^{+0.020+0.011}_{-0.019-0.011}$ & $-0.009^{+0.009+0.002}_{-0.010-0.003}$ & -1 (fixed)  \\ 
SNe+BAO+CMB & $0.285^{+0.020+0.010}_{-0.020-0.010}$ & $-0.010^{+0.010+0.006}_{-0.011-0.004}$ & $-1.001^{+0.069+0.080}_{-0.073-0.082}$  \\ 
			      \tableline
\end{tabular}
\end{center}
\caption{Fit results on cosmological parameters $\om$, $\Omega_k$ and $w$. The parameter values are followed by their statistical ($\sigma_{\rm stat}$) and systematic ($\sigma_{\rm sys}$) uncertainties. 
The parameter values and their statistical errors were obtained from minimizing the $\chi^2$ of Eq.\ \ref{eq:biased_chi}. The fit to the SNe data alone results in a $\chi^2$ of 310.8 for 303 degrees of freedom with a $\Delta\chi^2$ of less than one for the other fits.  The systematic errors were obtained from fitting with extra nuisance parameters according Eq. \ref{eq:nuisance} and subtracting from the resulting error, $\sigma_{\rm w/ sys}$, the statistical error: $\sigma_{\rm sys} = (\sigma_{\rm w/sys}^2 -\sigma_{\rm stat}^2)^{1/2}$.}
\label{tb:results}
\end{table*}


\begin{figure}[htp]
\begin{center}
\includegraphics[width=0.4\textwidth]{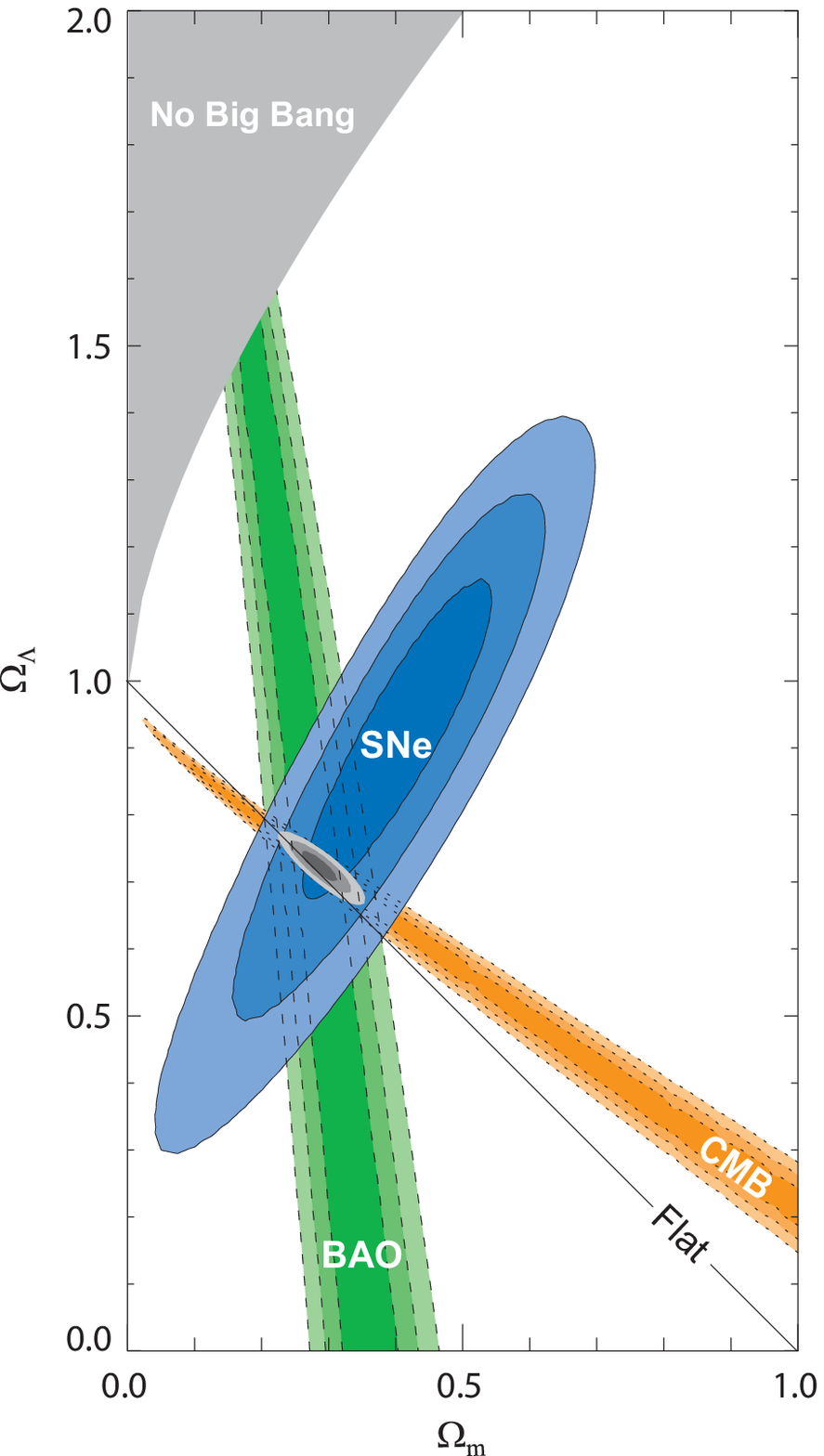}
\caption{68.3~\%, 95.4~\% and 99.7\% confidence level contours on 
$\ola$ and $\om$  
obtained from CMB, BAO and the Union SN set, 
as well as their combination (assuming $w=-1$). 
}
\label{fig:ok}
\end{center}
\end{figure}


The results quoted so far were derived assuming a flat Universe. Allowing for  
spatial curvature $\Omega_k$, our constraints from
combining SNe, CMB and BAO are  consistent with a flat \lcdm{} Universe (as seen in Table \ref{tb:results}). 
Fig. \ref{fig:ok} shows the corresponding constraints in the $\om-\ola$ plane.



Finally, one can attempt to investigate constraints on a redshift 
dependent equation of state (EOS) parameter $w(z)$.  Initially we 
consider this in terms of 
\begin{equation} 
w(z)=w_0+w_a\frac{z}{1+z}, 
\end{equation} 
shown by \cite{linder03} to provide excellent approximation to a wide 
variety of scalar field and other dark energy models.  Later, we 
examine other aspects of time variation of the dark energy EOS.  Assuming 
 a flat Universe and combining the Union set with constraints from CMB, we 
obtain constraints on $w_0$, the present value of the EOS, and $w_a$, giving a 
measure of its time variation, as 
shown in Fig.\ \ref{fig:waw}.  (A cosmological constant has 
$w_0=-1$, $w_a=0$.)  Due to degeneracies within the EOS and 
between the EOS and the matter density $\om$, the SN 
dataset alone does not give appreciable leverage on the dark 
energy properties.  By adding other measurements, the degeneracies can be 
broken and currently modest cosmology constraints obtained.

Fig.\ \ref{fig:waw} (left) shows the combination of the SN data with either the CMB constraints or the BAO constraints.  
  The results are similar; note that including either one results in a sharp cut-off at $w_0 + w_a = 0$, from the physics as mentioned in regards to Eq.~\ref{eq:rcmb}.  Since $w(z \gg 1) = w_0 + w_a$ in this parameterization, any model with more positive high-redshift $w$ will not yield a matter-dominated early Universe, altering the sound horizon in conflict with observations.

Note that BAO do not provide a purely 
``low'' redshift constraint, because implicit 
within the BAO data analysis, and hence the constraint, is that the high 
redshift Universe was matter dominated (so the sound horizon at 
decoupling is properly calculated).  Thus, one cannot avoid the 
issue of modeling how the dark energy EOS behaves at high redshifts 
by using this constraint rather than the CMB. (We differ here from  \citeauthor{riess06}~\citeyear{riess06}, who treat BAO as a low-redshift constraint.) 
SN data are especially useful in
  constraining $w(z)$ because there is no dependence at all on the high
  redshift behavior, unlike CMB and BAO data.

As one might expect, because of the different orientations of the 
confidence contours and the different physics 
that enters, combining both the CMB and BAO constraints with the SN data 
clears up the degeneracies somewhat, as seen in Fig.\ \ref{fig:waw}, with and without systematics.  Inclusion of curvature does not
substantially increase the contours.

\begin{figure}[hbt]
\begin{center}
\includegraphics[width=0.5\textwidth]{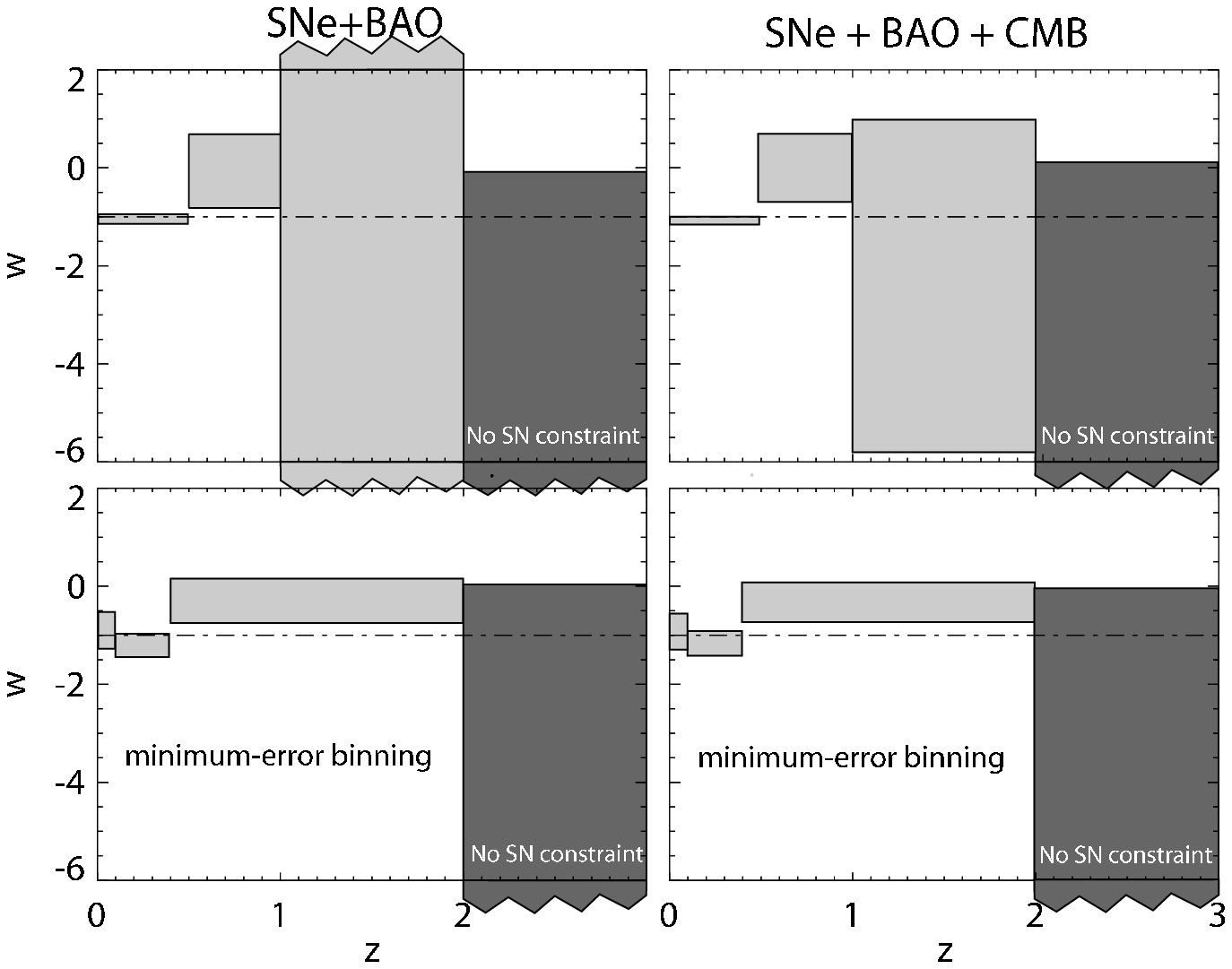}
\caption{68~\% constraints on $w(z)$, where $w(z)$ is assumed to be constant over each redshift bin.  
The left column combines the Union  SN set with BAO constraints only, while the right column includes also constraints from the CMB. 
The top row illustrates the fact that only extremely weak constraints on the equation of state exist at $z>1$.  The bottom row shows a different binning that minimizes the mean bin error.  Note that for $z>2$ (dark gray-''No SN constraint'') only upper limits exist, basically enforcing matter domination, coming from either CMB data or, in the case without CMB data, from requiring substantial structure formation (a linear growth factor within a factor of 10 of that observed).
}
\label{fig:wbins}
\end{center}
\end{figure}

We emphasize that the wall in $w_0$-$w_a$ space is not imposed a priori and does not represent a breakdown 
of the parameterization, but a real physical effect from violating 
early matter domination.  Nevertheless, we can ask what limits could 
be put on the early dark energy behavior -- either its presence or 
its equation of state -- if we do not use the $w_0$-$w_a$ parameterization. 
A simple, but general model for $w(z)$ creates a 
series of redshift bins and 
assumes $w$ is constant over each bin.  The constraints from this are shown in Fig.\ \ref{fig:wbins}.  Note that the data points are correlated.  

\citet{riess06} made a somewhat similar investigation with the emphasis on 
the impact of the highest redshift SNe.  A difference to the work of 
\citet{riess06}  is that  we do not decorrelate the constraints in the different redshift bins. While this implies that the bin-wise constraints shown in Fig.\ 17 are correlated, it ensures that the $w$-constraints shown
for a given bin are confined to the exact redshift range of the bin. If instead one applies a decorrelation procedure, some of the tight constraints from lower  redshifts feed through to higher redshifts (i.e. $z>1$). See \citet{deplbin} 
for general discussion of this issue. Unlike
\citet{riess06}, we additionally place a $w$ bin at higher redshift than the SN
data ($z>2$),  to account for the expansion history of the 
  early Universe, and do not fix $w$ in this bin.  The Riess ``strong'' prior has a
fourth bin for $z > 1.8$, but fixes $w=-1$.  The
``strongest'' prior does not have a fourth bin. Forcing either of these 
behaviors on 
  the $z>2$ Universe results in unfairly tight constraints and the danger of bias \citep{wcon,deplbin}; in failing to 
  separate the SN bins from those of the CMB and BAO essentially 
  the entire constraint in the redshift $z \gsim 1$ bin is from the CMB 
(see also \citeauthor{wright07}~\citeyear{wright07}).

Consider the top row of Fig.\ \ref{fig:wbins}.  These results are for bins with $z < 0.5, 0.5 < z < 1.0, 1.0 < z
< 2.0$ and $z>2.0$.  The only constraint that can be concluded from the
highest redshift bin is that $w_{[2,\infty]} \lsim 0$, but this constraint comes entirely from CMB and BAO, which requires that the early Universe is 
matter-dominated (see the above discussion of the 
wall in the $w_a-w_0$ plane). 
We then look at the $z=1-2$ bin for constraints on $w$ which would be due to 
the $z>1$ SNe and we find essentially no constraint. 

The lowest redshift bin is constrained to $w_{[0,0.5]} \approx -1\pm0.1$.  
The next bin is compatible with -1, but the central value is high.  This 
deviation from -1
seems to be due to the unexpected brightness (by about 0.1 magnitudes) of
the Hubble data at $z> 1$ (see Fig.\
\ref{fig:hubble_diagram}). (Recall that $w$ at some $z$ influences distances at larger redshifts.)
We clearly see that to be 
sensitive to appreciable deviations from $w=-1$ such as 0.1 mags at 
$z \sim 1$, which is 
key to constraining theories of dark energy, one requires better statistics for the very high-redshift  supernovae (and comparably 
good systematics).



Given that the  strongest constraints on $w$ are 
contained in the first bin, one might attempt to search for a redshift 
dependence of $w$ at lower redshifts by changing the borders of the bins. 
The smallest errors are obtained roughly with the binning $z < 0.1, 0.1 < z < 0.4, 0.4 < z < 2.0$, and $2.0 < z$.  These constraints are shown in the bottom row of Fig.\ \ref{fig:wbins}.  The results are similar to the results from the other binning, with the lowest two bins centered around $w = -1$ and the next bin centered around a more positive value. No significant redshift dependence 
is observed. Note the tight limit on the $0.4<z<2$ bin 
  is {\it not\/} saying $w(z>1)\approx-1$, even approximately, 
since the leverage on $w(z)$ 
  is coming from the $0.4<z<1$ part of the bin (this illustrates the 
  importance of considering multiple binnings).

To sum up, even in combination with current BAO and CMB data, current SN data sets cannot tell us whether 
an energy density component other than matter existed at $z>1$,  and cannot tell us whether such a component if it existed had an equation of 
state with negative pressure.  In the future, 
however, SN data that achieves Hubble diagram accuracy of 0.02 mag out 
to $z=1.7$ will be able to address these questions and provide 
independent checks of the $z>1$ Universe. 

Note that while constraints on a possible  
redshift dependency of $w$ have been shown in Figures \ref{fig:waw} 
and \ref{fig:wbins}, we do not present values for the projected,
one-dimensional constraints for several reasons. First, the 
bounds are still very weak and as a result the error bars show 
highly non-gaussian errors (as visible in Fig.\ \ref{fig:waw}). In addition, our  
treatment of systematic errors has not been optimized for a redshift 
dependent $w$ and a potential systematic redshift dependence of the distance 
modulus is only partially taken into account. As a consequence, the resulting (already large) 
systematic errors on $w(z)$ would be underestimated.

In this analysis so far we have not excluded any SNe based on extreme values of stretch or color, therefore including also the peculiar class of under-luminous  1991bg-like  SNe  that are typically associated with small stretch values. After unblinding, in an effort to study the robustness of our results, we have introduced a stretch cut, $s>0.6$, to eliminate SN1991bg-like SNe from the sample. The most significant consequence of this cut came with the removal of SN 1995ap, a supernova in the \citet{riess98} sample.  By itself the removal of this one supernova can change the cosmological fit parameters in the $\om-\ola$ and $\om-w$ planes by nearly 1$\sigma$ along the more degenerate contour axis 
(and away from a flat Universe). 
However, without SN 1995ap, the test for tension between data sets that we applied in Section 4.4 would show the \citet{riess98} dataset to be a 3.5$\sigma$ outlier and one would be forced, unless the tension can be resolved otherwise, 
to remove the data set from the compilation.
The net result of the $s > 0.6$ cut would then be a 0.25$\sigma$ 
change in $w,~\Omega_M$ and $\Omega_\Lambda$ in the direction of the more degenerate contour axis. The results presented in this paper are based on the sample 
without the stretch cut; however, 
since the parameters along the direction of the degeneracy are well 
constrained once CMB or BAO data are added, the combined constraints essentially do not depend on whether or not the stretch cut is applied.

\section{Conclusion}
\label{sec:conclusion}

The cosmological parameter constraints from the Union SN Ia compilation 
shown in Figures \ref{fig:cosmo_sys}, \ref{fig:omegam_w}, \ref{fig:waw}  and \ref{fig:wbins} reflect the current best 
knowledge of the world's Type Ia supernova datasets.   Specifically, in addition to the older data, they 
include the new datasets of 
nearby Hubble-flow SNe Ia we presented in this paper, the recent 
large, homogeneous, high-signal-to-noise SNLS and ESSENCE 
datasets published by \citet{astier05} and \citet{essence_m_07} 
as well as the high redshift supernovae in \cite{riess04,riess06}.
   Equally important is that a number of outstanding analysis issues have been
 addressed that improve the reliability and reduce the biases of the current 
Union SN Ia 
compilation, and should stand us in good stead for future compilations.  
  We are making the ingredients and results of the Union 
compilation available at the associated web site\footnote{http://supernova.lbl.gov/Union} 
and we 
intend 
to provide occasional updates to this as new information becomes available.

Several conclusions can be drawn from the new larger SCP Union 
SN Ia compilation that could not be approached with smaller datasets. In particular the large statistics can be used to  address systematic
uncertainties in novel ways.

We test for evolution by subdividing the sample into low-stretch 
and high-stretch SNe. According to recent evidence 
\citep{sullivan06}
these two samples might be dominated by different progenitor systems \citep{sb05,mannucci06}, which 
are likely to show different evolution. Hence 
performing consistent but independent cosmology fits for 
the two sub-samples provides a powerful test for potential 
evolutionary effects. The resulting cosmological fit parameters are found to 
be consistent. This comparison is particularly meaningful, as the 
statistical uncertainties from the subsamples are comparable 
to the total ($stat+sys$) uncertainties obtained from the full sample.

With the larger Union dataset, it is possible to begin to examine 
the rate of true outliers from the Hubble-plot fit. It appears that the 
current selection criteria for SNe Ia can find very homogeneous sets of 
supernovae, but
 not perfectly homogeneous sets. With these criteria, there are apparently 
true outliers, at the percent level for the SNLS sample and up to 10\% for 
other samples. 
The analysis performed here was made robust 
to outliers, reducing the associated  error on 
cosmological parameters to a level 
comparable to other sources of systematic error. 

Compilations 
offer the chance to test for observer dependent systematic effects, i.e. 
tension between the datasets. The blind analysis performed here is an 
  important element in rigorous estimation of systematics. While in general we find a high degree of 
consistency 
between samples,  we see modest tension when comparing the
slope of the Hubble-residuals as a function of redshift, $d\mu/dz$. 
For the present compilation, our cosmology results are expected to hold within the quoted systematic uncertainties. 
However, once  
the homogeneous datasets get larger---and the
 systematic errors dominate over the statistical ones for the different sets---such tests will  
become even more important, as they allow one to perform cross-checks with different datasets 
calibrated in different ways. Future data samples can  be added to the Union set, by first  
blinding the data and then performing a 
diagnostic analysis similar to the one performed here. Only after 
any inconsistencies can be resolved, would the new data be unblinded.

We proposed a 
scheme to incorporate both sample dependent and common 
systematic errors. 
We showed in Section 
\ref{sec:systematics} that systematic errors can 
be approached by treating the systematics as a normal distribution 
of a parameterized systematic term. We find that the combination of SNe constraints with 
CMB constraints, due to their larger complementarity with SNe data, 
results in smaller systematic errors 
than the combination with BAO constraints. Adding  BAO, CMB and SNe 
constraints leads to yet 
smaller statistical error bars, while the error bars including systematics do not improve.  

The robustness of the detection of the accelerating expansion of the 
Universe is continually increasing as improved systematics analysis 
is reinforced by larger SN data sets.  The current knowledge of the nature 
of dark energy is still modest, however, with the uncertainty on the 
assumed-constant equation of state only under 10\% {\it if\/} multiple 
probes are combined.
The current ``world'' estimate presented here employing the full set of current
 SN data, plus other measurements,  gives a best constraint 
of  $w=-0.969^{+0.059}_{-0.063}{\rm (stat)}^{+0.063}_{-0.066} {\rm (sys)}$ on a constant EOS parameter $w$ 
at 68.3\% confidence level.  
However, allowing for time variation in the dark energy equation of state 
further opens the possibilities for the physics driving the acceleration,
 consistent with all current observations.  In particular, present SN data 
sets do not have the sensitivity to answer the questions of whether dark 
energy persists to $z>1$, or whether it had negative pressure then. 

On the positive side, with the more sophisticated analyses and tests 
carried out here, we still have encountered no limits to the potential 
use of future, high accuracy SN data as cosmological probes.  New 
data sets for nearby, moderate, and high redshift well-characterized 
SNe Ia are forthcoming and we expect realistic, robust constraints to 
catch up with our optimistic hopes on understanding the accelerating 
Universe. 

\acknowledgements
This work is based on
observations made with:
the Lick and Keck Observatories;
the Cerro Tololo Inter-American Observatory 4-m Blanco Telescope;
the Yale/AURA/Lisbon/OSU (YALO) 1-m Telescope at Cerro Tololo Inter-American Observatory;
the Apache Point Observatory 3.5-meter telescope, which is owned and operated by the Astrophysical
Research Consortium;
the WIYN Observatory,  owned and operated by the WIYN Consortium, which consists of the University of Wisconsin, Indiana University, Yale University, and the National Optical Astronomy Observatory (NOAO);
the Isaac Newton Telescope, which  is operated on the island of La Palma by the Isaac Newton Group in the Spanish Observatorio del Roque de los Muchachos of the Instituto de Astrof'sica de Canarias;
the Nordic Optical Telescope, operated on the
island of La Palma jointly by Denmark, Finland, Iceland, Norway, and
Sweden, in the Spanish Observatorio del Roque de los Muchachos of the
Instituto de Astrofisica de Canarias;
and
the MDM Observatory 2.4-m Hiltner Telescope.
The authors wish to thank the telescope
allocation committees and the observatory staffs
for their support for the extensive supernova search campaign and followup observations that  contributed to the results reported here.
In particular, we wish to thank C. Bailyn and S. Tourtellotte for assistance with YALO observations,  D. Harner for obtaining WIYN data, and D. Folha and S. Smartt for the INT 2.5-m service observing.
For their efforts in the coordinated supernova search, we wish to acknowledge the NEAT search team (E. Helin, S. Pravdo, D. Rabinowitz,and K. Lawrence) at JPL   and the Spacewatch program at the University of Arizona (which includes R. S. McMillan, T. Gehrels, J.A. Larsen, J. L. Montani, J. V. Scotti, N. Danzl, and A. Gleason).
We also wish to thank B. Schmidt, A. Filippenko, M. Schwartz, A. Gal-Yam, D. Maoz for providing us with early announcements of supernova candidates.

This work was supported in
part by the Director, Office of Science, Office of High Energy and
Nuclear Physics,  U.S. Department of Energy,
through contract DE-AC02-05CH11231.
This research used
resources of the National Energy Research Scientific Computing Center, which is supported by the Office of Science of the U.S. Department of Energy under Contract No. DE-AC02-05CH11231.
The use of Portuguese time for  the YALO telescope was supported by Funda\c{c}\~ao para a Ci\^encia e Tecnologia, Portugal and by Project  PESO/ESO/P/PRO/1257/98.

M.K. acknowledges support from the Deutsche Forschungsgemeinschaft (DFG).
P.E.N. acknowledges support from the US Department of Energy Scientific Discovery through Advanced Computing program under contract DE-FG02-06ER06-04.
A.M.M. acknowledges financial support
from Funda\c{c}\~ao para a Ci\^encia e Tecnologia (FCT), Portugal,
through project PESO/P/PRO/15139/99.

\bibliography{references}
\clearpage
\begin{appendix}
\section{Instruments and color terms}
\label{sec:instruments}

\begin{table}[htbp]
\begin{center}
\begin{small}
\begin{tabular}{ l c c c c}
\tableline
\tableline
Telescope \& Instrument & $c_{\rm b}^{\rm bv}$& $c_{\rm v}^{\rm bv}$& $c_{\rm r}^{\rm vr}$& $c_{\rm i}^{\rm vi}$ \\
\tableline
CTIO 1.5m SITE2K6 &  -0.095 (0.003) &   0.029 (0.001) &   0.028 (0.002) &   0.018 (0.001) \\
CTIO 1.5m TK1 &  -0.017 (0.002) &   0.037 (0.001) &   0.026 (0.002) &   0.015 (0.001) \\
CTIO 0.9m TK2 &  -0.097 (0.005) &   0.016 (0.002) &   0.006 (0.004) &   0.022 (0.002) \\
DANISH DFOSC &   0.133 (0.002) &   0.033 (0.001) &   0.067 (0.001) &  -0.000 (0.3001) \\
JKT Tek1 &   0.055 (0.006) &   0.020 (0.004) &   0.030 (0.008) &   0.053 (0.007) \\
LICK 1m DEWAR2 &   0.094 (0.007) &  -0.026 (0.007) &  -0.052 (0.007) &   0.018 (0.004) \\
LICK 1m DEWAR5 &   0.226 (0.033) &  -0.073 (0.016) &  -0.106 (0.007) &   0.034 (0.002) \\
YALO ANDICAM &   0.094 (0.002) &  -0.035 (0.001) &   0.373 (0.002) &  -0.041 (0.001) \\
ESO 3.6m EFOSC &   0.048 (0.005) &   0.048 (0.002) &   0.048 (0.002) &  -0.010 (0.001) \\
KPNO 2.1m T1KA &   0.103 (0.006) &  -0.021 (0.002) &   0.029 (0.007) &   0.023 (0.002) \\
CFHT STIS2 &   0.105 (0.009) &   0.002 (0.009) &  -0.079 (0.005) &   0.043 (0.002) \\
WIYN S2KB &   0.059 (0.019) &  -0.002 (0.019) &  -0.018 (0.035) &   0.016 (0.015) \\
MARLY  &  - &  - &  -0.296 (0.003) &  -0.017 (0.002) \\
\tableline
\end{tabular}
\end{small}
\caption{Color term for instruments and bands.\label{tb:beta_term}
}
\end{center}
\end{table}

\begin{table}[htbp]
\begin{center}
\begin{small}
\begin{tabular}{l c c c c}
\tableline
Instrument & $\Delta \lambda_{\rm B}$(\AA) & $\Delta \lambda_{\rm V}$(\AA) & $\Delta \lambda_{\rm R}$(\AA) & $\Delta \lambda_{\rm I}$(\AA) \\
\tableline
CTIO 1.5m SITE2K6&         -20&           0&         -30&          80\\
CTIO 1.5m SITE2K6 (small)&         -20&           0&         -30&          90\\
CTIO 1.5m TK1&          10&         -10&         -40&          90\\
CTIO 0.9m TK2&         -10&         -10&         -10&         -10\\
DANISH DFOSC&         100&          30&          30&          50\\
JKT Tek1&           0&         -20&         -50&          60\\
LICK 1 m DEWAR5 (small)&          90&         -10&         -30&         -40\\
LICK 1m DEWAR2&          90&          10&         -10&           0\\
LICK 1m DEWAR5&         100&         -30&         -40&           0\\
YALO ANDICAM&          10&         -20&        -340&         -50\\
ESO 3.6m EFOSC&          20&         -20&           0&          50\\
KPNO 2.1m T1KA&         -30&         -30&         -40&         -60\\
CFHT STIS2&         -70&         -10&        -100&         -50\\
\tableline
\end{tabular}
\end{small}
\caption{The applied shifts in Angstroms of the pass-bands, which are needed to reproduce 
the colors of the observed standard and field stars.\label{tb:shift}}  
\end{center}
\end{table}


\section{Tertiary standard star catalog}
\label{sec:field_stars}
\LongTables
\def\arraystretch{1}
\begin{deluxetable*}{ c c c c c c c }
\tablecaption{Coordinates and magnitudes of  tertiary calibration stars. \label{tb:fieldstars}}
\tablehead{
\colhead{$\#$} & 
\colhead{Ra} & 
\colhead{Dec} & 
\colhead{V} & 
\colhead{B-V} & 
\colhead{V-R} & 
\colhead{V-I}
}
\startdata
 \multicolumn{7}{c}{SN 1999aa} \\
1 & 08:27:37.43 & 21:31:18.8 & 14.495(0.003) & 0.5145(0.006) & 0.3005(0.005) & 0.6109(0.006)\\
2 & 08:27:41.27 & 21:25:01.9 & 14.978(0.006) & 0.6927(0.013) & 0.4188(0.004) & 0.2622(0.008)\\
3 & 08:28:02.64 & 21:33:56.1 & 14.817(0.005) & 0.9347(0.014) & 0.4969(0.004) & 0.1772(0.012)\\
4 & 08:27:38.16 & 21:29:54.5 & 15.115(0.004) & 0.6558(0.008) & 0.4143(0.004) & 0.7879(0.006)\\
5 & 08:27:47.99 & 21:33:01.1 & 15.550(0.007) & 0.5243(0.010) & 0.3240(0.005) & 0.6872(0.007)\\
6 & 08:27:44.64 & 21:31:11.6 & 15.382(0.004) & 0.7130(0.010) & 0.4010(0.004) & 0.7809(0.007)\\
7 & 08:27:48.47 & 21:33:20.1 & 15.575(0.008) & 0.5614(0.016) & 0.3434(0.006) & 0.7287(0.008)\\
\enddata
\end{deluxetable*}


\section{Lightcurves from the Nearby Supernova Campaign}
\begin{deluxetable*}{cccccc}
\tablewidth{0pt}
\tablecaption{B,V,R,I magnitudes, uncertainties are given in brackets. \label{tb:nb99_ltcv}}
\tablehead{
\colhead{JD} & 
\colhead{Telescope} & 
\colhead{B} &
\colhead{V} & 
\colhead{R} & 
\colhead{I}
}
\startdata
\multicolumn{6}{c}{ SN 1999aa} \\ 
 221.81&LICK 1m DEWAR2&  15.828~(   0.032) &   - &   - &   -\\
  222.67&YALO  &  15.642~(   0.018) &  15.680~(   0.028) &  15.689~(   0.040) &  15.717~(   0.105)\\
  223.67&YALO  &  15.462~(   0.020) &   - &  15.486~(   0.036) &  15.514~(   0.083)\\
  225.65&YALO  &  15.211~(   0.017) &  15.260~(   0.028) &  15.276~(   0.030) &  15.312~(   0.025)\\
  227.73&LICK 1m DEWAR2&  15.006~(   0.016) &  15.060~(   0.024) &  15.080~(   0.018) &   -\\
  229.62&YALO  &  14.924~(   0.017) &  14.965~(   0.026) &  15.092~(   0.030) &   -\\
  232.61&YALO  &  14.908~(   0.017) &  14.913~(   0.028) &  15.062~(   0.030) &  15.253~(   0.025)\\
\enddata
\end{deluxetable*}

\begin{deluxetable*}{lccccccc}
\tablewidth{0pt}
\tablecaption{SNe of the Union compilation}
\tablehead{
\colhead{Name} & 
\colhead{$z$} & 
\colhead{$\mbmax$} & 
\colhead{$s$} & 
\colhead{$c$} & 
\colhead{$\mu$} & 
\colhead{ref.} & 
\colhead{cut}
}
\startdata
1993ag & 0.0500 & 17.79(0.05) & 0.91(0.02) & 0.09(0.02) & 36.77(0.15) & 1 & \\ 
1993o & 0.0529 & 17.61(0.05) & 0.90(0.01) & -0.01(0.02) & 36.82(0.15) & 1 & \\ 
1993h & 0.0251 & 16.74(0.09) & 0.68(0.01) & 0.21(0.01) & 35.17(0.17) & 1 & \\ 
1993b & 0.0701 & 18.38(0.09) & 0.99(0.03) & 0.04(0.04) & 37.57(0.15) & 1 & \\ 
1992bs & 0.0627 & 18.18(0.05) & 1.00(0.02) & -0.03(0.02) & 37.55(0.15) & 1 & \\ 
1992br & 0.0876 & 19.40(0.11) & 0.65(0.04) & 0.03(0.05) & 38.19(0.16) & 1 & \\ 
1992bp & 0.0786 & 18.28(0.03) & 0.87(0.02) & -0.04(0.02) & 37.52(0.15) & 1 & \\ 
1992bo & 0.0172 & 15.75(0.13) & 0.74(0.01) & 0.03(0.01) & 34.65(0.19) & 1 & \\
\enddata
\tablecomments{Explanation of cuts: o: $3\sigma$ outlier, p: insufficient 
early data, c: no color, d: too few data points, f: fit not converged}
\tablerefs{$^1$Hamuy et al. (1996), $^2$Krisciunas et al. (2005), $^3$Riess et al. (1996), 
$^4$Jha et al. (2006), $^5$this work, $^6$Riess et al. (1998) + HZT,  $^7$Perlmutter et al. (1999), 
$^8$Tonry et al. (2003), $^9$Barris et al. (2003), $^10$Knop et al. (2003), $^11$Riess et al. (2007), 
$^12$Astier et al. (2006), $^13$Miknaitis et al. (2007).}
\end{deluxetable*}
\newpage
\begin{deluxetable}{ccccccc}
\tablewidth{0pt}
\tablecaption{The photometric lightcurve data from the 42 SNe of \citet{perl99}.\label{p99_ltcv}}
\tablecomments{$^1$The observations were made in the Bessell $R$ and $I$ band and the data is presented as flux, with a common zero-point of $zp=30$ magnitudes. 
Multiple data points for a given night were combined into a single data point. 
Since reference images where used to subtract the host galaxy light, the data points are correlated. We recommend using the  
data from the web-link ``http://supernova.lbl.gov/Union'', which includes the covariance matrix.
More information about the SNe can be found in \citet{perl99}.}

\tablehead{
\colhead{} & 
\colhead{$R$} &
\colhead{} & 
\colhead{} & 
\colhead{} & 
\colhead{$I$} & 
\colhead{} \\
\colhead{MJD} & 
\colhead{flux$^1$} &
\colhead{$\sigma({\rm flux})$} &
\colhead{} & 
\colhead{MJD} &
\colhead{flux$^1$} &
\colhead{$\sigma({\rm flux})$} \\
\colhead{} & 
\colhead{($zp=30$~mag)} &
\colhead{} &
\colhead{} & 
\colhead{} & 
\colhead{($zp=30$~mag)} &
\colhead{} 
}
\startdata
\multicolumn{7}{c}{ SN 1997s ($z=0.612 $)}\\ 
50431.85 & 9.89e+01 & 5.71e+01&& 50459.04 & 5.84e+02 & 2.20e+02 \\ 
 50432.82 & 1.11e+02 & 5.85e+01&& 50465.96 & 3.68e+02 & 3.64e+02 \\ 
 50454.82 & 6.49e+02 & 4.05e+01&& 50466.92 & 7.34e+02 & 2.29e+02 \\ 
 50459.01 & 5.49e+02 & 4.47e+01&& 50480.86 & 4.28e+02 & 9.23e+01 \\ 
 50462.82 & 6.82e+02 & 4.05e+01&& 50489.79 & 9.54e+02 & 1.16e+02 \\ 
 50465.82 & 5.92e+02 & 7.19e+01&& 50513.78 & 4.54e+02 & 1.85e+02 \\ 
 50480.83 & 2.18e+02 & 3.04e+01&& 50514.77 & 6.36e+02 & 1.06e+02 \\
\enddata
\end{deluxetable}

\end{appendix}


\end{document}